\documentclass[twocolumn]{aastex631}
\usepackage[caption=false]{subfig}
\usepackage{multirow}
\usepackage{amsmath}

\newcommand{\HI}{H$\,${\sc i}}

\newcommand{\HISD}{$\Sigma_{\scriptsize{\textnormal{H}\,\textnormal{\sc{i}}}}$}
\DeclareMathSymbol{:}{\mathord}{operators}{"3A}

\received{December 20, 2024}
\revised{March 4, 2025}
\accepted{ }

\submitjournal{}

\shortauthors{Archer et al.}

\begin{document}

\title{Stellar Populations and Molecular Gas Composition in the Low-Metallicity Environment of WLM}

\correspondingauthor{Haylee N. Archer}
\email{harcher@lowell.edu}

\author[0000-0002-8449-4815]{Haylee N. Archer}
\affiliation{School of Earth and Space Exploration, Arizona State University, Tempe, AZ 85287, USA}
\affiliation{Lowell Observatory, 1400 W Mars Hill Rd, Flagstaff, AZ 86001, USA}

\author[0000-0002-3322-9798]{Deidre A. Hunter}
\affiliation{Lowell Observatory, 1400 W Mars Hill Rd, Flagstaff, AZ 86001, USA}

\author[0000-0002-1723-6330]{Bruce G.\ Elmegreen}
\affiliation{Katonah, NY 10536, USA}

\author[0000-0001-9162-2371]{Leslie K. Hunt}
\affiliation{INAF, Osservatorio Astrofisico di Arcetri, Largo E Fermi 5, 50125 Firenze Italy}

\author[0000-0003-3351-0878]{Rosalia O'Brien}
\affiliation{School of Earth and Space Exploration, Arizona State University, Tempe, AZ 85287, USA}

\author[0000-0002-7758-9699]{Elias Brinks}
\affiliation{Centre for Astrophysics Research, University of Hertfordshire, College Lane, Hatfield AL10 9AB, UK}

\author[0000-0002-8736-2463]{Phil Cigan}
\affiliation{United States Naval Observatory, 3450 Massachusetts Ave NW, Washington, DC 20392, USA}

\author[0000-0002-5307-5941]{Monica Rubio}
\affiliation{Departamento de Astronom\'{i}a, Universidad de Chile, Casilla 36-D, Santiago, 8320000, Chile}

\author[0000-0001-8156-6281]{Rogier A. Windhorst}
\affiliation{School of Earth and Space Exploration, Arizona State University, Tempe, AZ 85287, USA}

\author[0000-0003-1268-5230]{Rolf A. Jansen}
\affiliation{School of Earth and Space Exploration, Arizona State University, Tempe, AZ 85287, USA}

\author[0000-0003-0384-0681]{Elijah P. Mathews}
\affiliation{Department of Astronomy \& Astrophysics, The Pennsylvania State University, University Park, PA 16802, USA}
\affiliation{Institute for Computation \& Data Sciences, The Pennsylvania State University, University Park, PA 16802, USA}
\affiliation{Institute for Gravitation and the Cosmos, The Pennsylvania State University, University Park, PA 16802, USA}



\begin{abstract}
We investigate the stellar populations and molecular gas properties of a star-forming region within the dwarf irregular (dIrr) galaxy WLM. Low-metallicity dIrrs like WLM offer a valuable window into star formation in environments that are unlike those of larger, metal-rich galaxies such as the Milky Way. In these conditions, carbon monoxide (CO), typically used to trace molecular clouds, is more easily photodissociated by ultraviolet (UV) radiation, leading to a larger fraction of CO-dark molecular gas, where H$_2$ exists without detectable CO emission, or CO-dark gas in the form of cold H$\,$\textsc{i}. Understanding the molecular gas content and the stellar populations in these star-forming regions provides important information about the role of CO-bright and CO-dark gas in forming stars.

Using \textit{HST} imaging across five WFC3/UVIS bands and CO observations from the Atacama Large Millimeter Array (ALMA), we examine stellar populations within and outside CO cores and the photodissociation region (PDR). Our findings indicate similar physical characteristics such as age and mass across the different environments. Assuming 2\% of molecular gas is converted to stars, we estimate the molecular gas content and determine that CO-dark gas constitutes a large fraction of the molecular reservoir in WLM. These results are consistent with molecular gas estimates using a previous dust-derived CO-to-H$_2$ conversion factor ($\alpha_{CO})$ for WLM. These findings highlight the critical role of CO-dark gas in low-metallicity star formation.

\end{abstract}



\section{Introduction} \label{sec:intro}
The study of star formation in low-metallicity dwarf galaxies provides valuable insights into the star-forming environments of the most numerous galaxy type in the universe. Wolf–Lundmark–Melotte (WLM) is a Local Group dwarf irregular (dIrr) galaxy located at a distance of approximately 980 kiloparsecs \citep[kpc,][]{Leaman_2012,Albers_2019,Lee_2021,Newman_2024}. With a total stellar mass of $1.62 \times 10^7\ M_\odot$ \citep{Zhang_2012} and a metallicity of 12 + $\log$(O/H) = 7.8 \citep[13\%~$Z_\odot$, ][]{Lee_2005}, WLM is characterized by low mass and low metallicity. The galaxy’s isolation, with large separations from both the Milky Way and M31, implies a low likelihood of past interactions with these systems \citep{Teyssier_2012,Albers_2019}. This combination of low mass, low metallicity, distance, and isolation makes WLM an ideal laboratory for studying star formation in undisturbed dwarf galaxies, providing insight into star-forming processes in a metal-poor environment.

Metallicity plays a critical role in star formation processes, as metals enhance gas cooling and help shield molecular gas from dissociating radiation \citep[e.g.,][]{drain_2007,fukui_2010,wakelan_2017,osman_2020}. In low-metallicity environments, like those found in dwarf galaxies, the reduced metal content limits gas cooling efficiency and molecular cloud shielding, which can impact star formation rates, the initial mass function (IMF), and feedback mechanisms from young stars \citep[e.g.,][]{Elmegreen_1989,brosch_1998,Hunter_1998,Leroy_2008,Chevance_2020_1,Hunter_2024}. These conditions may lead to different star formation dynamics, where molecular gas cooling, cloud collapse, and star formation proceed less efficiently compared to metal-rich environments.

\begin{figure*}[bth!]
\epsscale{1.16}\plotone{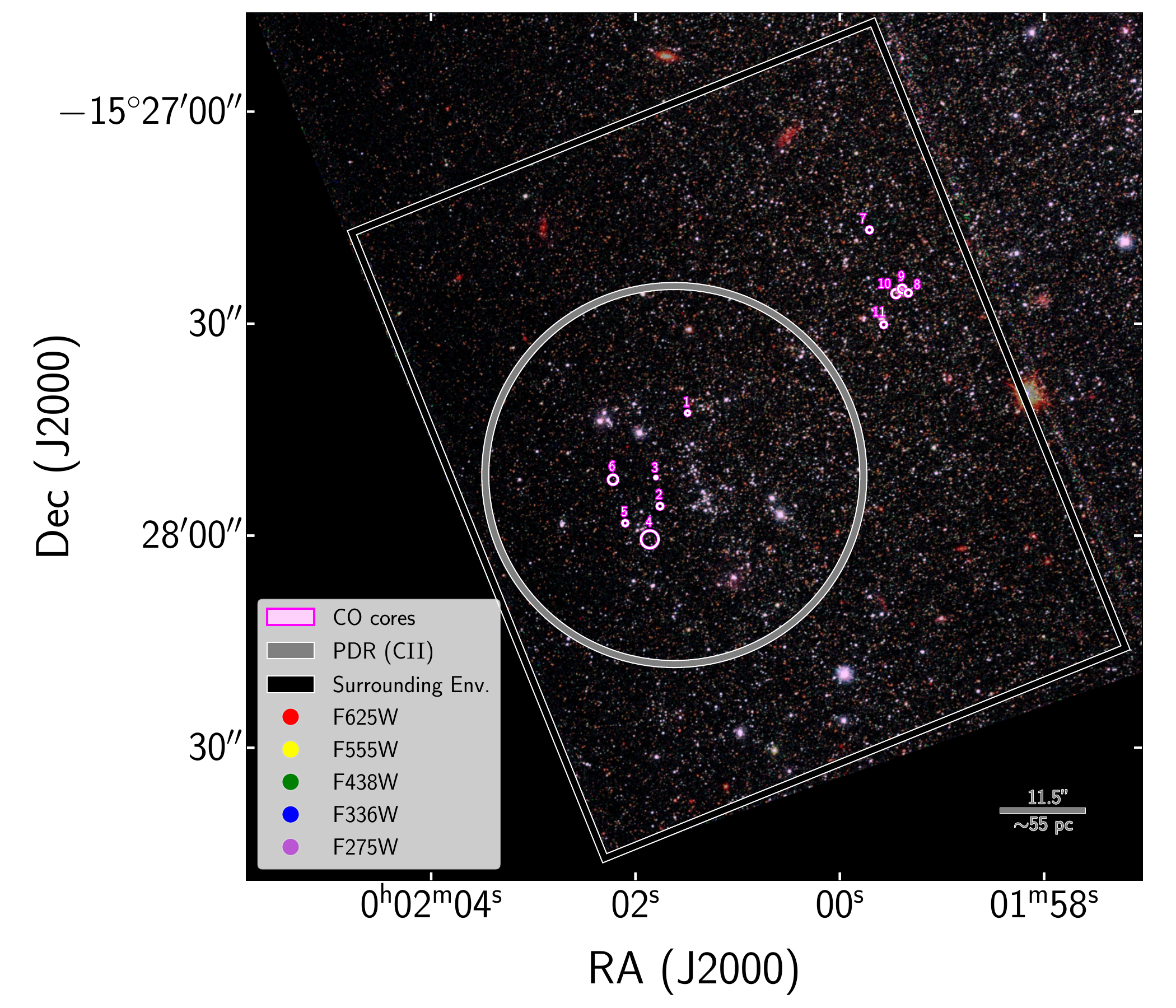}
\caption{Multicolor image combining the five \textit{HST} filters, with the outline of the PACS [C$\,${\sc ii}]-detected PDR from \citet{Cigan_2016} (large gray circle) and CO cores (smaller magenta circles) overlaid \citep{Rubio_2015}. The large black rectangle outlines the environment outside the PDR considered in this work. The legend shows the color assigned to each filter. We also include the 11\farcs5 (55 pc) PACS beam size, which is the resolution of the PDR, in the bottom right corner.
\label{fig:multicolor}}
\end{figure*}

\begin{figure}[bth!]
\epsscale{1.15}\plotone{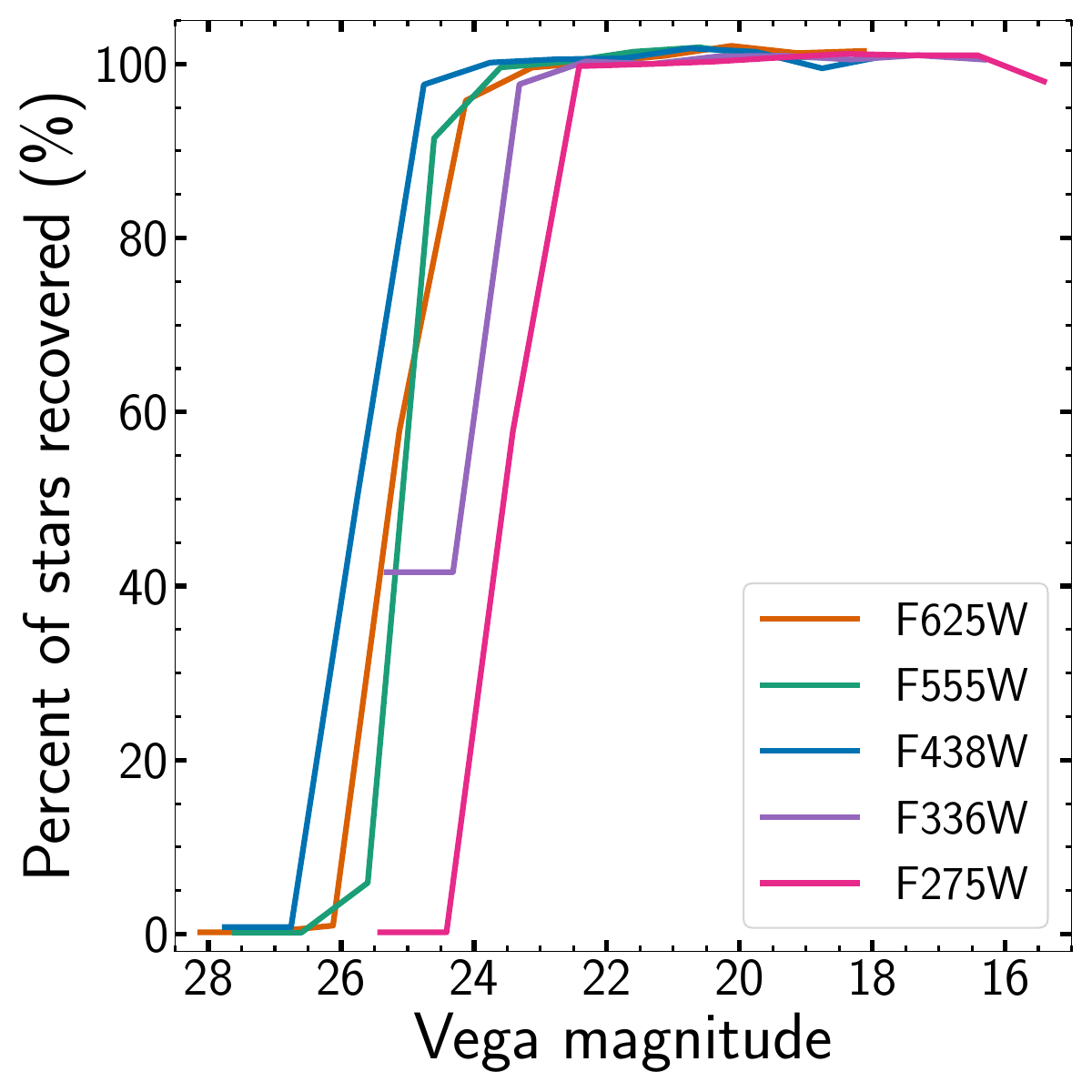}
\caption{Percent of fake stars recovered using \textsc{Daophot} as a function of Vega magnitude for the F625W (orange), F555W (green), F438W (blue), F336W (purple), and F275W (pink) \textit{HST} filters.
\label{fig:completeness}}
\end{figure}

\begin{figure*}[hbt!]
\captionsetup[subfloat]{farskip=-5pt}
\centering 
\subfloat{\includegraphics[width=0.33\linewidth]{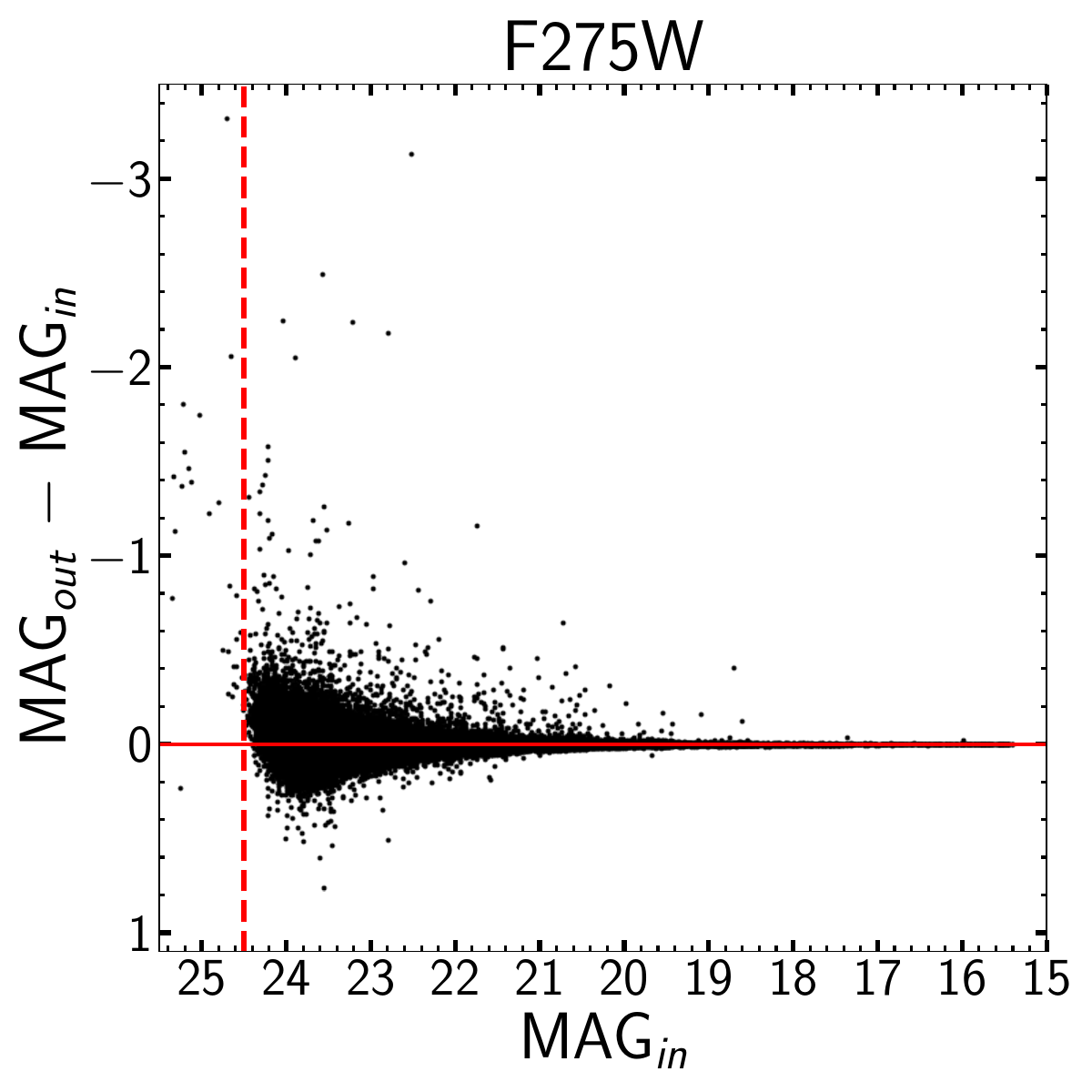}}\hspace{-0.69em}
\subfloat{\includegraphics[width=0.33\linewidth]{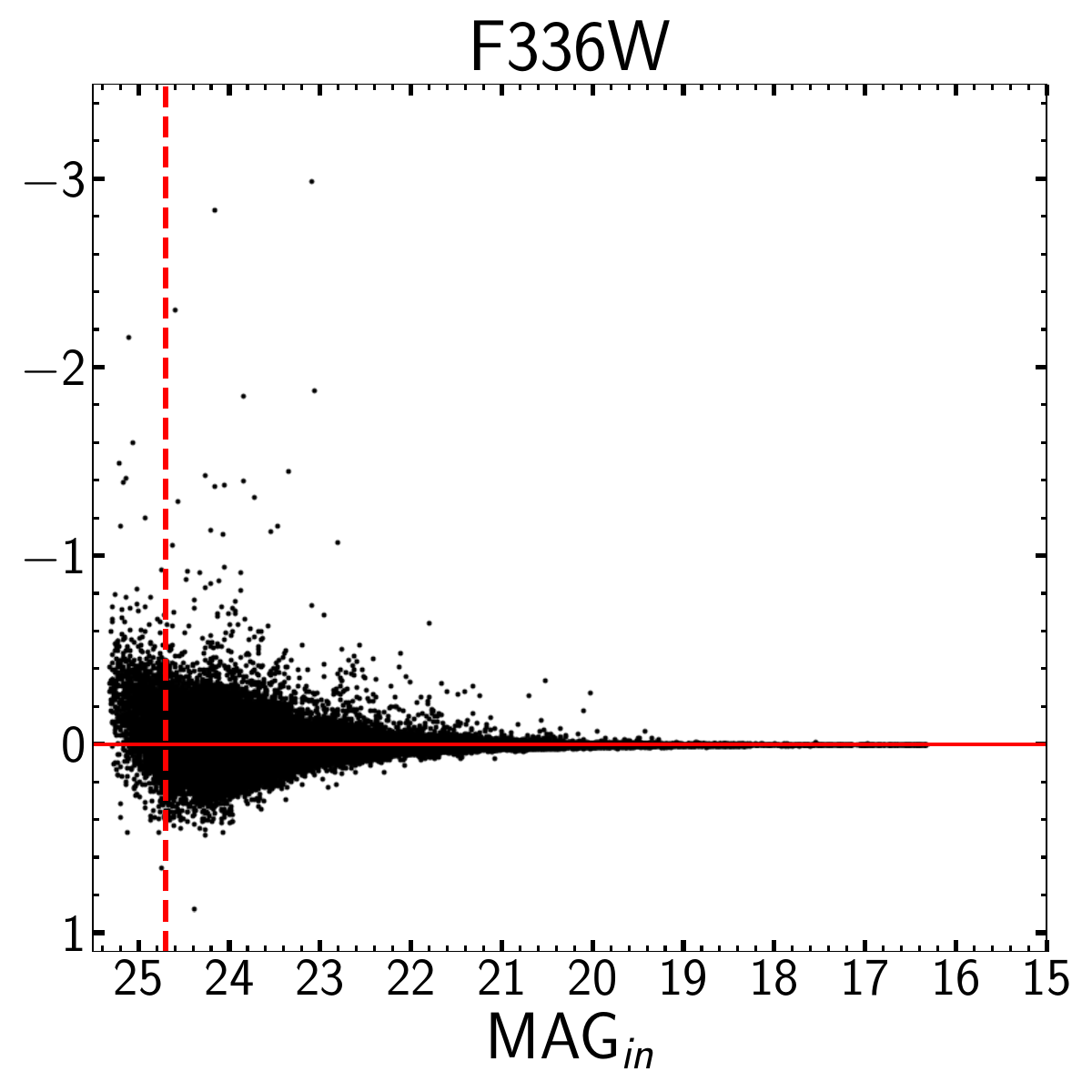}}

\subfloat{\includegraphics[width=0.33\linewidth]{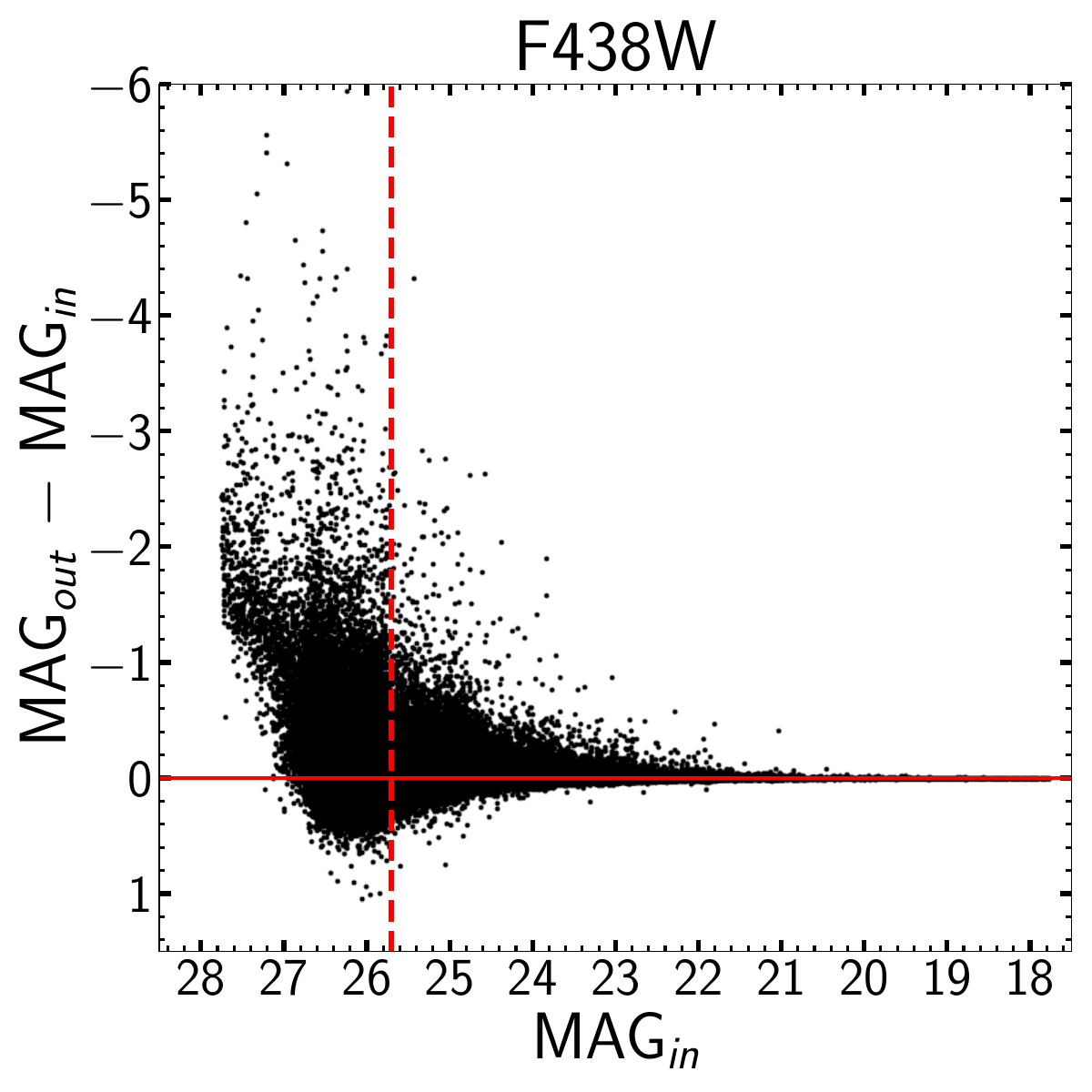}}\hspace{-0.67em}
\subfloat{\includegraphics[width=0.33\linewidth]{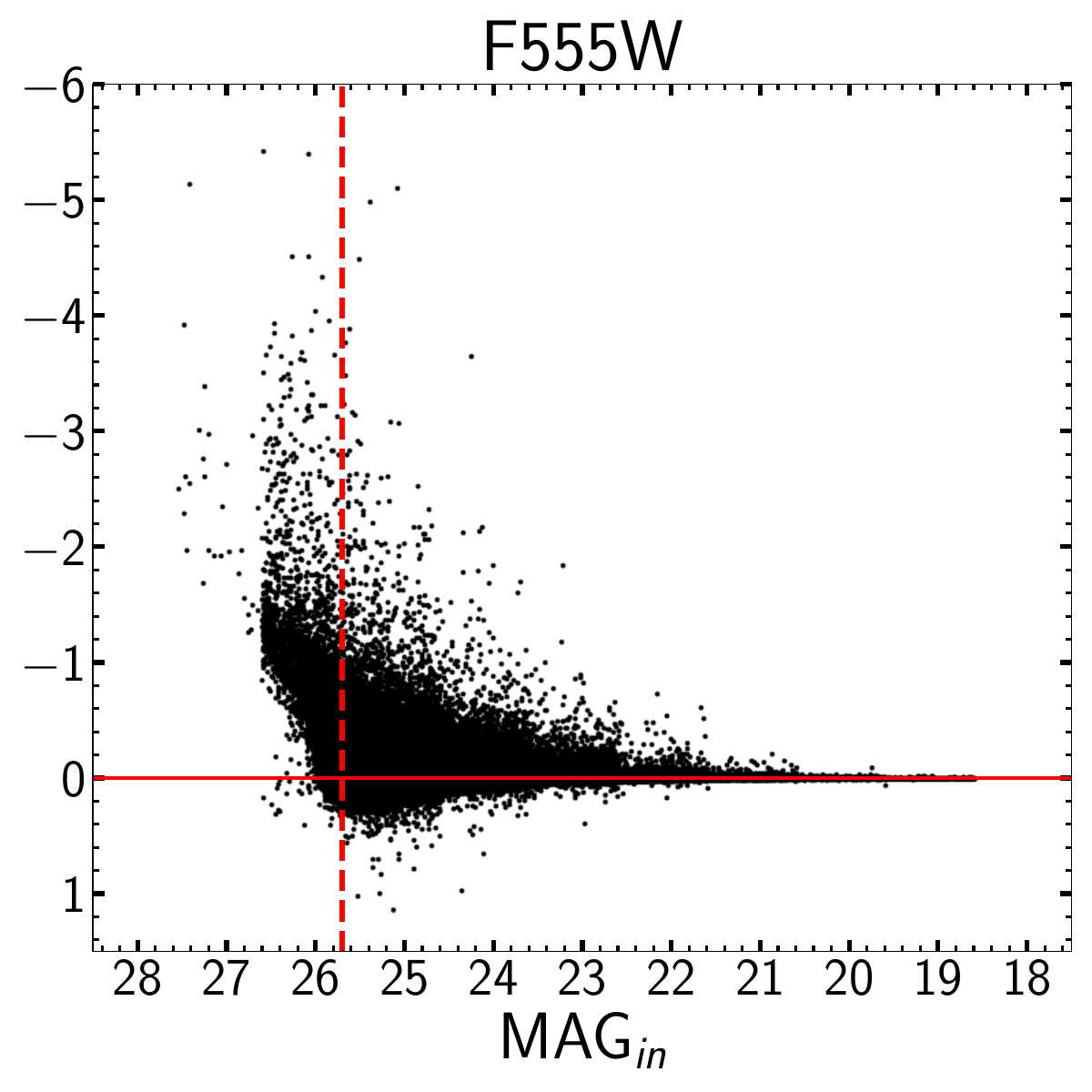}}\hspace{-0.67em}
\subfloat{\includegraphics[width=0.33\linewidth]{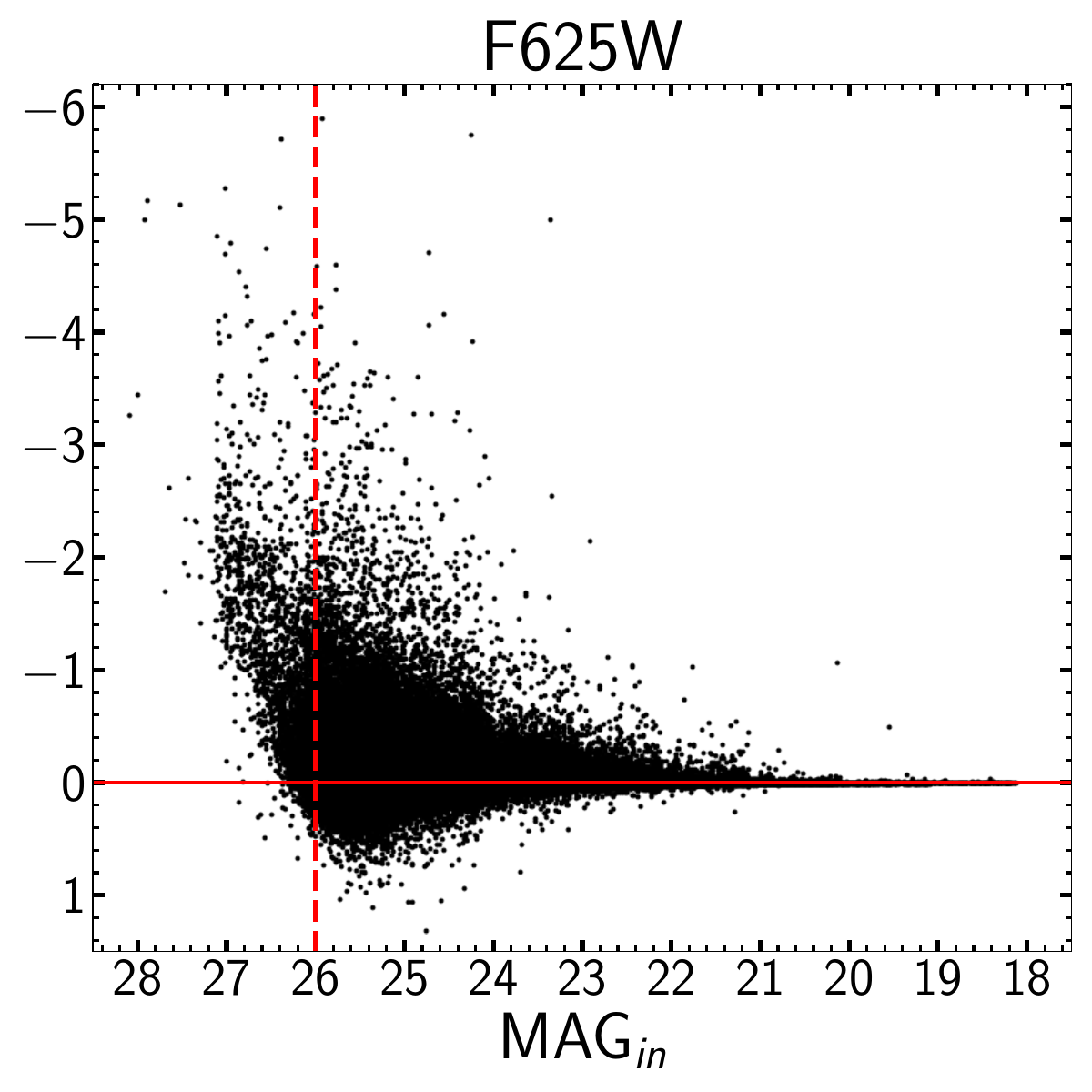}}
\caption{The difference between output and input magnitudes as a function of input magnitude for all artificial stars generated in the artificial star tests for each filter. The solid red horizontal line represents where input and output magnitudes are identical, while the dashed red vertical line marks the faintest magnitude in that filter observed in the final matched catalog of stars.
\label{fig:maginmagout}}
\end{figure*}

\begin{figure*}[bht!]
\captionsetup[subfloat]{farskip=-5pt}
\centering 
\subfloat{\includegraphics[width=0.48\linewidth]{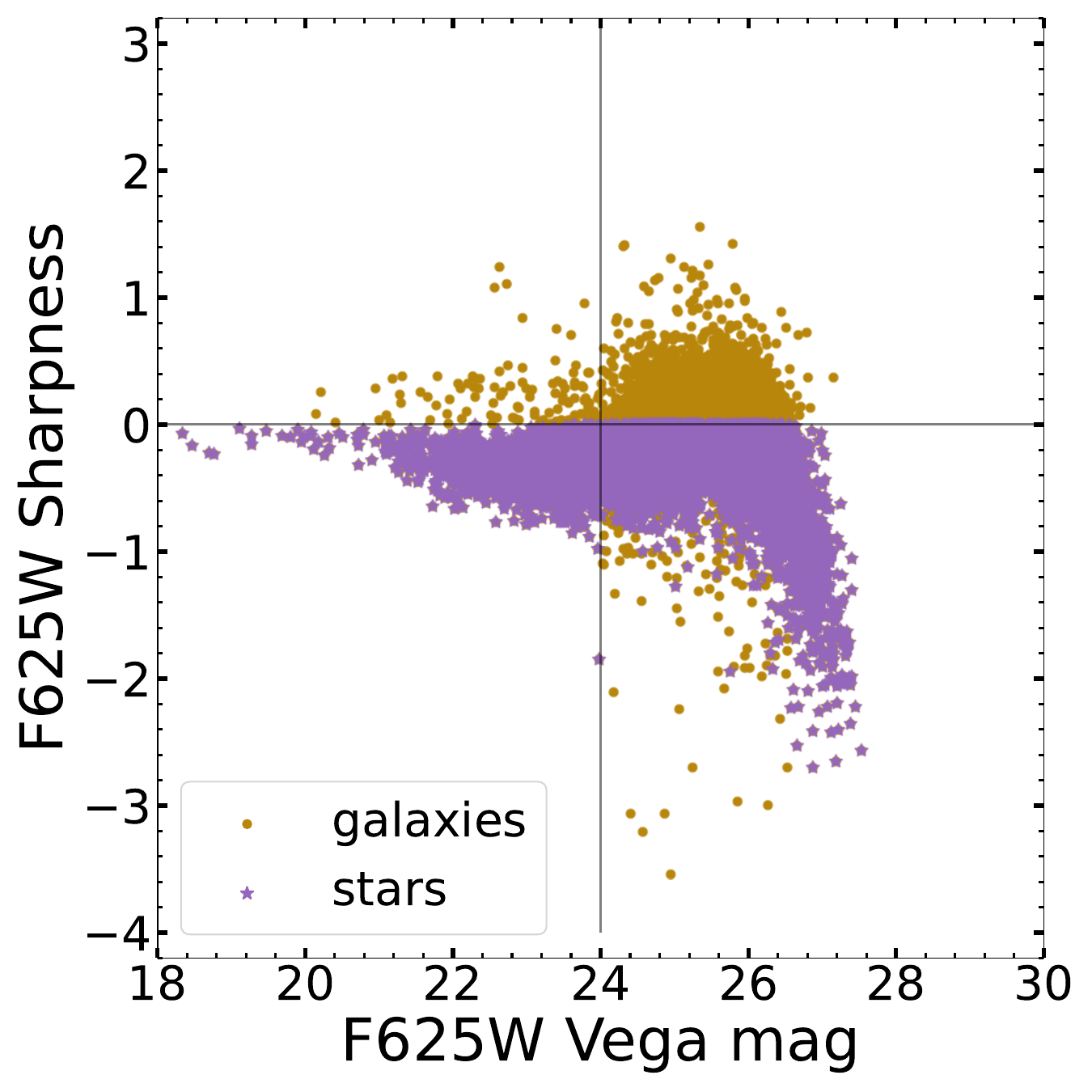}}
\vspace{-1.1em}
\subfloat{\includegraphics[width=0.48\linewidth]{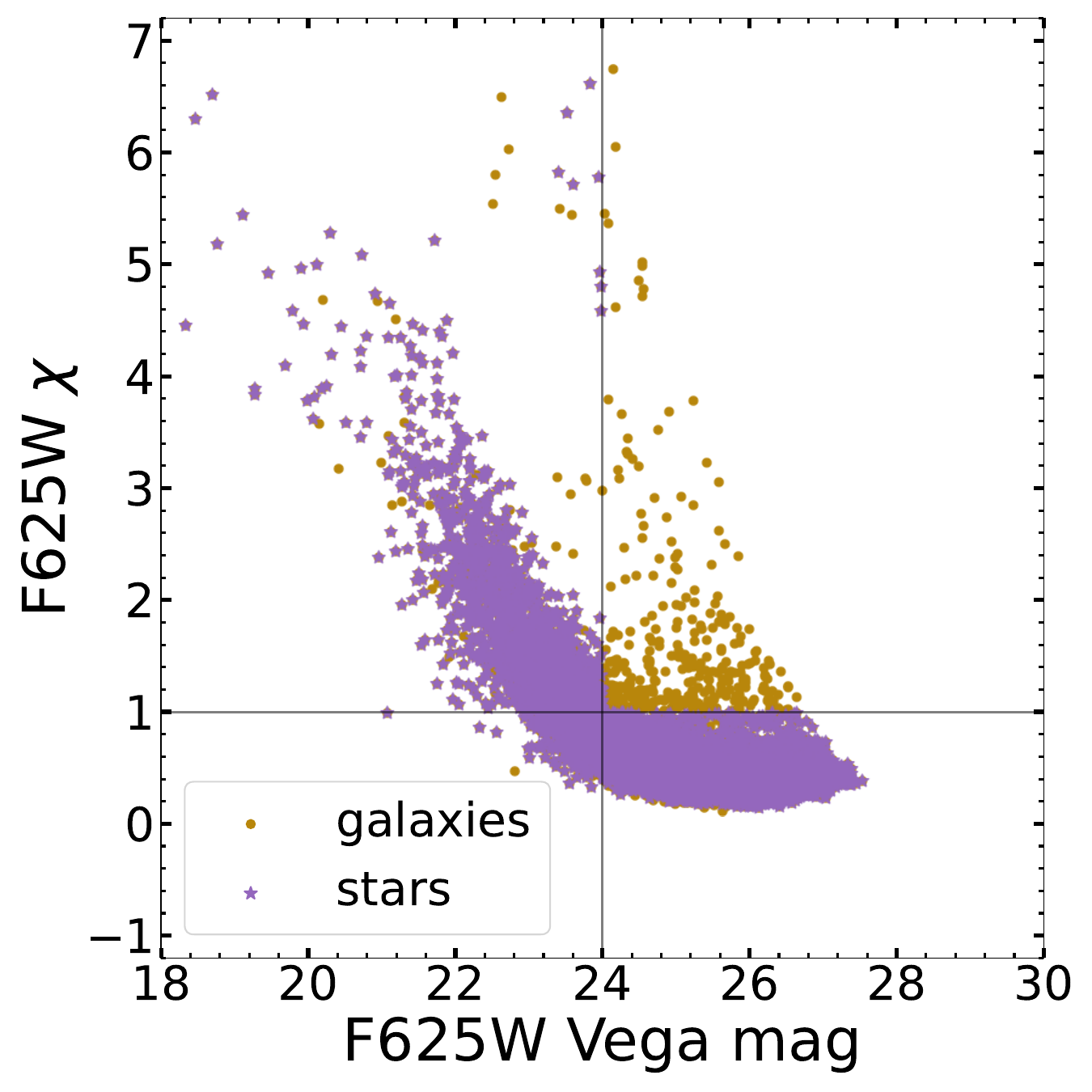}}
\caption{Sharpness (left) and $\chi$ (right) values as a function of Vega magnitude for all sources detected in the F625W filter. Gray vertical and horizontal lines are shown to demarcate which sources were stars or galaxies. Sources with a Vega magnitude brighter than 24 were determined to be stars if their sharpness was less than zero (bottom left quadrant in the sharpness plot), while sources with a Vega magnitude fainter than 24 were determined to be stars if both their sharpness was less than zero and their $\chi$ was less than one (bottom right quadrant in both plots). Sources determined to be galaxies are shown as gold points, while sources determined to be stars are shown as purple stars.
\label{fig:sharp_chi}}
\end{figure*}

One of the main challenges in studying molecular gas in low-metallicity environments is the detection of molecular hydrogen (H$_2$), the primary fuel for star formation. Unlike in higher-metallicity galaxies, where carbon monoxide (CO) serves as a reliable tracer for H$_2$, low-metallicity systems exhibit lower CO abundances due to a lack of shielding against photodissociating ultraviolet (UV) radiation \citep{Elmegreen_1980,Elmegreen_1989,Taylor_1998}. This results in a large fraction of CO-dark molecular gas, where H$_2$ is present without detectable CO emission \citep{Wolfire_2010}. Consequently, accurate assessment of molecular gas content in these environments requires alternative approaches, such as dust-based methods or [C$\,$\textsc{ii}] emission, to account for the significant CO-dark gas component \citep{planck_2011,Pineda_2014,Cormier_2017,Hunt_2023}.

This CO-dark gas may contribute extensively to the star-forming material, even though it is invisible in traditional CO surveys \citep{madden_2020,madden_2022}. By studying the stellar populations in and around these regions, we can gain insight into how star formation proceeds in areas with varying molecular gas visibility and density. The characteristics of these populations~--~such as their ages, masses, and spatial distribution~--~provide valuable clues about the role of CO-bright and CO-dark gas in forming stars and how the local environment influences star formation efficiency in metal-poor galaxies.

Following the discovery of CO(3–2) emission in two star-forming regions of WLM by \citet{Elmegreen_2013} using the Atacama Pathfinder Experiment (APEX) telescope, \citet{Rubio_2015} conducted pointed CO(1–0) observations of these regions with the Atacama Large Millimeter Array (ALMA). Their work produced the first detailed map of 10 CO cores in WLM, and Rubio et al.\ (in preparation) have since mapped most of the star forming area of WLM with ALMA CO(2–1) observations and detected an additional 35 cores. Surrounding six of the original 10 detected cores, [C$\,${\sc ii}] observations traced a photodissociation region (PDR) with a width five times larger than the cluster of CO cores, suggesting that molecular clouds at lower metallicities contain [C$\,${\sc ii}] that doesn't correspond to visible CO or \HI\ and more compact CO cores compared to those observed in the Milky Way \citep{Rubio_2015,Cigan_2016}.

In this work, we focus on the region defined by the PDR--the only area in WLM with [C$\,${\sc ii}] imaging--which contains six of the CO cores identified by \citep{Rubio_2015}. Studies of other low-metallicity dwarf galaxies have shown that most of the molecular gas reservoir is not well-traced by CO(1-0) but can instead be tracked using the [C$\,${\sc ii}] 158$\mu$m line \citep[e.g.,][]{raquena_2016, madden_2020, Ramambason_2024}. This motivated our choice to use the PDR to define the star-forming region. We compare the stellar populations within that region to those in the surrounding environment, which also contains five additional CO cores detected by Rubio et al.\ (in preparation), to understand their relationship to the CO cores and the PDR.

This paper is organized as follows. In Section \ref{sec:data}, we describe our data sources and processing techniques. Section \ref{sec:results} presents the results of our photometric analysis, stellar isochrone fitting, and molecular gas assessment, while Section \ref{sec:discussion} discusses the implications of these findings for understanding star formation and molecular gas in WLM and similar galaxies. Finally, Section \ref{sec:summary} provides a summary and conclusions of our study.

\begin{deluxetable}{ccc}
\tablecaption{\textit{HST} filter wavelengths and exposure times \label{tab:filters}}
\tablehead{\colhead{\textit{HST} Filter Name} & \colhead{Effective Wavelength} & \colhead{Exp Time} \\ 
\colhead{} & \colhead{(\AA)} & \colhead{(s)} } 
\startdata
F275W & 2709.7 & 2220 \\
F336W & 3354.5 & 1230 \\
F438W & 4326.2 & 1760 \\
F555W & 5308.4 & 1125 \\
F625W & 6242.6 & 1050 \\
\enddata
\end{deluxetable}

\section{Data} \label{sec:data}
\subsection{HST GO \#17068}
We obtained near-ultraviolet (NUV) and optical images covering most of the star-forming area of WLM through the \textit{HST} GO program \#17068 \citep{archer_hst_prop}. Focusing on the star-forming region constrained by the [C$\,${\sc ii}]-detected PDR, this project acquired the WFC3/UVIS F275W, F336W, F438W, F555W, and F625W images of the region for detecting and analyzing the stellar population. The F275W and F336W ultraviolet (UV) filters were post-flashed with 20 e$^-$ to account for the charge transfer efficiency (CTE) degradation of the UVIS detector, and the calwfc3 pipeline implements the CTE-correction code of \citet{Anderson_2021}.  We include the effective wavelength and exposure times for each filter in Table~\ref{tab:filters}. The \textit{HST} images were processed to align the exposures, remove cosmic rays, subtract the background, and correct for geometric distortion using the DrizzlePac tasks TweakReg and AstroDrizzle \citep{hoffmann_2021}. We utilized the standard calibrated flc files for WFC3/UVIS, and the pixel scales were kept at their default values of 0\farcs04. Figure~\ref{fig:multicolor} shows a multicolor image combining all five WFC3/UVIS filters, with the PDR, surrounding environment, and CO cores overlaid. All \textit{HST} data can be found in MAST: \dataset[10.17909/xyhn-3z68]{http://dx.doi.org/10.17909/xyhn-3z68}.

\begin{deluxetable*}{LLLLL}
\centerwidetable
\tablecaption{SED Free Parameters \label{tab:priors}}
\tablehead{\colhead{Parameter} & \colhead{Symbol} & \colhead{Unit} & \colhead{} & \colhead{Prior Distribution} }
\startdata
\textrm{Initial mass} & \log_{10}M_{\textrm{ini}} & M_\odot & & \textrm{Kroupa}\tablenotemark{\scriptsize{a}} \textrm{ IMF prior (see Equation \ref{eqn:kroupa}}) \\
\textrm{Stellar age} & \log_{10}t & \textrm{yr} & & \textrm{Constant SFH prior (see Equation \ref{eqn:sfh})} \\
\textrm{Optical dust attenuation} & A_V & \textrm{mag} & & \textrm{Normal}\,(\mu=0.3,\sigma=1.0)\textrm{, truncated to the range}(0, 4) \\
\textrm{Luminosity distance} & \log_{10}d_\textrm{L} & \textrm{pc} &  & \textrm{Normal}\,(\mu=5.9934, \sigma=0.0132)\textrm{, truncated to the range }(5.9273, 6.0598) \\
\enddata
\tablenotetext{\scriptsize{a}}{\citet{kroupa_2002}}
\end{deluxetable*}

\subsubsection{Photometry}
Crowded field photometry was performed individually on all five \textit{HST} UVIS images using the Image Reduction and Analysis Facility (\textsc{IRAF}) \citep{Tody_1986} routine \textsc{Daophot}, derived from the \citet{stetson_1987} version. 
To determine the completeness limit for star detection in our crowded field photometry, we conducted a series of artificial star tests on a band-by-band basis using \textsc{Daophot}. First, we took the total number of stars detected in the image and divided them into magnitude bins. For each bin, we generated a set of artificial (or fake) stars with magnitudes corresponding to that bin and random positions distributed across the entire field, excluding the edges. The number of fake stars inserted in each bin was set to 10\% of the total stars originally detected in that magnitude range. These fake stars were then added to the image, and we assessed whether \textsc{Daophot} could retrieve them. This process was repeated 200 times for each of the five images, allowing us to build robust statistics on the detection efficiency at different magnitudes for the different filters. From this, we determined the percentages of stars recovered as a function of magnitude for each filter on a band-by-band basis, shown in Figure~\ref{fig:completeness}. The scatter in the artificial star tests is for each filter is shown in Figure \ref{fig:maginmagout}.

To remove background galaxies, we used the \textsc{Daophot} output parameters: \textsc{sharpness}, a goodness-of-fit statistic indicating how much broader the object's profile appears compared to the PSF, and chi ($\chi$), the ratio of observed pixel-to-pixel deviation from the profile fit to the expected noise based on Poisson and readout noise.
\citet{annunziatella_2013} found that plotting sharpness and $\chi$ against magnitude clearly separates stars and galaxies, with stars having a sharpness below zero and galaxies showing higher $\chi$ values at fainter magnitudes. Due to the overlap of stars and galaxies in sharpness and $\chi$ at fainter magnitudes, we applied different criteria for sources with Vega magnitudes brighter and fainter than 24. Sources brighter than 24 were classified as stars if their sharpness is less than zero, while sources fainter than 24 were classified as stars if both their sharpness is less than zero and $\chi$ is less than one. Although sharpness and $\chi$ were obtained for all five filters, we used F625W values for their clearer population separation. Figure~\ref{fig:sharp_chi} illustrates sharpness and $\chi$ values as a function of Vega magnitude for sources detected in the F625W filter. We do not observe a distinct separation between the populations in color and, therefore, do not use color as a criterion for star-galaxy classification. 
Using single filters to accomplish star-galaxy separation is justified based on a comparison between our Figure \ref{fig:maginmagout} and the deeper data of \citet[][panels 2-5 in their Figure 10a]{Windhorst_2011} in WFC3 and ACS filters very similar to ours. Our Figure \ref{fig:maginmagout} suggests approximate completeness limits of $\sim$24-26 mag in F275W to F625W, respectively. To the equivalent depth in the filters from the deeper images of \citet{Windhorst_2011}, the large majority 
of unresolved objects are stars, while almost all galaxies to our shallower depths will be resolved with FWHM$>$0.1--0.2\arcsec. In addition, the stellar density in our WLM fields is far higher than the star counts in the \citet{Windhorst_2011} GOODS-S field at high galactic latitude. The fraction of truly compact galaxies with FWHM$<$0.2\arcsec\ to our shallower detection limits is therefore very small. Hence, we do not need to use color for reliable star-galaxy separation.

We first created individual catalogs of stars detected in each of the five filters. To construct a combined catalog of stars detected across all five filters, we performed step-by-step matching, beginning with the UV filters (F275W and F336W), as these are expected to have the shallowest detection limits. Next, we sequentially matched this initial catalog with detections in the F438W, F555W, and F625W filters, combining results at each step. The matching process was carried out using the \textsc{KDTree.query$\textunderscore$radius} function from the \textsc{scikit-learn} Python library. A matching radius of 0\farcs018 was adopted, which was determined by measuring the positional offsets of a small sample of stars identified by eye across multiple filters. The stars in the resulting catalog were examined to ensure there were no spurious detections on diffraction spikes included in the sample. The same methodology was used to create a combined catalog of stars detected across all but the F275W filter.

\subsection{SED fitting}
To relate physical stellar properties to observed filter magnitudes and photometric uncertainties, we use the CMD 3.8\footnote{Available at \href{http://stev.oapd.inaf.it/cgi-bin/cmd}{http://stev.oapd.inaf.it/cgi-bin/cmd}.} tool, which collects the PARSEC 1.2S \citep{Bressan_2012,chen_2014,tang_2014,Chen_2015}, and COLIBRI S\_37 \citep{marigo_2017,pastorelli_2019,pastorelli_2020} stellar evolutionary tracks onto a mass/age grid, fixing stellar metallicity to $Z_\textrm{ini} = 0.0026$. For each mass/age grid point, CMD provides model fluxes for each of the HST filters used in this work. To generate model fluxes between grid points, we interpolate the model fluxes linearly in $M_\textrm{ini}$ and $\log_{10}{t}$, allowing flux to be generated for any arbitrary mass or age within the range given by CMD. For stellar masses above the maximum mass present in the grid for a given stellar age, we set the flux to $M_\textrm{Vega} = 999.99$ as we do not model stellar remnants. Finally, we apply dust attenuation using an SMC extinction curve \citep{gordon_2003} in addition to luminosity distance, as follows,

\begin{equation}
m_\textrm{predict}(\theta) = m_\textrm{interp}(M_\textrm{ini},t) + 5\log_{10}d_L + A_Vk_\lambda,
\end{equation}
where $m_\textrm{interp}$ is the flux\footnote{All magnitudes given in this work are Vega magnitudes.} predicted by the isochrone table interpolation, $d_L$ is the luminosity distance, $A_V$ is the dust attenuation, and $k_\lambda$ specifies the dust curve and varies by filter,
\begin{equation}
    k_\lambda = \begin{cases}3.625&\textrm{F275W},\\1.672&\textrm{F336W},\\1.374&\textrm{F438W},\\1.000&\textrm{F555W},\\0.801&\textrm{F625W}.\end{cases}
\end{equation}

Our four free parameters and their priors are listed in Table \ref{tab:priors}. For the initial stellar mass, we assume a \citet{kroupa_2002} IMF prior, with prior probability given as
\begin{equation}\label{eqn:kroupa}
\footnotesize{    \ln p_\textrm{M}(M_\textrm{ini}) = \begin{cases}(1-\alpha_0)\ln10\log_{10}{M_\textrm{ini}} & M_\textrm{ini} \leq M_1, \\ \Psi_2 + (1-\alpha_1)\ln10\log_{10}{M_\textrm{ini}} & M_1 < M_\textrm{ini} \leq M_2, \\ \Psi_3 + (1-\alpha_2)\ln10\log_{10}{M_\textrm{ini}} & M_\textrm{ini} > M_2, \end{cases}}
\end{equation}
where $\Psi_2 = (\alpha_1-\alpha_0)\ln10\log_{10}{M_1}$ and $\Psi_3 = Q_2 + (\alpha_2-\alpha_1)\ln10\log_{10}{M_2}$, and the $\alpha_i$ and $M_i$ values are adopted from \citet{kroupa_2002}. Additionally, we assume a uniform prior in stellar age $t$,
\begin{equation}\label{eqn:sfh}
    \ln p_{t}(\log_{10}{t}) = (\ln{10})\log_{10}{t},
\end{equation}
which is equivalent to assuming a constant SFH prior, consistent with the choice made by \citet{Gordon_2016}, who also employed Bayesian inference for SED fitting of stars in M31. It should be noted that neither of these two priors are normalized, since Markov Chain Monte Carlo (MCMC) samplers generally only require a probability function that is \emph{proportional} to the true posterior probability. For the optical dust attenuation $A_V$, we adopt the Normal distribution prior from the Prospector-$\alpha$ physical model \citep{leja_19}. Finally, we adopt a Normal distribution prior for the luminosity distance, with mean $\sim\!985\ \textrm{kpc}$ and standard deviation of $\sim\!30\ \textrm{kpc}$ to account for the varying distance estimates found in the literature \citep[e.g.][]{Leaman_2012,Albers_2019,Lee_2021,Newman_2024}, and truncated to $\pm5\sigma$.

Using the flux predicted by the interpolation scheme, we compute the likelihood of the observed fluxes $\mu_i$ and their uncertainties $\sigma_i$ for each filter $i$ given the model $\theta$ using a multivariate normal distribution,
\begin{equation}
    \ln p(\mu,\sigma\vert\theta) = -\frac{1}{2}\sum_{i}\left(\frac{m_{\textrm{predict},i}(\theta)-\mu_i}{\sigma_i}\right)^2.
\end{equation}
Finally, we compute the non-normalized posterior likelihood as
\begin{align}
\begin{split}
    \ln p(\theta\vert\mu,\sigma) = &\ln p(\mu,\sigma\vert\theta) + \ln p_M(M_\textrm{ini}) + \\ &\ln p_t(t) + \ln p_d(d_L) + \ln p_{A_V}(A_V).
\end{split}
\end{align}
We set the initial position for the sampler at the \emph{maximum a posteriori} (MAP) location, which is estimated using the Adam optimizer \citep{kingma_2017} with $\alpha = 10^{-2}$, run for $10^5$ iterations with the Optim.jl Julia package \citep{mogensen_2018}. Compared to providing a random or zero initial position vector, the MAP location helps the sampler explore the primary mode in the posterior and avoid getting stuck proposing stellar remnant solutions, which may provide a zero gradient since those solutions are fixed at $M_\textrm{Vega} = 999.99$ without varying. Once the MAP location is found, we then adapt the step size and mass matrix for the No-U-Turn sampler \citep[NUTS,][]{hoffman_2011} as implemented in the AdvancedHMC.jl package \citep{xu_2020} using the windowed adaptation scheme from Stan \citep{stan_2024}, assuming a dense mass matrix and a target acceptance rate of 80\%. We run the sampler for 4000 adaptation iterations, after which the mass matrix and step size are frozen. Finally, after adaptation, we use NUTS to draw 4000 samples from the posterior. We estimate each parameter's value as the median (50th percentile) of the parameter's marginalized posterior distribution. The associated uncertainty is quantified as half the difference between the 84th and 16th percentiles: (P$_{84}-$P$_{16}$)/2.

\begin{deluxetable}{L|CCCR@{ +/--}D}
\tablecaption{Locations, radii, and masses of the CO cores \label{tab:COinfo}}
\tablehead{\colhead{CO} & \colhead{RA} & \colhead{DEC} & \colhead{Radius} & \multicolumn{3}{c}{$M_{vir}$} \\
\colhead{Core} & \colhead{(deg)} & \colhead{(deg)} & \colhead{(pc)} & \multicolumn{3}{c}{($M_\odot$)} }
\decimals
\startdata
\phantom{1}1\tablenotemark{\scriptsize{a}} & 0.5062 & -15.462 & 1.7 & 1000 & 700 \\
\phantom{1}2\tablenotemark{\scriptsize{a}} & 0.5073 & -15.466 & <\! 1 & <\! 400 & 300 \\
\phantom{1}3\tablenotemark{\scriptsize{a}} & 0.5075 & -15.464 & 2.2 & 1100 & 700 \\
\phantom{1}4\tablenotemark{\scriptsize{a}} & 0.5078 & -15.467 & 6.0 & 10900 & 3200 \\
\phantom{1}5\tablenotemark{\scriptsize{a}} & 0.5086 & -15.466 & 2.0 & 6900 & 5400 \\
\phantom{1}6\tablenotemark{\scriptsize{a}} & 0.5092 & -15.464 & 3.4 & 1400 & 800 \\
\phantom{1}7 & 0.4988 & -15.455 & 1.9 & 2100 & 1300 \\
\phantom{1}8 & 0.4972 & -15.457 & 1.9 & 900 & 600 \\
\phantom{1}9 & 0.4975 & -15.457 & 2.5 & 3900 & 1300 \\
10 & 0.4977 & -15.457 & 3.0 & 1800 & 1000 \\
11 & 0.4983 & -15.458 & 1.7 & 1600 & 1200 \\
\enddata
\tablenotetext{a}{From Table 1 of \citet{Rubio_2015}.}
\end{deluxetable}

\begin{deluxetable*}{R|ccD@{ +/--}DD@{ +/--}DD@{ +/--}DD@{ +/--}DD@{ +/--}D}
\tablecaption{Right ascension, declination, and Vega magnitudes for the five \textit{HST} filters for stars inside the PDR and projected inside the CO cores \label{tab:inPDRinCO}}
\tablehead{\colhead{} & \colhead{RA} & \colhead{DEC} & \multicolumn{4}{c}{F275W} & \multicolumn{4}{c}{F336W} & \multicolumn{4}{c}{F438W}& \multicolumn{4}{c}{F555W} & \multicolumn{4}{c}{F625W} \\ 
\colhead{} & \colhead{(deg)} & \colhead{(deg)} & \multicolumn{4}{c}{(Vega mag)} & \multicolumn{4}{c}{(Vega mag)} & \multicolumn{4}{c}{(Vega mag)} & \multicolumn{4}{c}{(Vega mag)} & \multicolumn{4}{c}{(Vega mag)} } 
\decimals
\startdata
1 & 0.5077 & -15.467 & 23.07 & 0.07 & 23.10 & 0.05 & 24.04 & 0.04 & 24.04 & 0.04 & 23.84 & 0.06 \\
2 & 0.5079 & -15.467 & 22.86 & 0.06 & 23.23 & 0.06 & 24.26 & 0.04 & 24.38 & 0.04 & 24.30 & 0.05 \\
\enddata
\end{deluxetable*}

\begin{deluxetable*}{R|ccD@{ +/--}DD@{ +/--}DD@{ +/--}DD@{ +/--}DD@{ +/--}D}
\tablecaption{Right ascension, declination, and Vega magnitudes for the five \textit{HST} filters for sources inside the PDR and outside the CO cores \label{tab:inPDRoutCO}}
\tablehead{\colhead{} & \colhead{RA} & \colhead{DEC} & \multicolumn{4}{c}{F275W} & \multicolumn{4}{c}{F336W} & \multicolumn{4}{c}{F438W}& \multicolumn{4}{c}{F555W} & \multicolumn{4}{c}{F625W} \\ 
\colhead{} & \colhead{(deg)} & \colhead{(deg)} & \multicolumn{4}{c}{(Vega mag)} & \multicolumn{4}{c}{(Vega mag)} & \multicolumn{4}{c}{(Vega mag)} & \multicolumn{4}{c}{(Vega mag)} & \multicolumn{4}{c}{(Vega mag)} } 
\decimals
\startdata
1 & 0.5122 & -15.468 & 23.09 & 0.08 & 23.43 & 0.07 & 24.03 & 0.05 & 23.99 & 0.05 & 23.95 & 0.06 \\
2 & 0.5131 & -15.465 & 23.66 & 0.07 & 23.35 & 0.06 & 23.42 & 0.05 & 22.68 & 0.05 & 21.91 & 0.05 \\
3 & 0.5130 & -15.464 & 20.43 & 0.04 & 20.86 & 0.04 & 22.29 & 0.05 & 22.45 & 0.04 & 22.25 & 0.05 \\
4 & 0.5113 & -15.466 & 20.99 & 0.04 & 21.44 & 0.05 & 22.89 & 0.04 & 23.04 & 0.05 & 23.06 & 0.07 \\
5 & 0.5109 & -15.467 & 21.70 & 0.04 & 22.04 & 0.05 & 22.63 & 0.04 & 22.71 & 0.04 & 22.61 & 0.06 \\
6 & 0.5095 & -15.471 & 22.24 & 0.05 & 22.65 & 0.05 & 23.66 & 0.04 & 23.86 & 0.04 & 23.74 & 0.06 \\
7 & 0.5097 & -15.470 & 22.60 & 0.05 & 22.90 & 0.05 & 24.08 & 0.06 & 24.14 & 0.05 & 24.00 & 0.05 \\
8 & 0.5121 & -15.461 & 22.93 & 0.06 & 22.26 & 0.05 & 21.72 & 0.04 & 21.55 & 0.04 & 21.26 & 0.06 \\
9 & 0.5111 & -15.463 & 21.46 & 0.05 & 21.35 & 0.05 & 21.44 & 0.06 & 21.6 & 0.05 & 21.32 & 0.06 \\
10 & 0.5100 & -15.466 & 22.28 & 0.05 & 22.60 & 0.06 & 23.82 & 0.05 & 24.02 & 0.05 & 23.83 & 0.06 \\
\enddata
\tablecomments{Only a portion of this table is shown here to demonstrate its form and content. A machine-readable version is available for all 443 sources detected inside the PDR and outside the CO cores.}
\end{deluxetable*}

\begin{deluxetable*}{R|ccD@{ +/--}DD@{ +/--}DD@{ +/--}DD@{ +/--}DD@{ +/--}D}
\tablecaption{Right ascension, declination, and Vega magnitudes for the five \textit{HST} filters for sources outside the PDR and projected inside the CO cores \label{tab:outPDRinCO}}
\tablehead{\colhead{} & \colhead{RA} & \colhead{DEC} & \multicolumn{4}{c}{F275W} & \multicolumn{4}{c}{F336W} & \multicolumn{4}{c}{F438W}& \multicolumn{4}{c}{F555W} & \multicolumn{4}{c}{F625W} \\ 
\colhead{} & \colhead{(deg)} & \colhead{(deg)} & \multicolumn{4}{c}{(Vega mag)} & \multicolumn{4}{c}{(Vega mag)} & \multicolumn{4}{c}{(Vega mag)} & \multicolumn{4}{c}{(Vega mag)} & \multicolumn{4}{c}{(Vega mag)} } 
\decimals
\startdata
1 & 0.4975 & -15.457 & 21.51 & 0.05 & 21.43 & 0.04 & 22.59 & 0.04 & 22.31 & 0.07 & 22.24 & 0.04 \\
2 & 0.4976 & -15.457 & 22.88 & 0.08 & 23.38 & 0.07 & 24.45 & 0.07 & 24.24 & 0.06 & 24.13 & 0.06 \\
\enddata
\end{deluxetable*}

\begin{deluxetable*}{R|ccD@{ +/--}DD@{ +/--}DD@{ +/--}DD@{ +/--}DD@{ +/--}D}
\tablecaption{Right ascension, declination, and Vega magnitudes for the five \textit{HST} filters for sources outside the PDR and outside the CO cores \label{tab:outPDRoutCO}}
\tablehead{\colhead{} & \colhead{RA} & \colhead{DEC} & \multicolumn{4}{c}{F275W} & \multicolumn{4}{c}{F336W} & \multicolumn{4}{c}{F438W}& \multicolumn{4}{c}{F555W} & \multicolumn{4}{c}{F625W} \\ 
\colhead{} & \colhead{(deg)} & \colhead{(deg)} & \multicolumn{4}{c}{(Vega mag)} & \multicolumn{4}{c}{(Vega mag)} & \multicolumn{4}{c}{(Vega mag)} & \multicolumn{4}{c}{(Vega mag)} & \multicolumn{4}{c}{(Vega mag)} } 
\decimals
\startdata
1 & 0.5151 & -15.464 & 22.55 & 0.05 & 22.93 & 0.06 & 23.95 & 0.04 & 24.04 & 0.04 & 23.79 & 0.05 \\
2 & 0.5124 & -15.471 & 22.68 & 0.06 & 22.85 & 0.05 & 23.89 & 0.04 & 24.05 & 0.04 & 23.86 & 0.09 \\
3 & 0.5128 & -15.470 & 23.39 & 0.06 & 23.29 & 0.05 & 24.07 & 0.05 & 23.98 & 0.04 & 23.92 & 0.07 \\
4 & 0.5123 & -15.470 & 23.81 & 0.09 & 23.29 & 0.07 & 23.57 & 0.08 & 23.08 & 0.06 & 22.70 & 0.06 \\
5 & 0.5123 & -15.471 & 22.74 & 0.06 & 22.95 & 0.06 & 23.97 & 0.04 & 24.00 & 0.04 & 23.81 & 0.05 \\
6 & 0.5102 & -15.476 & 21.77 & 0.04 & 22.07 & 0.04 & 23.46 & 0.05 & 23.47 & 0.04 & 23.22 & 0.06 \\
7 & 0.5103 & -15.474 & 24.51 & 0.18 & 23.93 & 0.09 & 23.69 & 0.04 & 23.49 & 0.04 & 23.24 & 0.05 \\
8 & 0.5095 & -15.476 & 23.87 & 0.11 & 23.13 & 0.07 & 22.89 & 0.05 & 22.61 & 0.04 & 22.17 & 0.06 \\
9 & 0.5111 & -15.471 & 20.76 & 0.04 & 21.13 & 0.04 & 22.35 & 0.04 & 22.44 & 0.04 & 22.53 & 0.04 \\
10 & 0.5103 & -15.473 & 22.67 & 0.05 & 22.8 & 0.05 & 23.36 & 0.05 & 23.40 & 0.04 & 23.18 & 0.05 \\
\enddata
\tablecomments{Only a portion of this table is shown here to demonstrate its form and content. A machine-readable version is available for all 566 sources detected outside the PDR and outside the CO cores.}
\end{deluxetable*}

\subsection{CO Cores and [C$\,${\textsc{ii}}]}
In Cycle 1, \citet{Rubio_2015} used ALMA to image two star-forming regions in WLM, focusing on CO(1-0) emissions, and detected 10 CO cores. The beam size for these observations was $0\farcs9 \times 1 \farcs3$. Of the 10 detected cores, six were located in the PDR, referred to as Region~B in \citet{Elmegreen_2013}, WLM-SE region in \citep{Rubio_2015}, and Region~1 in \citet{archer_2022}, which is the primary focus of this paper. The masses and locations of these six CO cores, labeled as 1 through 6 in Figure~\ref{fig:multicolor}, can be found in Table 1 of \citet{Rubio_2015} as regions SE-1 through SE-6. An additional 35 CO cores were detected using CO(2-1) observations at 1\arcsec resolution (4.8 pc at WLM distance) with ALMA Cycle 6 (Rubio et al.\ in prep), all of which were detected outside the PDR as the survey did not include it. Five of these 35 CO cores were included when examining the environment surrounding the PDR to compare stellar populations inside the CO cores and outside the PDR to stellar populations inside the CO cores and inside the PDR. The locations, radii, and virial masses of the 11 CO cores included in this work can be found in Table~\ref{tab:COinfo}.

The [C$\,${\sc ii}] 158$\mu$m image was obtained using the PACS spectrometer aboard Herschel for LITTLE THINGS \citep{Cigan_2016}. The beam size for the PACS [C$\,${\sc ii}] was 11\farcs5 (shown in Figure~\ref{fig:multicolor}), which imaged the targeted region in WLM with a diameter of 54\arcsec, and showed [C$\,${\sc ii}] filling the entire region. We acknowledge that any clouds smaller than 11\farcs5 would be unresolved in our analysis. Additionally, since the PACS pointing was the only one available for WLM, the [C$\,${\sc ii}] may extend beyond the region defined as the PDR boundary in this study.

\begin{deluxetable*}{R|ccD@{ +/--}DD@{ +/--}DD@{ +/--}DD@{ +/--}D}
\tablecaption{Right ascension, declination, and Vega magnitudes for the four \textit{HST} filters for the star inside the PDR and projected inside the CO cores not detected in the F275W filter \label{tab:inPDRinCO_no275}}
\tablehead{\colhead{} & \colhead{RA} & \colhead{DEC} & \multicolumn{4}{c}{F336W} & \multicolumn{4}{c}{F438W}& \multicolumn{4}{c}{F555W} & \multicolumn{4}{c}{F625W} \\ 
\colhead{} & \colhead{(deg)} & \colhead{(deg)} & \multicolumn{4}{c}{(Vega mag)} & \multicolumn{4}{c}{(Vega mag)} & \multicolumn{4}{c}{(Vega mag)} & \multicolumn{4}{c}{(Vega mag)} } 
\decimals
\startdata
1 & 0.5062 & -15.462 & 24.08 & 0.10 & 25.09 & 0.06 & 25.35 & 0.07 & 24.89 & 0.08 \\
\enddata
\end{deluxetable*}

\begin{deluxetable*}{R|ccD@{ +/--}DD@{ +/--}DD@{ +/--}DD@{ +/--}D}
\tablecaption{Right ascension, declination, and Vega magnitudes for the four \textit{HST} filters for sources inside the PDR and outside the CO cores not detected in the F275W filter \label{tab:inPDRoutCO_no275}}
\tablehead{\colhead{} & \colhead{RA} & \colhead{DEC} & \multicolumn{4}{c}{F336W} & \multicolumn{4}{c}{F438W}& \multicolumn{4}{c}{F555W} & \multicolumn{4}{c}{F625W} \\ 
\colhead{} & \colhead{(deg)} & \colhead{(deg)} & \multicolumn{4}{c}{(Vega mag)} & \multicolumn{4}{c}{(Vega mag)} & \multicolumn{4}{c}{(Vega mag)} & \multicolumn{4}{c}{(Vega mag)} }
\decimals
\startdata
1 & 0.5128 & -15.470 & 23.18 & 0.05 & 23.98 & 0.05 & 23.91 & 0.04 & 23.86 & 0.07 \\
2 & 0.5123 & -15.470 & 23.18 & 0.07 & 23.48 & 0.08 & 23.01 & 0.06 & 22.65 & 0.06 \\
3 & 0.5150 & -15.462 & 22.52 & 0.05 & 23.49 & 0.04 & 23.65 & 0.04 & 23.31 & 0.05 \\
4 & 0.5111 & -15.471 & 21.01 & 0.04 & 22.26 & 0.04 & 22.37 & 0.04 & 22.47 & 0.04 \\
5 & 0.5110 & -15.471 & 19.97 & 0.07 & 21.41 & 0.04 & 21.50 & 0.04 & 21.62 & 0.06 \\
6 & 0.5101 & -15.472 & 22.79 & 0.05 & 23.81 & 0.04 & 23.83 & 0.04 & 23.80 & 0.06 \\
7 & 0.5134 & -15.462 & 23.90 & 0.09 & 23.67 & 0.04 & 23.48 & 0.04 & 23.22 & 0.07 \\
8 & 0.5095 & -15.472 & 23.26 & 0.07 & 23.10 & 0.04 & 23.05 & 0.04 & 22.88 & 0.05 \\
9 & 0.5103 & -15.466 & 23.78 & 0.09 & 23.61 & 0.04 & 23.57 & 0.04 & 23.25 & 0.05 \\
10 & 0.5113 & -15.463 & 23.89 & 0.09 & 23.36 & 0.04 & 22.47 & 0.04 & 21.95 & 0.07 \\
\enddata
\tablecomments{Only a portion of this table is shown here to demonstrate its form and content. A machine-readable version is available for all 144 sources detected inside the PDR and outside the CO cores.}
\end{deluxetable*}

\begin{deluxetable*}{R|ccD@{ +/--}DD@{ +/--}DD@{ +/--}DD@{ +/--}D}
\tablecaption{Right ascension, declination, and Vega magnitudes for the four \textit{HST} filters for sources outside the PDR and outside the CO cores not detected in the F275W filter \label{tab:outPDRoutCO_no275}}
\tablehead{\colhead{} & \colhead{RA} & \colhead{DEC} & \multicolumn{4}{c}{F336W} & \multicolumn{4}{c}{F438W}& \multicolumn{4}{c}{F555W} & \multicolumn{4}{c}{F625W} \\ 
\colhead{} & \colhead{(deg)} & \colhead{(deg)} & \multicolumn{4}{c}{(Vega mag)} & \multicolumn{4}{c}{(Vega mag)} & \multicolumn{4}{c}{(Vega mag)} & \multicolumn{4}{c}{(Vega mag)} }
\decimals
\startdata
1 & 0.5112 & -15.473 & 24.06 & 0.09 & 23.72 & 0.04 & 22.90 & 0.04 & 22.35 & 0.06 \\
2 & 0.5166 & -15.456 & 23.83 & 0.09 & 23.71 & 0.04 & 23.07 & 0.04 & 22.40 & 0.05 \\
3 & 0.5094 & -15.476 & 23.70 & 0.08 & 22.91 & 0.05 & 21.87 & 0.07 & 21.12 & 0.06 \\
4 & 0.5086 & -15.476 & 23.51 & 0.08 & 23.10 & 0.04 & 22.27 & 0.05 & 21.70 & 0.07 \\
5 & 0.5094 & -15.474 & 23.79 & 0.09 & 23.67 & 0.05 & 23.21 & 0.04 & 22.94 & 0.05 \\
6 & 0.5132 & -15.455 & 24.33 & 0.13 & 24.46 & 0.06 & 23.86 & 0.04 & 23.24 & 0.06 \\
7 & 0.5128 & -15.454 & 23.66 & 0.07 & 23.58 & 0.04 & 22.91 & 0.04 & 22.30 & 0.04 \\
8 & 0.5046 & -15.477 & 22.92 & 0.06 & 21.17 & 0.05 & 19.52 & 0.04 & 18.70 & 0.05 \\
9 & 0.5107 & -15.454 & 23.79 & 0.08 & 23.33 & 0.04 & 22.81 & 0.04 & 22.08 & 0.05 \\
10 & 0.5094 & -15.455 & 24.08 & 0.08 & 23.12 & 0.04 & 22.01 & 0.04 & 21.24 & 0.05 \\
\enddata
\tablecomments{Only a portion of this table is shown here to demonstrate its form and content. A machine-readable version is available for all 78 sources detected outside the PDR and outside the CO cores.}
\end{deluxetable*}

\begin{figure*}[p!]
\captionsetup[subfloat]{farskip=-5pt}
\centering 
\subfloat{\includegraphics[width=0.48\linewidth]{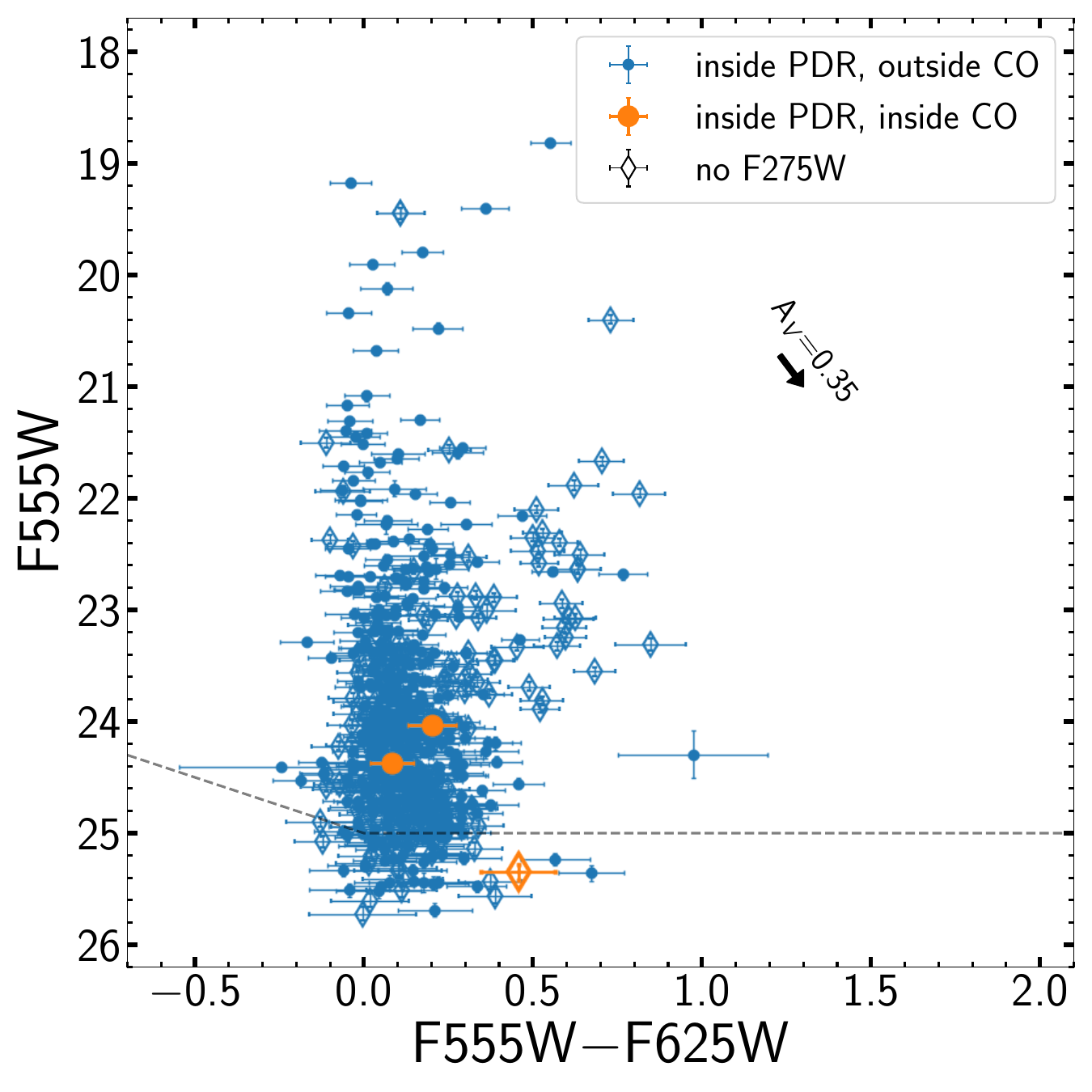}}
\subfloat{\includegraphics[width=0.48\linewidth]{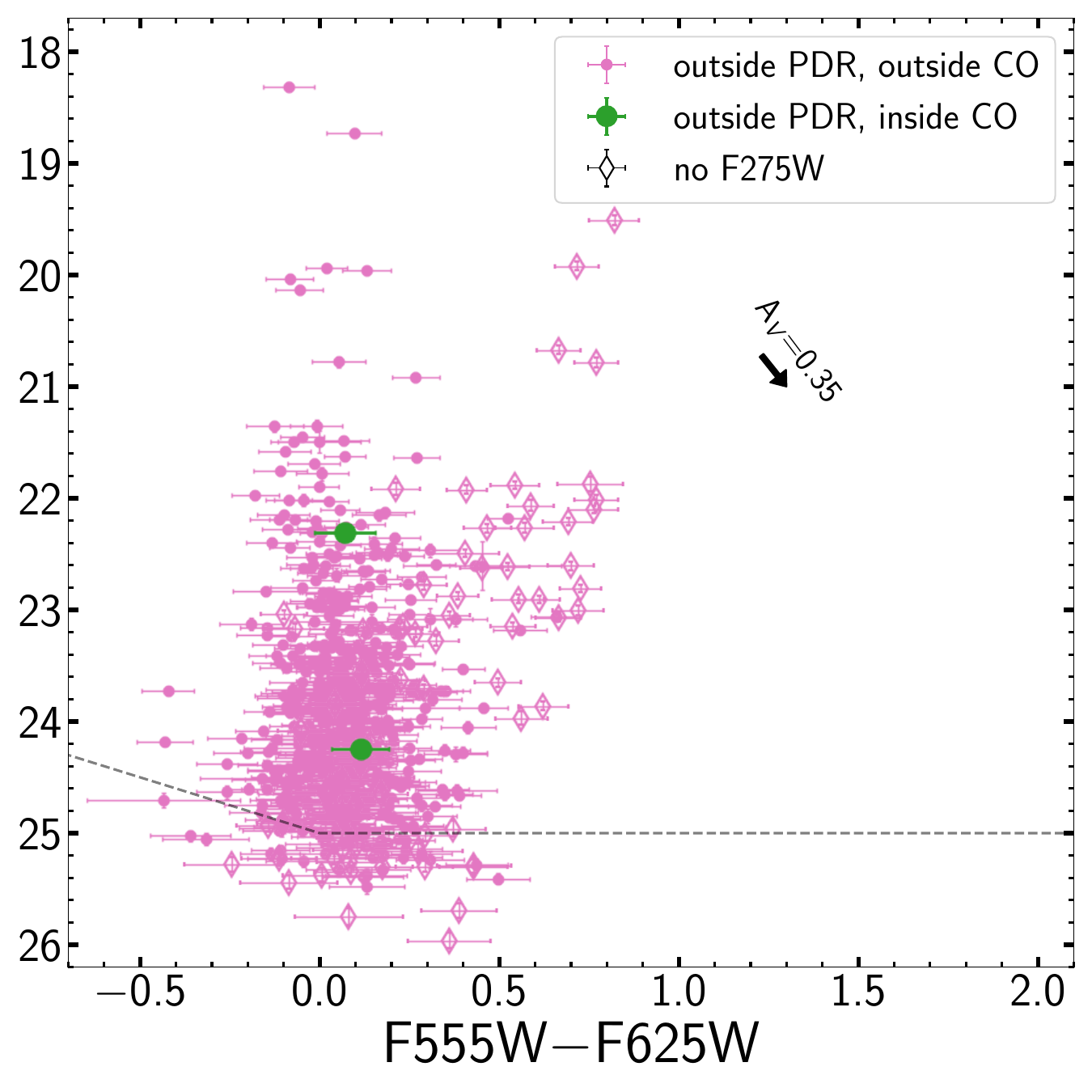}}\\
\subfloat{\includegraphics[width=0.48\linewidth]{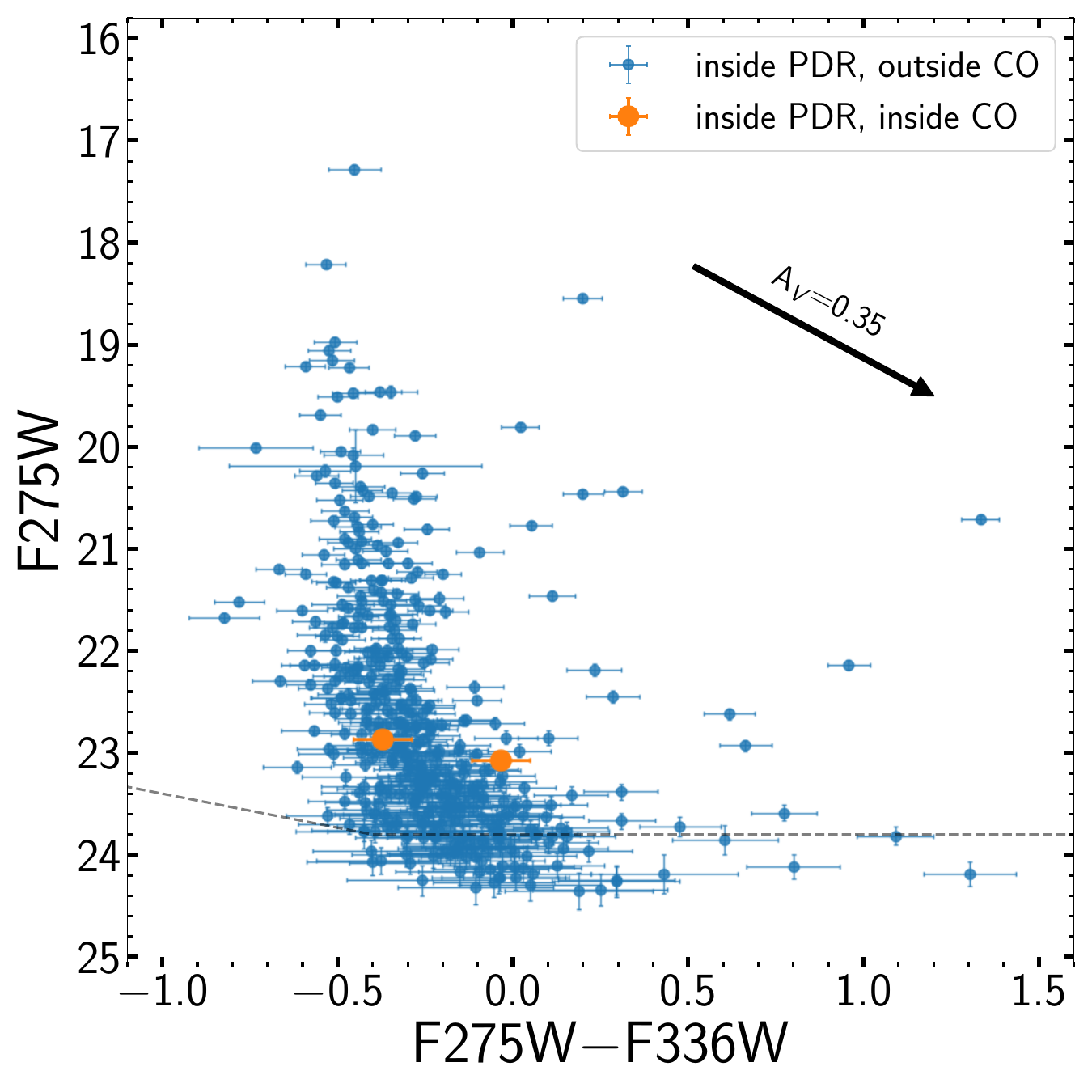}}
\subfloat{\includegraphics[width=0.48\linewidth]{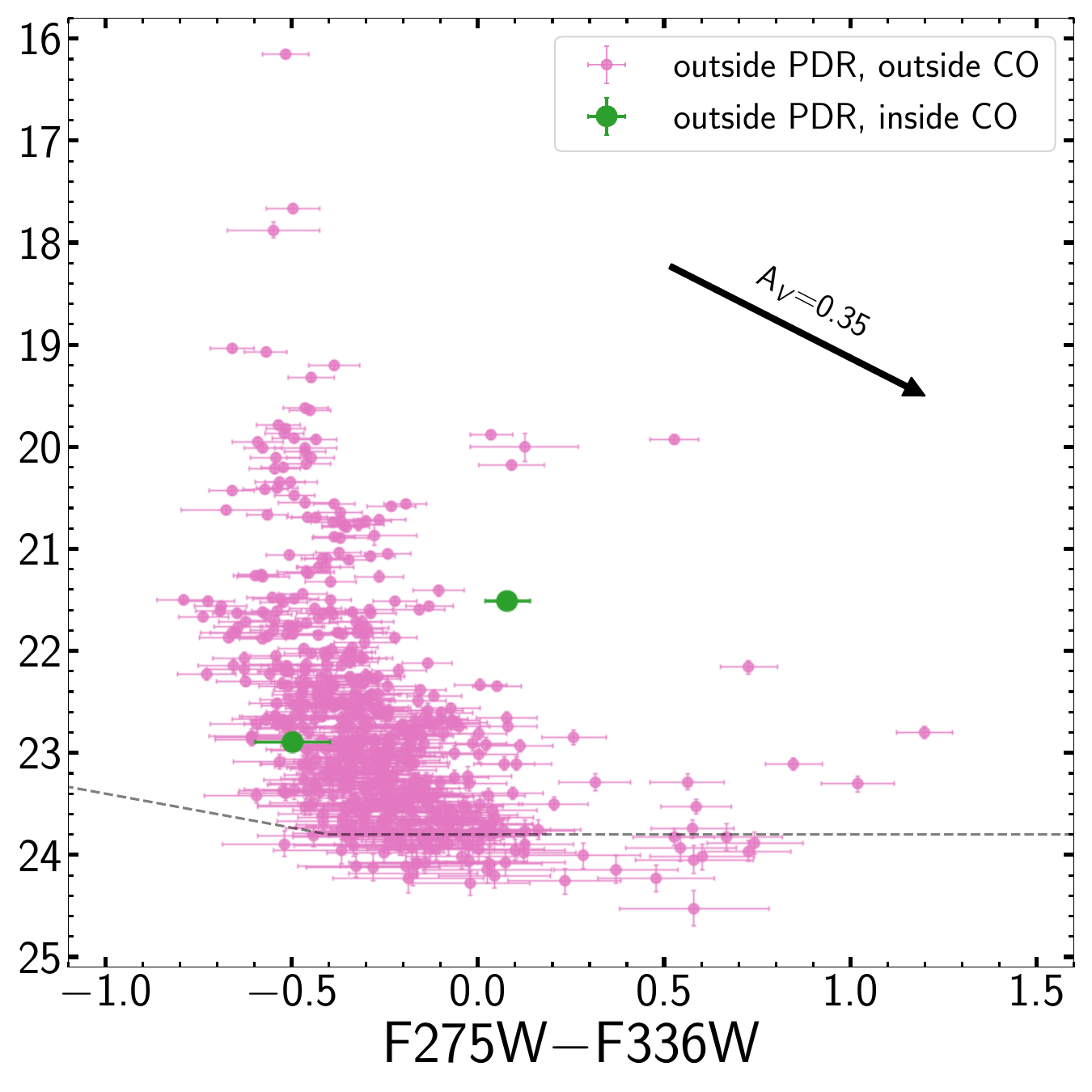}}
\caption{Top left: F555W vs F555W--F625W color-magnitude diagram for stars inside the PDR and outside the CO cores (blue) and stars inside the PDR and projected inside the CO cores (orange). Top right: F555W vs F555W--F625W color-magnitude diagram for stars outside the PDR and outside the CO cores (pink) and stars outside the PDR and projected inside the CO cores (green). Bottom left: F275W vs F275W--F6336W color-magnitude diagram for stars inside the PDR and outside the CO cores (blue) and stars inside the PDR and projected inside the CO cores (orange). Bottom right: F275W vs F275W--F336W color-magnitude diagram for stars outside the PDR and outside the CO cores (pink) and stars outside the PDR and projected inside the CO cores (green). Stars detected in all filters are represented by circles ($\circ$), while stars that were not detected in the F275W filter are represented by diamonds ($\diamondsuit$). The black arrow in each plot shows the reddening vector for $A_V$=0.35, the mean $A_V$ of stars in WLM measured by \citet{Wang_2022}, assuming SMC-like extinction. The errorbars in the upper left corner of each plot demonstrate the mean uncertainty associated with the data shown. The gray dashed line shows the $5\sigma$ point source detection limit for the given filters and exposure times \citep{Windhorst_2022}.
\label{fig:cmds}}
\end{figure*}

\begin{figure*}[htb!]
\captionsetup[subfloat]{farskip=-5pt}
\centering 
\subfloat{\includegraphics[width=0.5\linewidth]{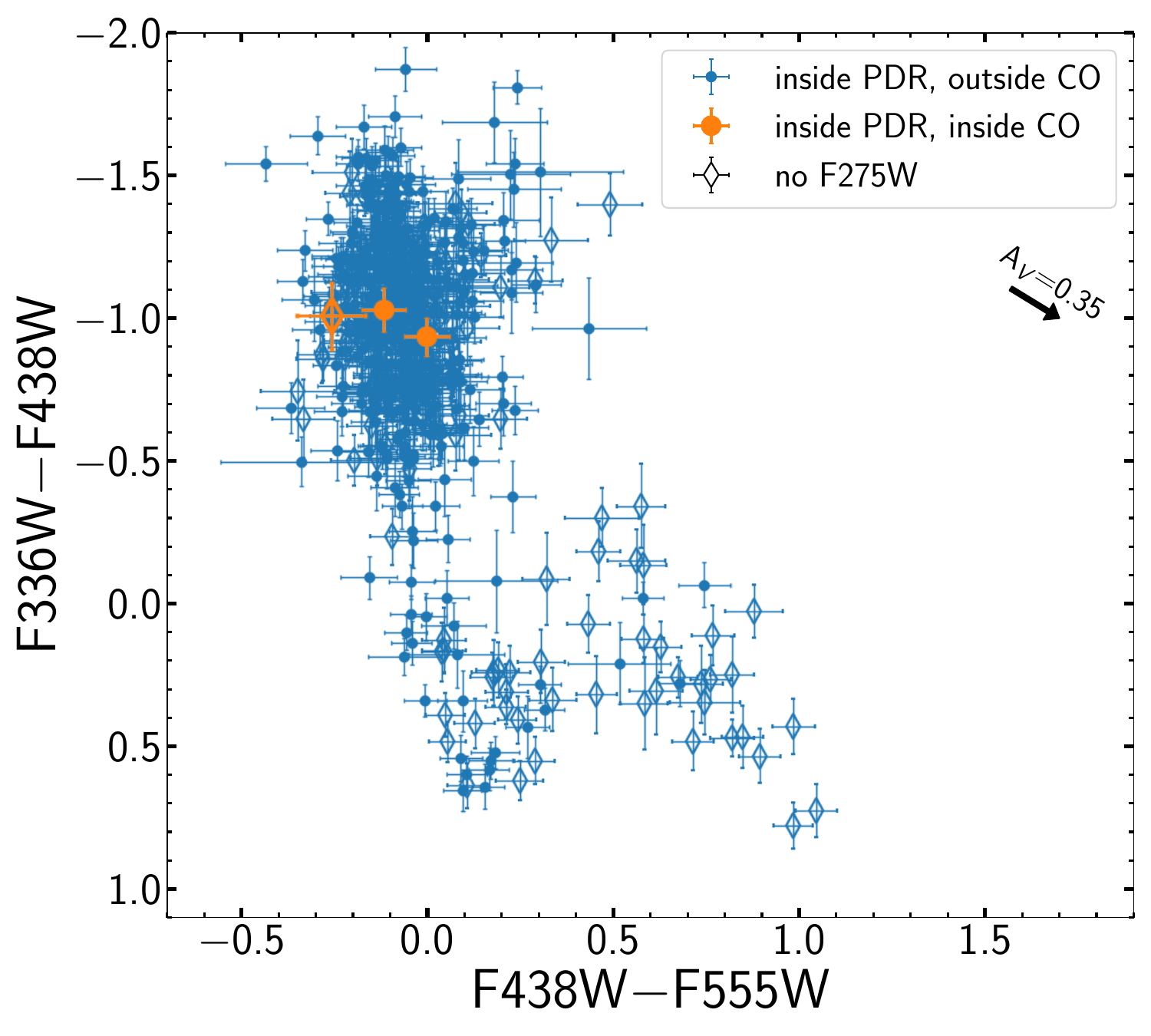}}\hspace{-0.6em}
\subfloat{\includegraphics[width=0.5\linewidth]{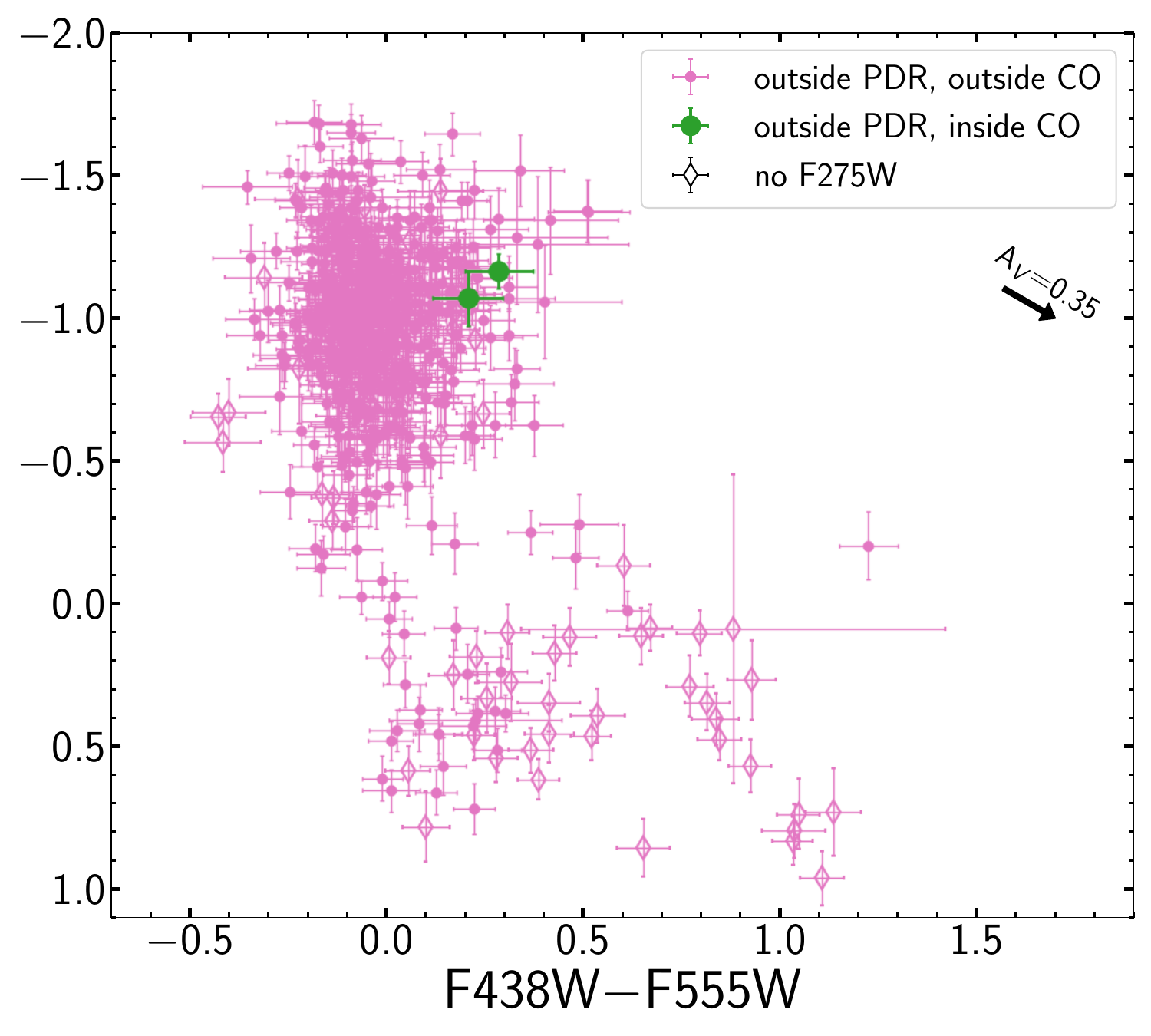}}
\caption{Left: F336W--F438W vs F438W--F555W color-color diagram for stars inside the PDR and outside the CO cores (blue) and stars inside the PDR and projected inside the CO cores (orange). Right: F336W--F438W vs F438W--F555W color-color diagram for stars outside the PDR and outside the CO cores (pink), and stars outside the PDR and projected inside the CO cores (green). Stars detected in all filters are represented by circles ($\circ$), while stars that were not detected in the F275W filter are represented by diamonds ($\diamondsuit$). The black arrow in each plot shows the reddening vector for $A_V$=0.35, the mean $A_V$ of stars in WLM measured by \citet{Wang_2022}, assuming SMC-like extinction. The errorbars in the upper left corner of each plot demonstrate the mean uncertainty associated with the data shown.
\label{fig:ccd}}
\end{figure*}

\begin{figure*}[htb!]
\captionsetup[subfloat]{farskip=-5pt}
\centering 
\subfloat{\includegraphics[width=0.5\linewidth]{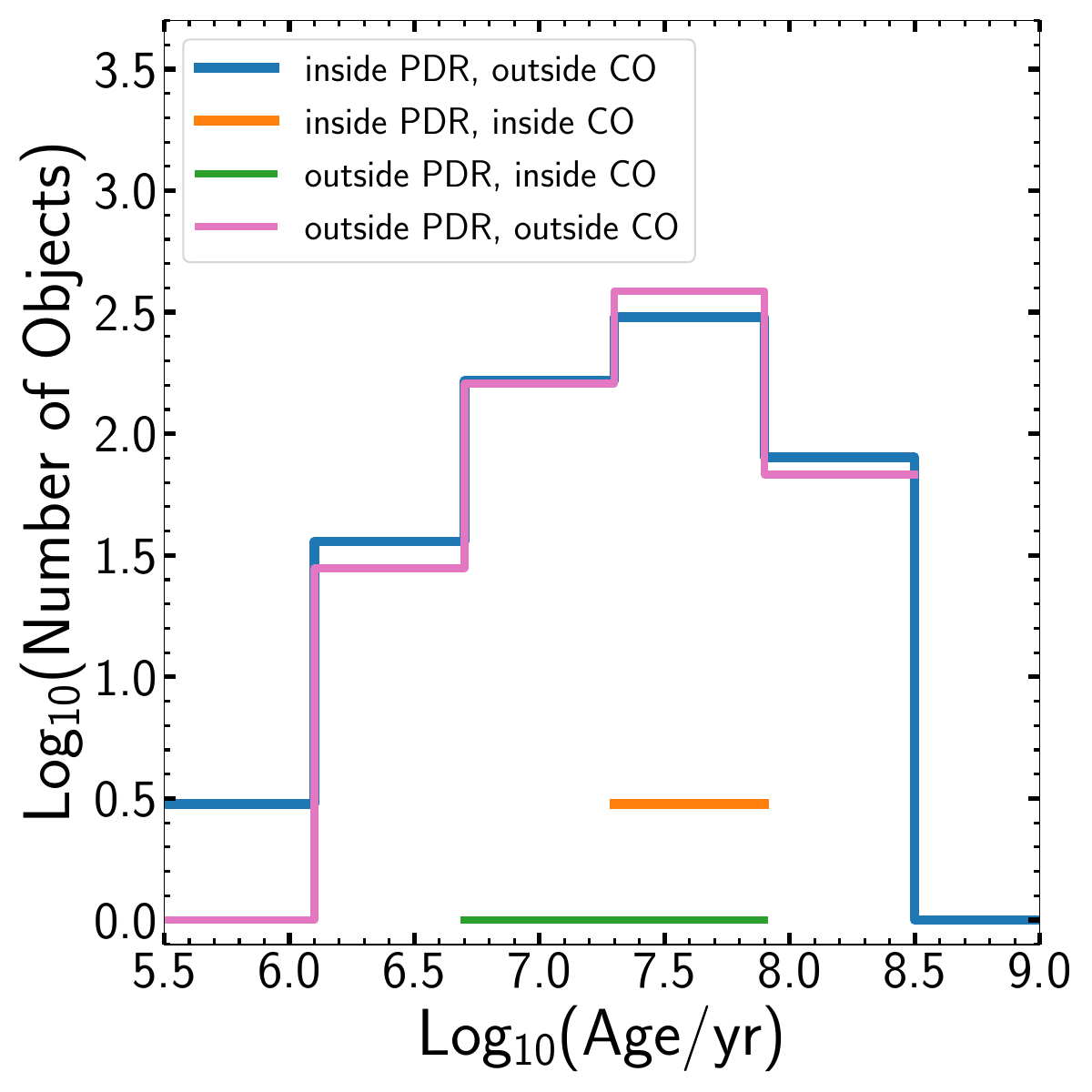}}\hspace{-0.6em}
\subfloat{\includegraphics[width=0.5\linewidth]{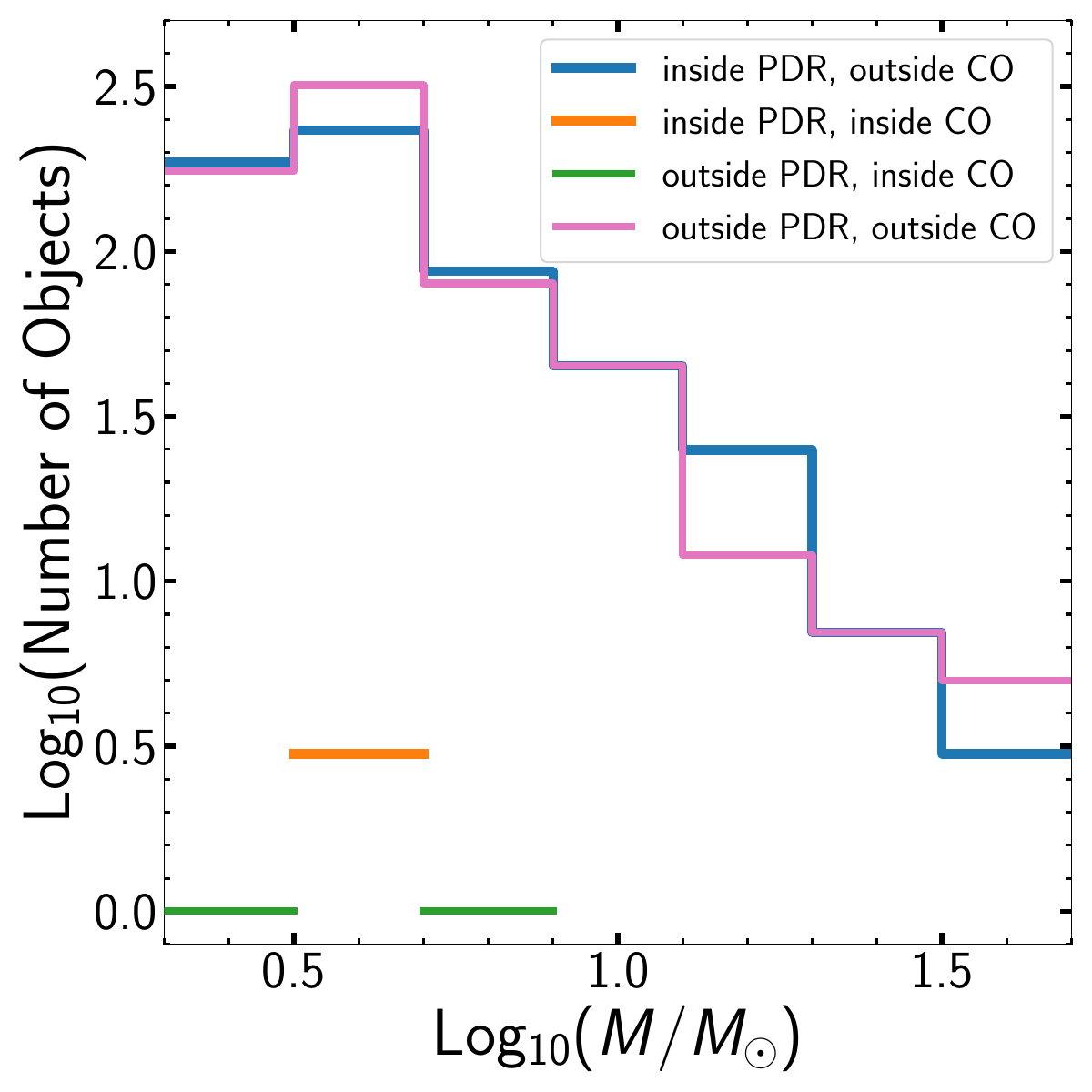}}\\
\subfloat{\includegraphics[width=0.5\linewidth]{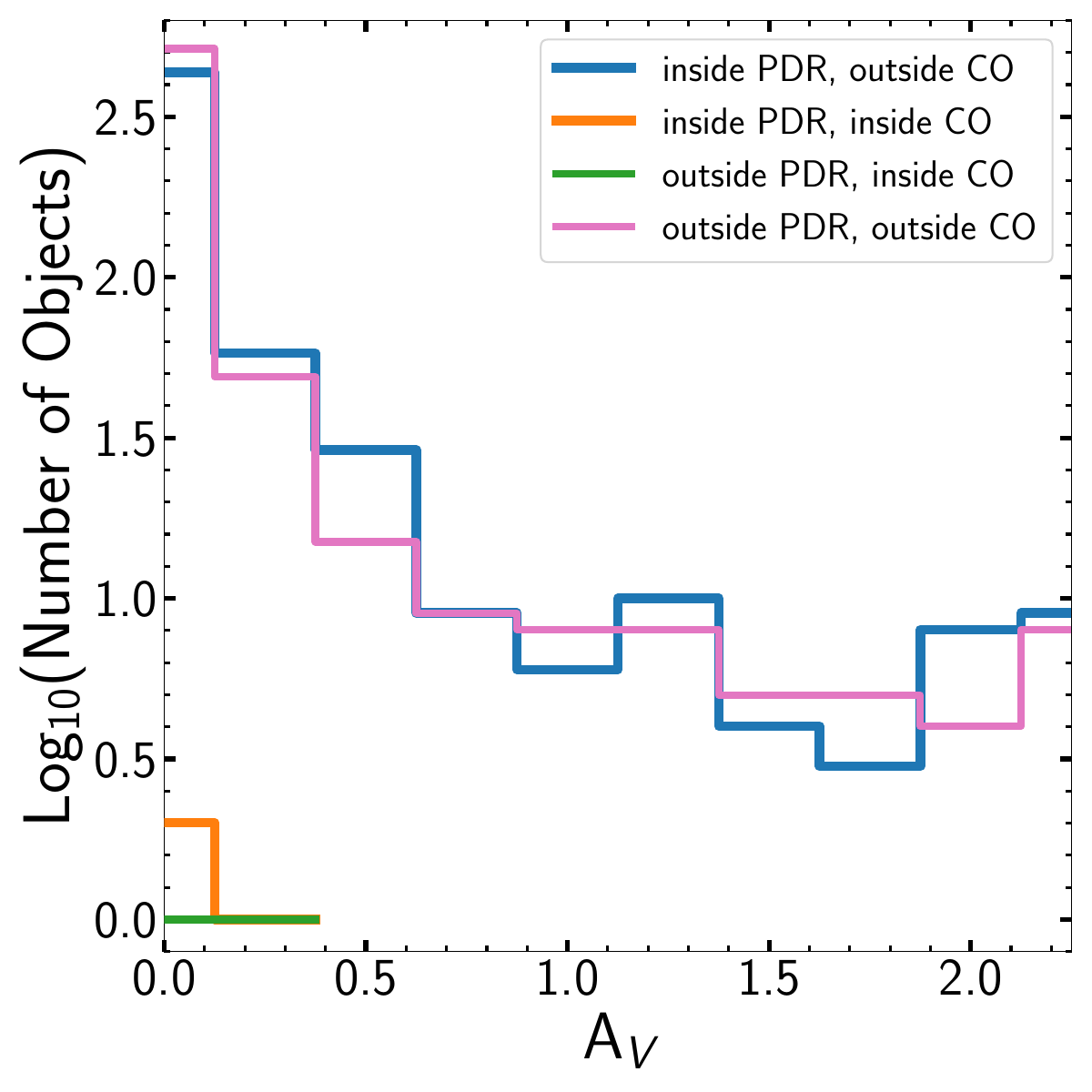}}
\caption{Histograms of the ages (top left), masses (top right), and $A_V$ (bottom) for stars inside the PDR and outside the CO cores (blue), stars inside the PDR and projected inside the CO cores (orange), stars outside the PDR and projected inside the CO cores (green), and stars outside the PDR and outside the CO cores (pink). Clusters of closely packed stars may not be resolved into individual components.
\label{fig:histograms}}
\end{figure*}

\begin{figure*}[htb!]
\captionsetup[subfloat]{farskip=-5pt}
\centering 
\subfloat{\includegraphics[width=0.5\linewidth]{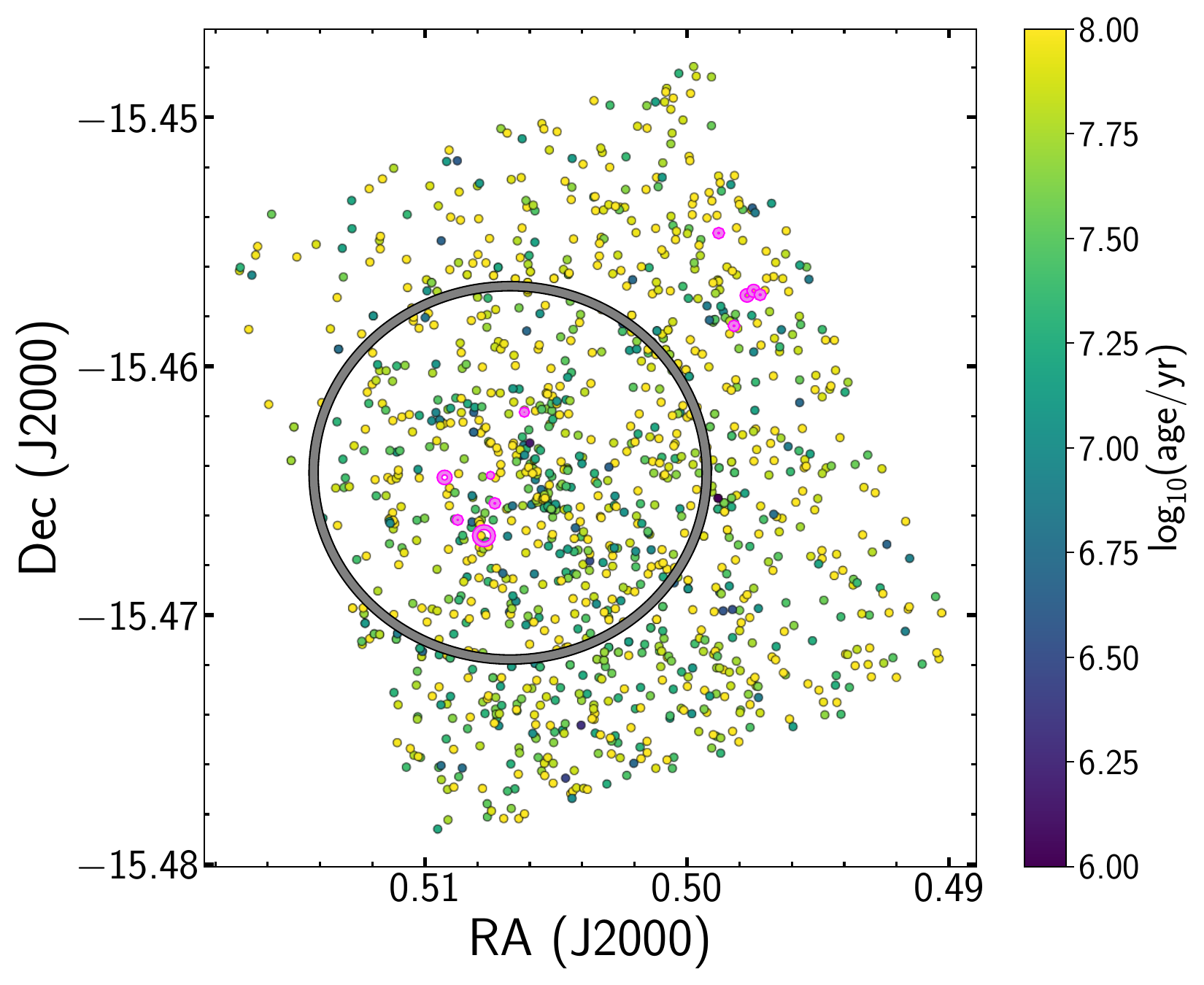}}\hspace{-0.4em}
\subfloat{\includegraphics[width=0.5\linewidth]{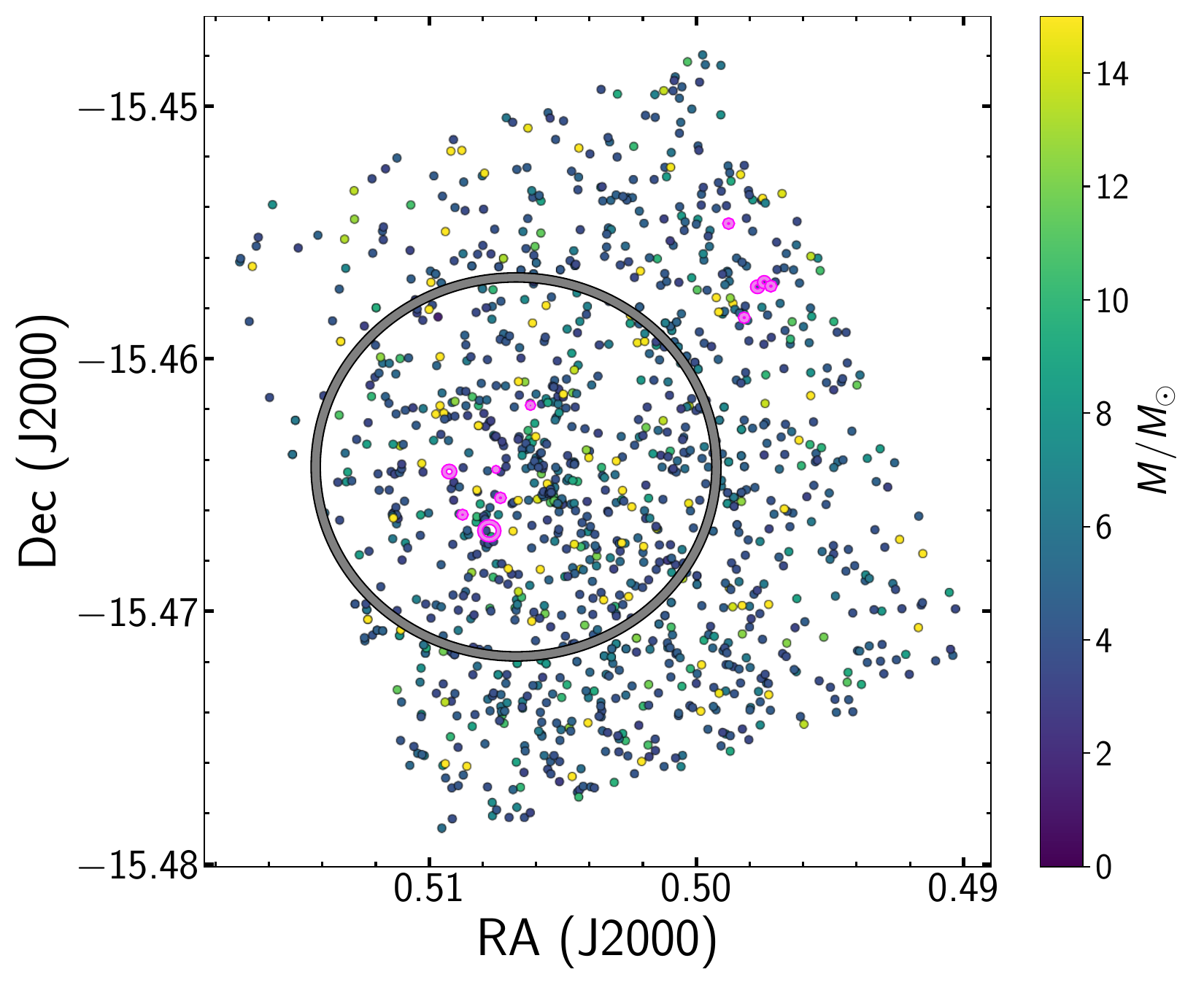}}\\
\subfloat{\includegraphics[width=0.5\linewidth]{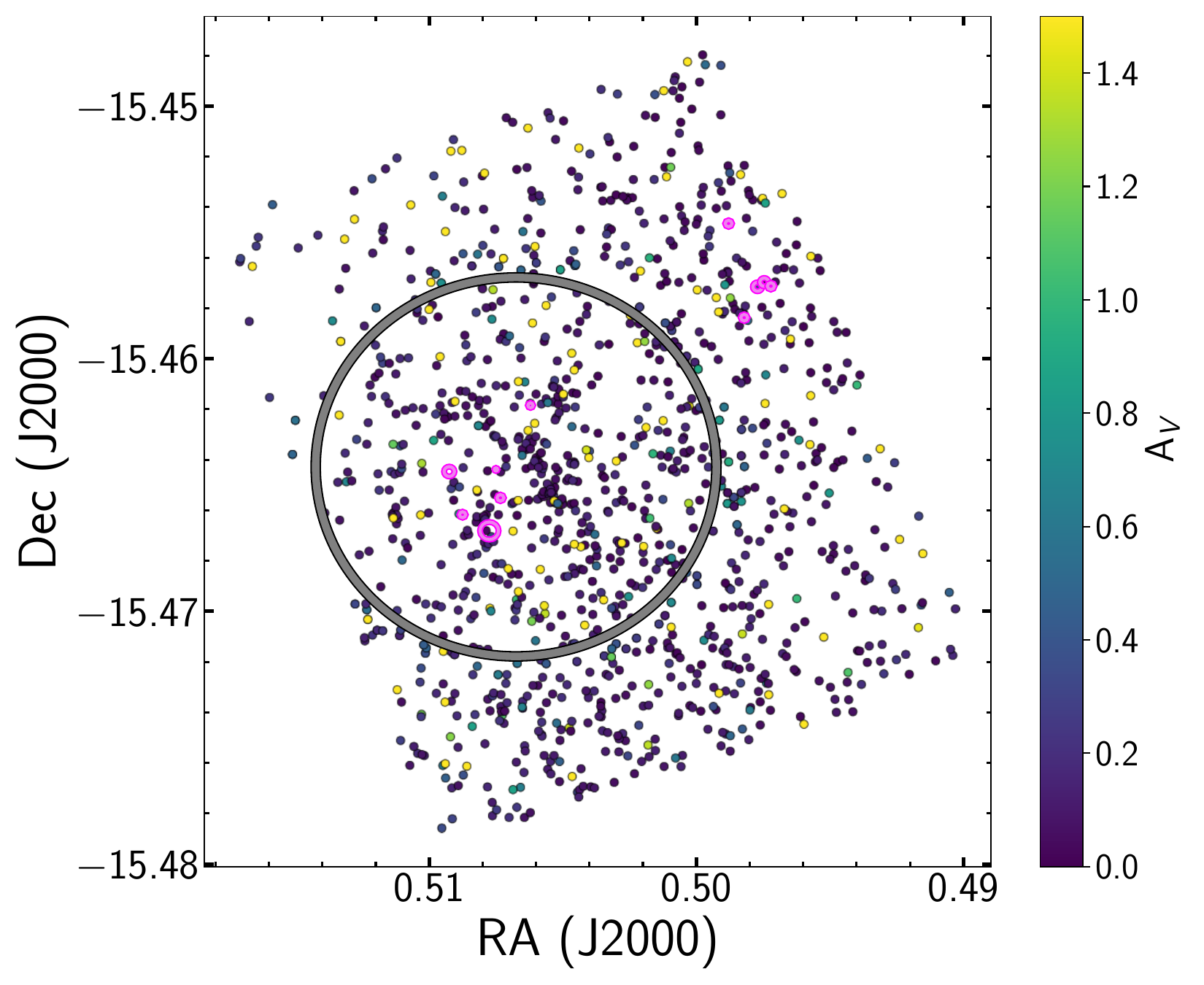}}
\caption{Plots showing the spatial distribution of stars, color-coded by their ages (top left), masses (top right), and $A_V$ (bottom). The large gray circle demarcates the PDR, while the smaller magenta circles show the locations and sizes of the CO cores.
\label{fig:spatial_sed}}
\end{figure*}
\section{Results} \label{sec:results}
\subsection{Photometry}
We separated the stars into four categories based on their coincidence with the PDR and CO cores: (1) stars inside the PDR and projected inside the CO cores, (2) stars inside the PDR and outside the CO cores, (3) stars outside the PDR and projected inside the CO cores, and (4) stars outside the the PDR and outside the CO cores. Only four stars are spatially coincident with the CO cores. To better constrain the SED, we include only stars detected in all five filters and in all but the F275W filter. Consequently, there may be stars within the CO that are excluded, as these stars would be embedded and not appear in the bluest filters. This limitation reduces the number of stars available for analysis in these regions. Additionally, some stars coincident with the CO cores may be located in front of the CO rather than within the cores themselves. Stars visible in the reddest \textit{HST} filter (F625W) but absent from the bluest filters are also detected in the F555W filter, further suggesting that the UVIS dataset does not capture embedded stars. Identifying such stars would require the unique high-resolution infrared capabilities of \textit{JWST}, particularly that of the Mid-Infrared Instrument (MIRI), as the MIRI filters are found to play a crucial role in distinguishing young stellar objects from cool, evolved red stars and background galaxies \citep{Peltonen_2024}. The JWST Resolved Stellar Populations Early Release Science Program \citep[e.g.][]{Weisz_2023,McQuinn_2024,Boyer_2024,Newman_2024} provide publicly available NIR photometric catalogs for WLM as part of the JWST Resolved Stellar Populations Early Release Science Program. However, the fields they targeted do not overlap with the region analyzed in this study.

Because the requirement of a detection in the F275W/F336W filters, a larger number of stars detected in the reddest filter, F625W, were excluded. After separating stars from galaxies using the sharpness and $\chi$ parameters, the total number of stars detected in F625W was 11,732, while the total number of detected stars in the F275W filter was 1,946. The resulting catalog after matching all five filters contains 1,013 stars, or around 10\% of the stars found in F625W, while the resulting catalog after matching all but the F275W filter includes an additional 223 stars more than the full five-filter catalog.

The right ascension (RA), declination (DEC), and apparent Vega magnitudes corresponding to the five \textit{HST} filters for stars in each of the four categories are included in Tables \ref{tab:inPDRinCO}, \ref{tab:inPDRoutCO}, \ref{tab:outPDRinCO}, \ref{tab:outPDRoutCO}, with the full Tables for \ref{tab:inPDRoutCO} and \ref{tab:outPDRoutCO} available in machine-readable format in the online materials. For stars not detected in the F275W filter, the right ascension (RA), declination (DEC), and apparent Vega magnitudes corresponding to the other four \textit{HST} filters are included in Table \ref{tab:inPDRinCO_no275} for stars inside the PDR and projected inside the CO cores, Table \ref{tab:inPDRoutCO_no275} for stars inside the PDR and outside the CO cores, and Table \ref{tab:outPDRoutCO_no275} for stars outside the PDR and outside the CO cores. The full Tables for \ref{tab:inPDRoutCO_no275} and \ref{tab:outPDRoutCO_no275} are available in machine-readable format in the online materials. No additional stars outside the PDR and projected inside the CO cores were detected after excluding the F275W filter. The sharpness and $\chi$ values for all detected filters of the sources determined to be stars are included in Appendix \ref{sec:append}. We find that all stars, irrespective of proximity to the PDR or CO cores, occupy the same color and magnitude ranges, which can be seen in the F555W vs F555W--F625W and F275W vs F275W--F336W color-magnitude diagrams (CMDs) in Figure~\ref{fig:cmds}, and the F336W--F438W vs F438W--F555W color-color diagrams in Figure~\ref{fig:ccd}. The scatter in color observed in the CMDs appears to be primarily due to color uncertainties. However, we also find that stars not detected in the F275W filter tend to appear redder in the F555W vs F555W–F625W CMDs, as expected. The redward shift of fainter objects in the F275W vs F275W–F336W color-magnitude diagram may be attributed to reddening or to undercorrected faint object fluxes resulting from the \citet{Anderson_2021} CTE correction applied in the pipeline, as demonstrated by \citet{Windhorst_2022}. The positions of stars on the color-color diagram in Figure~\ref{fig:ccd} are also consistent with the \textit{U–B} vs. \textit{B–V} color indices of main-sequence stars \citep{nicolet_1980, Bressan_2012, Choi_2016}. Additionally, stars not detected in the F275W filter are more frequently found in the redder region of the color-color diagrams in Figure \ref{fig:ccd}, aligning with the expected location of cooler main-sequence stars.

\begin{deluxetable}{R|D@{ +/--}DD@{ +/--}DD@{ +/--}D}
\tablecaption{Mass and age for stars inside the PDR and projected inside CO cores. \label{tab:inPDRinCO_sed}}
\tablehead{\colhead{} & \multicolumn{4}{c}{$\log_{10}(M/M_\odot)$} & \multicolumn{4}{c}{$\log_{10}$(age/yr)} & \multicolumn{4}{c}{$A_V$/mag}} 
\decimals
\startdata
1 & 0.62 & 0.03 & 8.07 & 0.07 & 0.20 & 0.06 \\
2 & 0.63 & 0.03 & 7.96 & 0.11 & 0.05 & 0.04 \\
\enddata
\end{deluxetable}

\begin{deluxetable}{R|D@{ +/--}DD@{ +/--}DD@{ +/--}D}
\tablecaption{Mass and age for stars inside the PDR and outside CO cores. \label{tab:inPDRawayCO_sed}}
\tablehead{\colhead{} & \multicolumn{4}{c}{$\log_{10}(M/M_\odot)$} & \multicolumn{4}{c}{$\log_{10}$(age/yr)} & \multicolumn{4}{c}{$A_V$/mag}}  
\decimals
\startdata
1 & 0.56 & 0.01 & 8.24 & 0.03 & 0.09 & 0.05 \\
2 & 0.90 & 0.01 & 7.56 & 0.01 & 0.69 & 0.02 \\
3 & 0.95 & 0.04 & 7.40 & 0.10 & 0.09 & 0.04 \\
4 & 0.95 & 0.05 & 7.12 & 0.34 & 0.07 & 0.03 \\
5 & 0.88 & 0.01 & 7.55 & 0.01 & 0.18 & 0.04 \\
6 & 0.67 & 0.02 & 7.93 & 0.06 & 0.03 & 0.03 \\
7 & 0.68 & 0.04 & 7.85 & 0.13 & 0.10 & 0.05 \\
8 & 1.00 & 0.01 & 7.37 & 0.01 & 0.17 & 0.03 \\
9 & 1.13 & 0.01 & 7.16 & 0.01 & 0.41 & 0.04 \\
10 & 0.75 & 0.05 & 7.62 & 0.22 & 0.10 & 0.05 \\
\enddata
\tablecomments{Only a portion of this table is shown here to demonstrate its form and content. A machine-readable version is available for all 433 sources detected inside the PDR and outside the CO cores.}
\end{deluxetable}

\begin{deluxetable}{R|D@{ +/--}DD@{ +/--}DD@{ +/--}D}
\tablecaption{Mass and age for stars outside the PDR and projected inside CO cores. \label{tab:awayPDRinCO_sed}}
\tablehead{\colhead{} & \multicolumn{4}{c}{$\log_{10}(M/M_\odot)$} & \multicolumn{4}{c}{$\log_{10}$(age/yr)} & \multicolumn{4}{c}{$A_V$/mag}}  
\decimals
\startdata
1 & 0.97 & 0.01 & 7.40 & 0.01 & 0.32 & 0.04 \\
2 & 0.58 & 0.02 & 8.15 & 0.07 & 0.04 & 0.04 \\
\enddata
\end{deluxetable}

\begin{deluxetable}{R|D@{ +/--}DD@{ +/--}DD@{ +/--}D}
\tablecaption{Mass and age for stars outside the PDR and outside CO cores. \label{tab:awayPDRawayCO_sed}}
\tablehead{\colhead{} & \multicolumn{4}{c}{$\log_{10}(M/M_\odot)$} & \multicolumn{4}{c}{$\log_{10}$(age/yr)} & \multicolumn{4}{c}{$A_V$/mag}}  
\decimals
\startdata
1 & 0.64 & 0.03 & 8.02 & 0.07 & 0.06 & 0.04 \\
2 & 0.68 & 0.03 & 7.90 & 0.10 & 0.13 & 0.05 \\
3 & 0.59 & 0.02 & 8.18 & 0.05 & 0.26 & 0.04 \\
4 & 0.97 & 0.01 & 7.40 & 0.01 & 1.00 & 0.03 \\
5 & 0.65 & 0.03 & 8.00 & 0.09 & 0.14 & 0.05 \\
6 & 0.82 & 0.04 & 7.56 & 0.14 & 0.14 & 0.04 \\
7 & 0.91 & 0.01 & 7.50 & 0.01 & 1.18 & 0.05 \\
8 & 0.91 & 0.01 & 7.51 & 0.01 & 0.61 & 0.04 \\
9 & 0.82 & 0.01 & 7.69 & 0.01 & 0.04 & 0.02 \\
10 & 0.78 & 0.01 & 7.75 & 0.01 & 0.26 & 0.04 \\
\enddata
\tablecomments{Only a portion of this table is shown here to demonstrate its form and content. A machine-readable version is available for all 566 sources detected outside the PDR and outside the CO cores.}
\end{deluxetable}

\subsection{Masses, Ages, and $A_V$}
Tables~\ref{tab:inPDRinCO_sed}, \ref{tab:inPDRawayCO_sed}, \ref{tab:awayPDRinCO_sed}, \ref{tab:awayPDRawayCO_sed}, \ref{tab:inPDRinCO_no275_sed}, \ref{tab:inPDRawayCO_no275_sed}, and \ref{tab:awayPDRawayCO_no275_sed} contain the masses, ages, and $A_V$ found using PARSEC for stars in each of the four categories, along with stars not detected in the F275W filter, with the full Tables \ref{tab:inPDRawayCO_sed}, \ref{tab:awayPDRawayCO_sed}, \ref{tab:inPDRawayCO_no275_sed}, and \ref{tab:awayPDRawayCO_no275_sed} included as machine-readable tables in the online materials. The small uncertainties in the inferred physical properties may result from the SED fitting process rather than reflecting genuinely low uncertainties. We include a corner plot illustrating the SED fit for a representative star from each category in the Appendix \ref{sec:sedfits}. Figure~\ref{fig:histograms} shows the histograms of the masses, ages, and $A_V$ of the stars, where we find that stars across all four categories exhibit similar mass, age, and $A_V$ distributions. The mean $A_V$ of all the stars detected is $\sim$0.34$\pm$0.06 mag, which is similar to the mean $A_V$ of stars in WLM measured by \citet{Wang_2022} found to be 0.35 mag. This mean extinction value is similar across the different categories. Stellar ages typically range from $\sim$1-100 Myr, with older stars likely belonging to the underlying disk population. We note that clusters of closely packed stars may not be fully resolved into individual components.

\begin{figure*}[htb!]
\captionsetup[subfloat]{farskip=-5pt}
\centering 
\subfloat{\includegraphics[width=\linewidth]{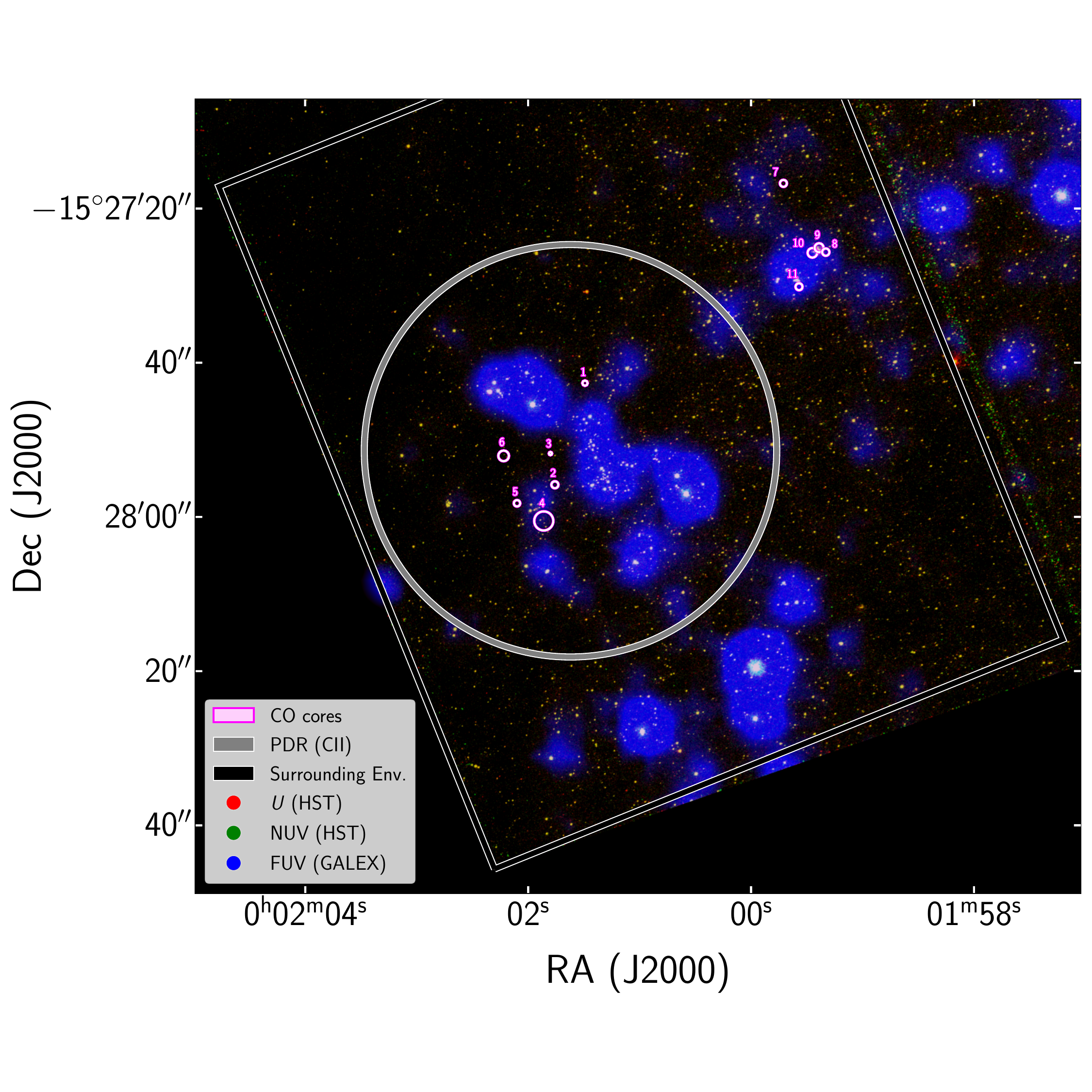}}
\caption{Three-color composite image combining the \textit{HST} 336W (red), \textit{HST} 275W (green), and GALEX FUV (blue) images of the region, highlighting how the ultraviolet clumps of star formation correspond to the structures and clusters of younger stars within the PDR shown in Figure~\ref{fig:spatial_sed}.
\label{fig:fuv}}
\end{figure*}

We also find no correlation is observed between the spatial locations of stars and their respective masses, ages, or $A_V$ as shown in Figure~\ref{fig:spatial_sed}. However, we identify some structure and clusters of younger stars near the center of the PDR, which align with regions bright in the far-ultraviolet (FUV). A three-color image of the region is shown in Figure~\ref{fig:fuv}, where red corresponds to the \textit{HST} F336W image, green corresponds to the \textit{HST} F275W image, and blue corresponds to the FUV image from the NASA Galaxy Evolution Explorer (GALEX\footnote[1]{GALEX was operated for NASA by the California Institute of Technology under NASA contract NAS5-98034.}) satellite \citep{galex,Zhang_2012}.

The mean velocity dispersion in the region is approximately 8 km/s \citep{iorio_2017}, indicating that stars could have been dispersed by nearly 82 pc over 10 Myr. This dispersion may explain the scattering of young stars observed outside the PDR, which has a radius of $\sim$130 pc. Additionally, Figure~\ref{fig:fuv} highlights ongoing star formation beyond the PDR, which may not be directly linked to the same star-forming event and could account for the young stars seen outside the PDR.

\subsection{Gas Mass}
To get a comprehensive view of the gas in our targeted region, we combined our \textit{HST} and CO data with extant \HI\ masses of the region. The \HI\ mass comes from converting the \HI\ surface density (\HISD) in Table 2 of \citet{archer_2022} to mass. The robust-weighted \HISD\ map was acquired with the Karl G.\ Jansky Very Large Array (VLA) for Local Irregulars That Trace Luminosity Extremes, The \HI\ Nearby Galaxy Survey (LITTLE THINGS), a multiwavelength survey of 37 nearby dIrr galaxies and 4 nearby Blue Compact Dwarf (BCD) galaxies \citep{Hunter_2012}. We include the mass of the \HI\ atomic gas for this region in Table~\ref{tab:masses}.

To determine the amount of CO-dark molecular gas in the region, we first found the total mass of young stars detected in our region. We estimated the number of detected disk stars in the PDR to be approximately 400. This number also approximately corresponds to the number of stars with ages greater 30 Myr. Only including stars younger than 30 Myr--the more recent star formation--we estimate the total stellar mass of young stars in our sample to be $\sim$2,000 $M_\odot$. The absence of low-mass stars in our sample due to completeness suggests that the total stellar mass is likely much greater than this estimate by a factor of 2 or 3, considering a standard IMF. \citet{krumholz_2012} find that approximately 1\% $\pm$ 2\% of the molecular gas is converted to stars per local free-fall time. For our region, spanning 260 pc in diameter and assuming a velocity dispersion comparable to the stellar dispersion of 8 km/s, the turbulence crossing time is 32 Myr, which is comparable to the selected age window for our stellar mass.  The timescale is also comparable to that of large-scale star formation in the LMC, which is $\sim$20 Myr on this scale from Figure 1 in \citet{Elmegreen_2000}.  Taking these timescales as the effective free fall time over the large PDR region considered here, and a conservative estimate of 2\% of the gas mass converting to stars in this time. Taking a conservative estimate of 2\%, we find the total star-forming gas mass in the PDR using our estimated stellar mass is then:

\begin{equation}
    2 \times 10^3\ M_\odot\ /\ 0.02 \approx 1 \times 10^5\ M_\odot.
\end{equation}
Accounting for the total virial mass of the six CO cores in the region, $\sim$20,000 $M_\odot$, we find the CO-dark molecular gas mass to be
\begin{equation}
    1 \times 10^5\ M_\odot - 2 \times 10^4\ M_\odot = 8 \times 10^4\ M_\odot,
\end{equation}
suggesting that approximately 80\% of the molecular gas mass is CO-dark. Assuming 1\% or 3\% of the molecular gas is converted to stars instead yields a CO-dark gas percentage of 90\% or 70\% respectively. The total CO virial mass, total stellar mass, total estimated molecular gas mass, and total estimated CO-dark gas mass, along with their associated uncertainties, are included in Table~\ref{tab:masses}.

\begin{deluxetable}{R|D@{ +/--}DD@{ +/--}DD@{ +/--}D}
\tablecaption{Mass and age for the star inside the PDR and projected inside CO cores not detected in the F275W filter. \label{tab:inPDRinCO_no275_sed}}
\tablehead{\colhead{} & \multicolumn{4}{c}{$\log_{10}(M/M_\odot)$} & \multicolumn{4}{c}{$\log_{10}$(age/yr)} & \multicolumn{4}{c}{$A_V$/mag}} 
\decimals
\startdata
1 & 0.62 & 0.06 & 7.71 & 0.41 & 0.34 & 0.17 \\
\enddata
\end{deluxetable}

\begin{deluxetable}{R|D@{ +/--}DD@{ +/--}DD@{ +/--}D}
\tablecaption{Mass and age for stars inside the PDR and outside CO cores not detected in the F275W filter. \label{tab:inPDRawayCO_no275_sed}}
\tablehead{\colhead{} & \multicolumn{4}{c}{$\log_{10}(M/M_\odot)$} & \multicolumn{4}{c}{$\log_{10}$(age/yr)} & \multicolumn{4}{c}{$A_V$/mag}}  
\decimals
\startdata
1 & 0.61 & 0.01 & 8.13 & 0.03 & 0.33 & 0.10 \\
2 & 1.20 & 0.03 & 7.09 & 0.04 & 2.55 & 0.09 \\
3 & 0.79 & 0.05 & 7.72 & 0.11 & 0.55 & 0.13 \\
4 & 0.90 & 0.04 & 7.52 & 0.08 & 0.14 & 0.10 \\
5 & 1.24 & 0.09 & 6.87 & 0.28 & 0.33 & 0.10 \\
6 & 0.72 & 0.05 & 7.83 & 0.14 & 0.36 & 0.15 \\
7 & 1.04 & 0.01 & 7.31 & 0.01 & 2.11 & 0.06 \\
8 & 1.07 & 0.01 & 7.26 & 0.01 & 1.88 & 0.05 \\
9 & 0.95 & 0.01 & 7.45 & 0.01 & 1.65 & 0.08 \\
10 & 0.91 & 0.01 & 7.52 & 0.01 & 1.15 & 0.07 \\
\enddata
\tablecomments{Only a portion of this table is shown here to demonstrate its form and content. A machine-readable version is available for all 144 sources detected inside the PDR and outside the CO cores.}
\end{deluxetable}

\begin{deluxetable}{R|D@{ +/--}DD@{ +/--}DD@{ +/--}D}
\tablecaption{Mass and age for stars outside the PDR and outside CO cores not detected in the F275W filter. \label{tab:awayPDRawayCO_no275_sed}}
\tablehead{\colhead{} & \multicolumn{4}{c}{$\log_{10}(M/M_\odot)$} & \multicolumn{4}{c}{$\log_{10}$(age/yr)} & \multicolumn{4}{c}{$A_V$/mag}}  
\decimals
\startdata
1 & 1.06 & 0.01 & 7.30 & 0.01 & 2.26 & 0.03 \\
2 & 1.29 & 0.01 & 6.96 & 0.01 & 3.07 & 0.06 \\
3 & 1.52 & 0.01 & 6.77 & 0.01 & 3.89 & 0.06 \\
4 & 1.52 & 0.01 & 6.76 & 0.01 & 3.61 & 0.06 \\
5 & 1.13 & 0.01 & 7.16 & 0.01 & 2.44 & 0.08 \\
6 & 1.12 & 0.05 & 7.17 & 0.09 & 2.70 & 0.13 \\
7 & 1.10 & 0.01 & 7.21 & 0.01 & 2.60 & 0.09 \\
8 & 2.02 & 0.01 & 6.44 & 0.01 & 4.00 & 0.01 \\
9 & 1.03 & 0.01 & 7.35 & 0.01 & 1.93 & 0.02 \\
10 & 1.63 & 0.01 & 6.66 & 0.01 & 3.99 & 0.01 \\
\enddata
\tablecomments{Only a portion of this table is shown here to demonstrate its form and content. A machine-readable version is available for all 78 sources detected outside the PDR and outside the CO cores.}
\end{deluxetable}

\section{Discussion} \label{sec:discussion}
Estimating molecular gas mass in low-metallicity galaxies like WLM remains a significant challenge due to the high fraction of CO-dark gas. \citet{Elmegreen_2013} estimated $\alpha_{\textrm{CO}}$ for WLM using dust mass inferred from 160$\mu$m emission from the Spitzer Local Volume Survey \citep{dale_2009} and 870$\mu$m emission from the APEX telescope. By adjusting the dust-to-gas ratio for WLM’s lower metallicity, they determined a dust-derived $\alpha_{\textrm{CO}}$ of $124 \pm 60\ M_\odot$ pc$^{-2}$ K$^{-1}$ km$^{-1}$ s. Using this $\alpha_{\textrm{CO}}$ value and the CO core luminosity in our region, the total H$_2$ mass would be: 
\begin{equation}
\begin{split}
    M_{gas} & = \alpha_{CO} \times L_{CO} \\
    & = (124 \pm 60\ M_\odot\ \textrm{pc}^{-2}\ \textrm{K}^{-1}\ \textrm{km}^{-1}\ \textrm{s})\\
    & \phantom{==} \times  (935 \pm 60\ \textrm{K km s}^{-1}\ \textrm{pc}^2) \\
    & \approx 115,900 \pm 5,700\ M_\odot,
\end{split}
\end{equation}
where $L_{CO}$ is the summed $L_{CO}$ values for the six CO cores in the region from \citet[][Table 1]{Rubio_2015}. Alternatively, computing the $\alpha_{CO}$ from our total molecular gas mass of $1\times10^5\ M_\odot$, we find: 
\begin{equation}
\begin{split}
    \alpha_{\textrm{CO}} & = M_{gas}/ L_{CO} \\
     & = \frac{(1 \times 10^5 \pm 5 \times 10^4 \ M_\odot)}{(935 \pm 60\ \textrm{K km s}^{-1}\ \textrm{pc}^2)} \\
    & \approx 100 \pm 50\ M_\odot\ \textrm{pc}^{-2}\ \textrm{K}^{-1}\ \textrm{km}^{-1}\ \textrm{s},
\end{split}
\end{equation}
which is consistent with the dust-derived $\alpha_{CO}$ found by \citet{Elmegreen_2013}.

The high fraction of CO-dark gas in WLM indicates that a substantial portion of the molecular gas available for star formation exists in a state not directly detectable via CO emission. Studies of other low metallicity galaxies such as the SMC, LMC, and the Dwarf Galaxy Survey (DGS) find 70\% to 100\% of the molecular hydrogen in low-metallicity galaxies ($Z\!=\!0.02$ to $0.6\ Z_\odot$) is CO-dark, increasing with lower metallicity \citep[e.g.][]{raquena_2016,Chevance_2020_2,madden_2020,Ramambason_2024}, which is consistent with our estimated CO-dark gas percentage. Similar to the tiny CO cores detected in WLM, \citet{Saldano_2023} find that the molecular mass associated with CO clouds in the SMC is primarily concentrated in low-mass clouds distributed throughout the galaxy. This reinforces the understanding that CO-bright regions correspond to the densest, most shielded parts of molecular clouds in low-metallicity environments, while CO-dark regions constitute a diffuse and widespread reservoir of H$_2$ \citep{Wolfire_2010, krumholz_2012, Bolatto_2013} or cold \HI\ \citep{Hu_2021, Hu_2022, Hu_2023}. These findings underscore the necessity of accounting for CO-dark gas when evaluating the star formation potential of galaxies, particularly in low-metallicity conditions. The agreement between the molecular gas mass inferred from dust measurements \citep{Elmegreen_2013} and that estimated by combining stellar mass with an assumed 2\% star formation efficiency is encouraging. If the dust-related total gas mass is assumed to be the most reliable, then the missing low mass stars suggest that the product of the efficiency per unit free fall time and the number of free fall times for star formation could be low by a factor of $\sim$2, which is the likely correction for stellar mass given a standard IMF. For example, the 30 Myr window for our evaluation of young stellar mass could represent two free fall times on this large scale, rather than one as assumed. 

\begin{deluxetable}{c|r}
\tablecaption{Gas masses in the targeted region \label{tab:masses}}
\tablehead{\colhead{Type} & \colhead{Mass} \\ 
\colhead{} & \colhead{($M_\odot$)} } 
\startdata
\HI\ & 1,620,000 $\pm$ \phantom{0,}600 \\
Stars & 2,000 $\pm$ \phantom{0,}300 \\
CO$_{vir}$ & 20,000 $\pm$ 6,000 \\
Total (molecular)\tablenotemark{\scriptsize{a}} & 100,000 $\pm$ 5,000\\
CO-dark \tablenotemark{\scriptsize{a}} & 80,000 $\pm$ 5,000\\
\enddata
\tablenotetext{a}{Assuming 2\% $\pm$ 1\% of molecular gas is converted to stars \citep{krumholz_2012}}
\end{deluxetable}

\section{Summary and Conclusions} \label{sec:summary}
In this study, we explored the stellar and gas characteristics within the nearby galaxy WLM using multi-wavelength \textit{HST} imaging and ALMA CO(1–0) and CO(2-1) observations. By employing photometry across five \textit{HST} filters ranging from 2709.7 to 6242.6~\AA, we classified stars and distinguished them from background galaxies, allowing us to analyze stellar masses, ages, and $A_V$ using the PARSEC isochrone models. Our results demonstrate that stars located within the PDR and the CO cores, as well as those outside these regions, exhibit similar distributions in age, mass, and optical depth, indicating a uniform stellar population across the observed area. 

To provide a comprehensive assessment of the gas content, we incorporated existing \HI\ data and estimated the total molecular gas mass, including contributions from CO-dark molecular gas. Our analysis revealed a significant fraction of CO-dark gas, emphasizing its critical role in molecular gas mass estimates that cannot rely solely on CO observations. Additionally, the dust-derived $\alpha_{\textrm{CO}}$ for WLM from \citet{Elmegreen_2013} yields a total molecular gas mass consistent with our estimate based on stellar mass and an assumed star formation efficiency of 2\%. However, the stellar mass estimate excludes lower-mass stars that were not detected in our sample. This agreement suggests that combining stellar mass with a 2\% star formation efficiency provides an alternative for estimating total molecular gas mass in star-forming regions when dust and CO data are unavailable, though both methods likely underestimate the actual molecular gas mass.

This work examines the molecular gas composition and star formation processes in low-metallicity environments. The results highlight the critical role of CO-dark gas in these systems. Expanding this analysis to a larger sample of star-forming regions within WLM and other low-metallicity galaxies could determine whether the high CO-dark gas content observed in this region is a common characteristic or a unique feature. Such investigations would enhance our understanding of the gas reservoirs that fuel star formation across diverse galactic environments and contribute to a more comprehensive framework for star formation in the local universe.

\section{Acknowledgements} \label{sec:acknowledgements}
This research is based on observations made with the NASA/ESA Hubble Space Telescope obtained from the Space Telescope Science Institute, which is operated by the Association of Universities for Research in Astronomy, Inc., under NASA contract NAS~5–26555. These observations are associated with program HST-GO-17068.

M.R. wishes to acknowledge support from ANID(CHILE) through Basal~FB210003.

This paper makes use of the following ALMA data:~ADS/JAO.ALMA\#2012.1.00208.S. ALMA is a partnership of ESO (representing its member states), NSF (USA), and NINS (Japan), together with NRC (Canada), MOST and ASIAA (Taiwan), and KASI (Republic of Korea), in cooperation with the Republic of Chile. The Joint ALMA Observatory is operated by ESO, AUI/NRAO and NAOJ. The National Radio Astronomy Observatory is a facility of the National Science Foundation operated under cooperative agreement by Associated Universities, Inc.

We thank the referee for their thorough review and constructive feedback, which significantly improved the quality and clarity of this manuscript.

Lowell Observatory sits at the base of mountains sacred to tribes throughout the region. We honor their past, present, and future generations, who have lived here for millennia and will forever call this place home.

\vspace{5mm}
\facilities{ALMA, HST(WFC3/UVIS)}


\software{IRAF~\citep{Tody_1986}, scikit-learn~\citep{sklearn}, AdvancedHMC.jl~\citep{xu_2020}, CMD~\citep{Bressan_2012,chen_2014,Chen_2015,tang_2014,marigo_2017,pastorelli_2019,pastorelli_2020}, Julia~\citep{bezanson_2017}, Matplotlib~\citep{hunter_2007}, NumPy~\citep{harris_2020}, Optim.jl~\citep{mogensen_2018}
}

\appendix
\section{Sharpness and Chi Parameters}\label{sec:append}
The sharpness and chi parameters for all objects determined to be stars are included in Tables \ref{tab:inPDRinCO_sharpChi}, \ref{tab:inPDRoutCO_sharpChi}, \ref{tab:outPDRinCO_sharpChi}, \ref{tab:outPDRoutCO_sharpChi}, \ref{tab:inPDRinCO_sharpChi_no275}, \ref{tab:inPDRoutCO_sharpChi_no275}, and \ref{tab:outPDRoutCO_sharpChi_no275}. The full Tables for \ref{tab:inPDRoutCO_sharpChi}, \ref{tab:outPDRoutCO_sharpChi}, \ref{tab:inPDRoutCO_sharpChi_no275}, and \ref{tab:outPDRoutCO_sharpChi_no275} are available in machine-readable format in the online materials.
\section{SED fits}\label{sec:sedfits}
\begin{figure*}[p]
\captionsetup[subfloat]{farskip=-5pt}
\centering 
\subfloat{\includegraphics[width=\linewidth]{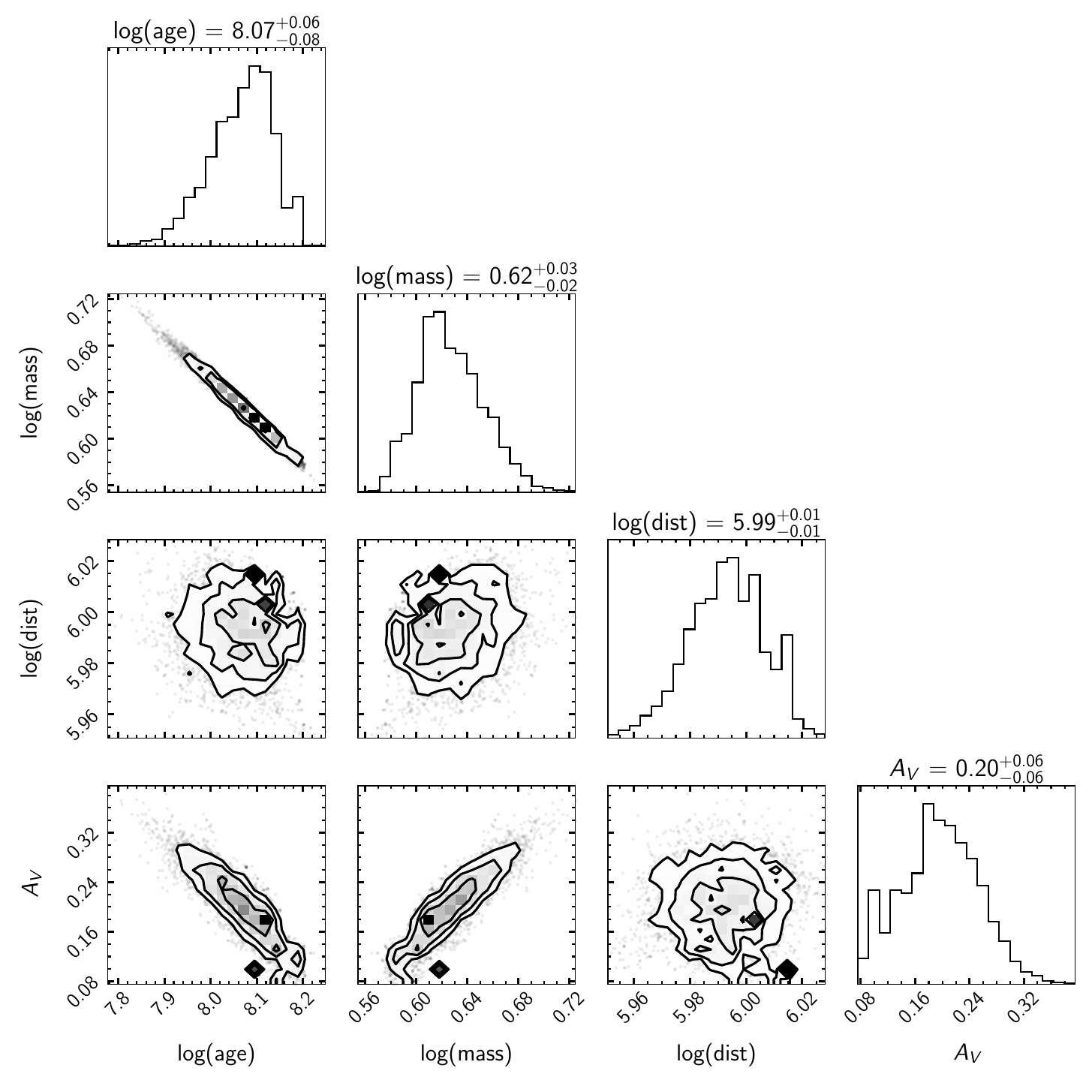}}
\caption{Corner plot of samples drawn from the posterior distribution for Star 1 in Table \ref{tab:inPDRinCO_sed} of stars inside the PDR and projected inside the CO cores demonstrating the degeneracies between mass, age, distance, and dust.
\label{fig:corner_inPDRinCO}}
\end{figure*}

\begin{figure*}[p]
\captionsetup[subfloat]{farskip=-5pt}
\centering 
\subfloat{\includegraphics[width=\linewidth]{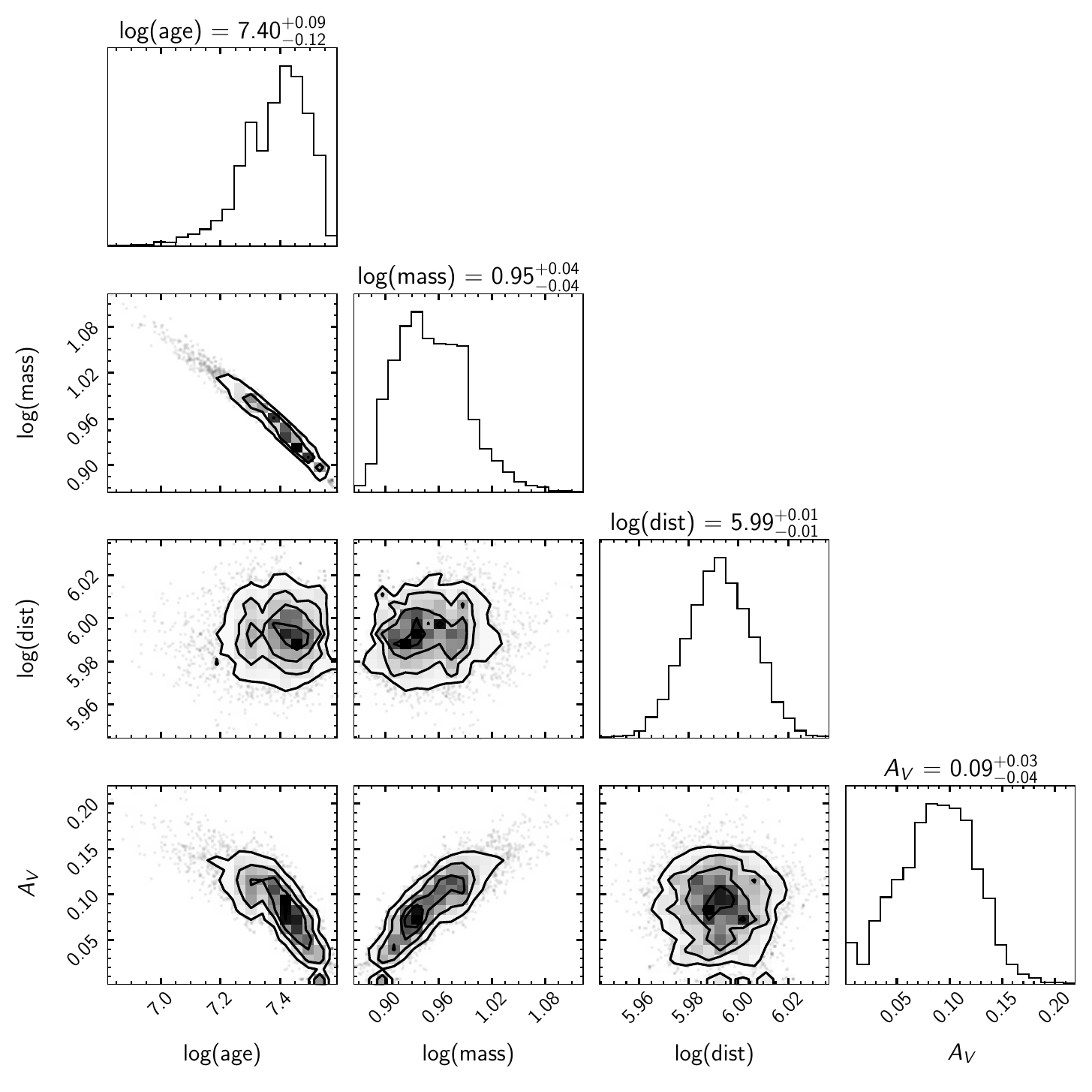}}
\caption{Corner plot of samples drawn from the posterior distribution for Star 3 in Table \ref{tab:inPDRawayCO_sed} of stars inside the PDR and away from the CO cores demonstrating the degeneracies between mass, age, distance, and dust.
\label{fig:corner_inPDRawayCO}}
\end{figure*}

\begin{figure*}[p]
\captionsetup[subfloat]{farskip=-5pt}
\centering 
\subfloat{\includegraphics[width=\linewidth]{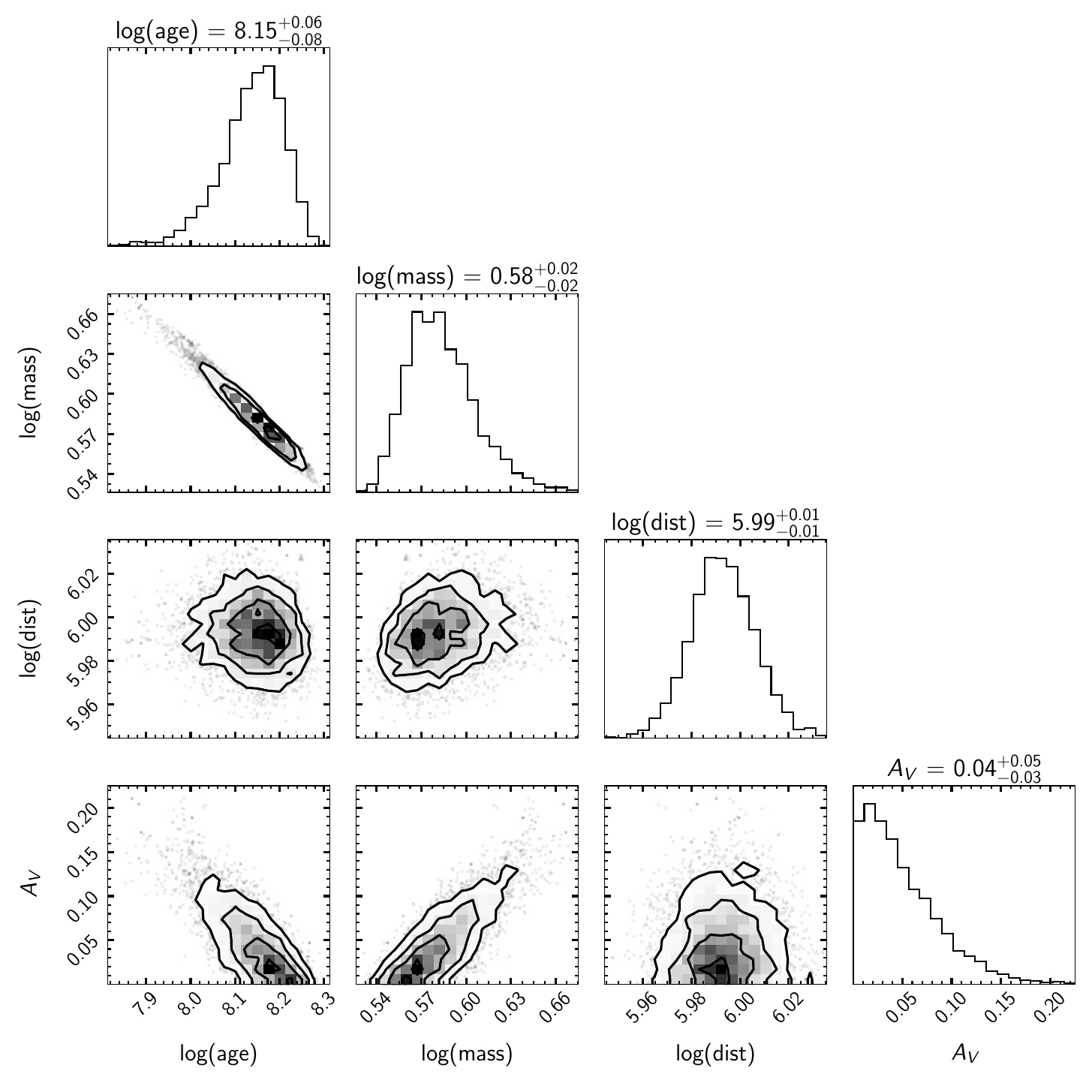}}
\caption{Corner plot of samples drawn from the posterior distribution for Star 2 in Table \ref{tab:awayPDRinCO_sed} of stars away from the PDR and projected inside the CO cores demonstrating the degeneracies between mass, age, distance, and dust.
\label{fig:corner_awayPDRinCO}}
\end{figure*}

\begin{figure*}[p]
\captionsetup[subfloat]{farskip=-5pt}
\centering 
\subfloat{\includegraphics[width=\linewidth]{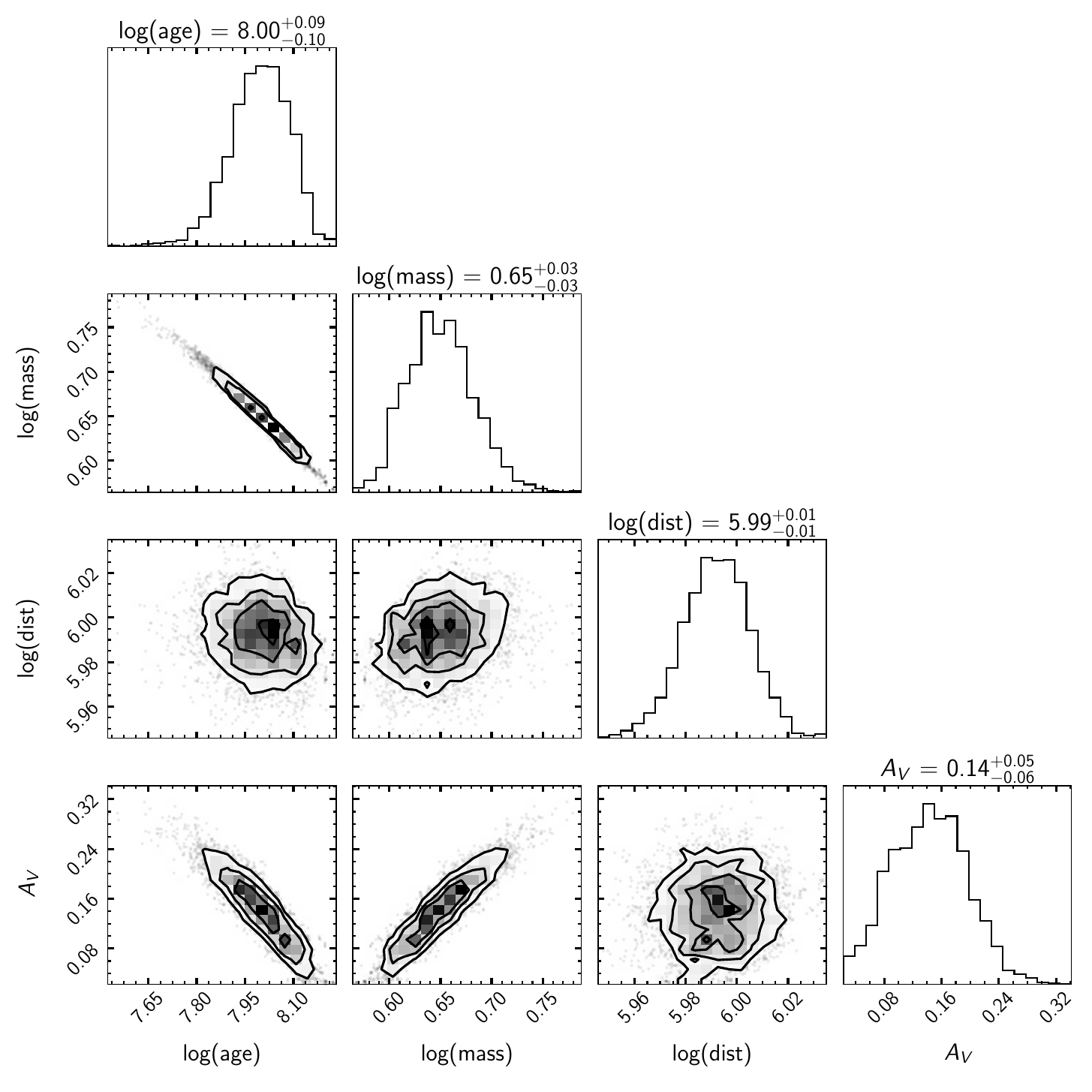}}
\caption{Corner plot of samples drawn from the posterior distribution for Star 5 in Table \ref{tab:awayPDRawayCO_sed} of stars away from the PDR and away from the CO cores demonstrating the degeneracies between mass, age, distance, and dust.
\label{fig:corner_awayPDRawayCO}}
\end{figure*}

\begin{figure*}[p]
\captionsetup[subfloat]{farskip=-5pt}
\centering 
\subfloat{\includegraphics[width=\linewidth]{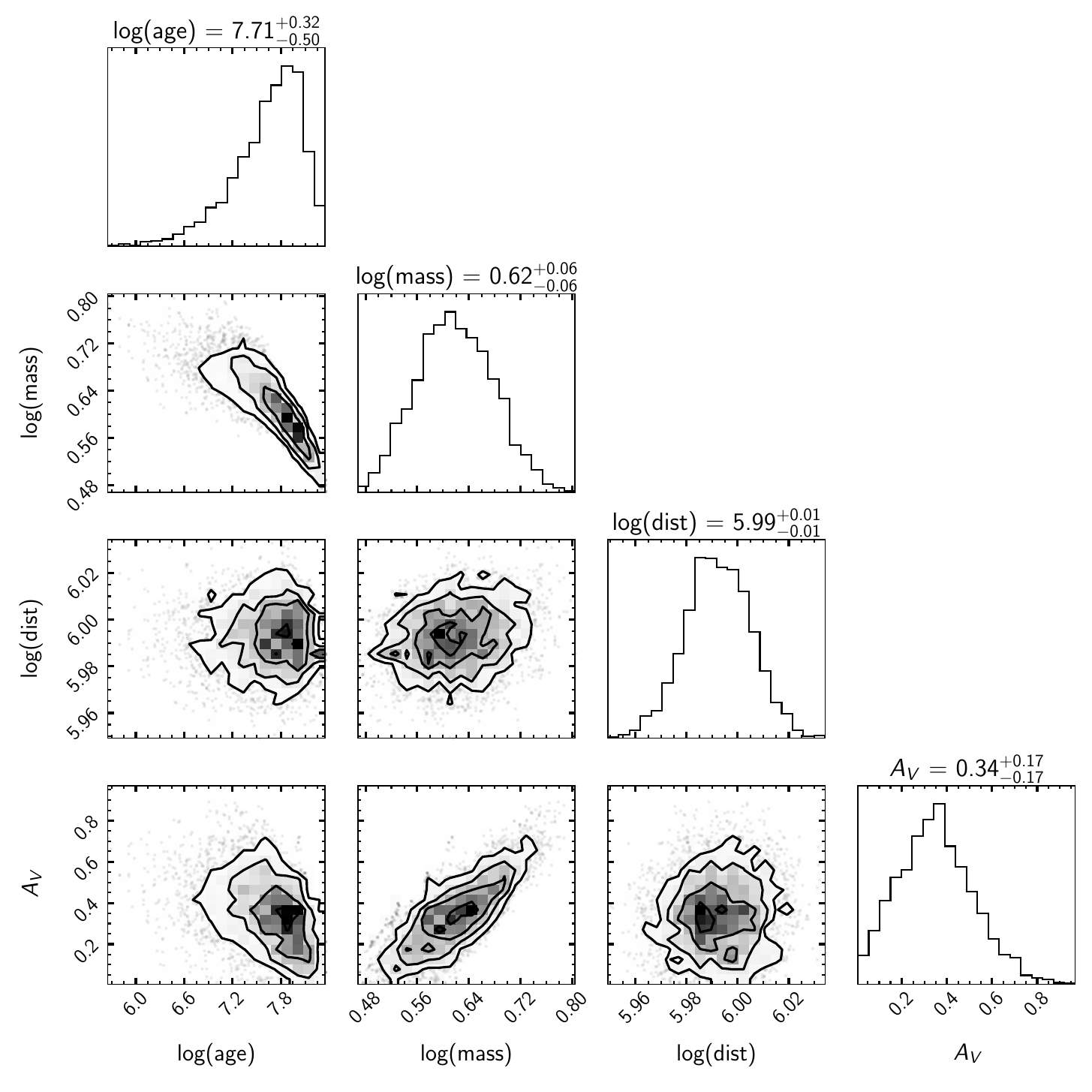}}
\caption{Corner plot of samples drawn from the posterior distribution for Star 1 in Table \ref{tab:inPDRinCO_no275_sed} of the star in the PDR and projected inside the CO cores not detected in the F275W filter demonstrating the degeneracies between mass, age, distance, and dust.
\label{fig:corner_inPDRinCO_no275}}
\end{figure*}

\begin{figure*}[p]
\captionsetup[subfloat]{farskip=-5pt}
\centering 
\subfloat{\includegraphics[width=\linewidth]{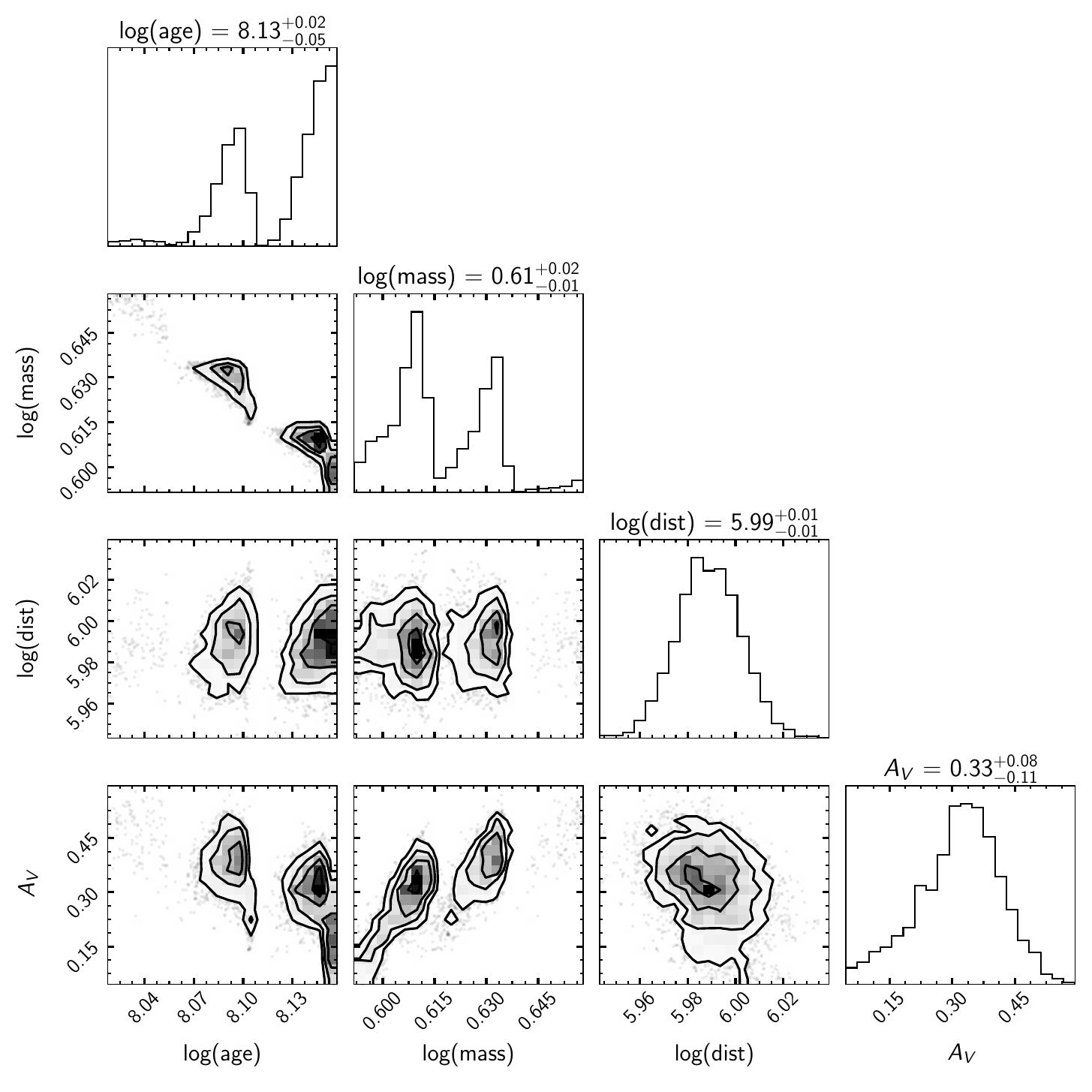}}
\caption{Corner plot of samples drawn from the posterior distribution for Star 1 in Table \ref{tab:inPDRawayCO_no275_sed} of stars in the PDR and away from the CO cores not detected in the F275W filter demonstrating the degeneracies between mass, age, distance, and dust.
\label{fig:corner_inPDRawayCO_no275}}
\end{figure*}

\begin{figure*}[p]
\captionsetup[subfloat]{farskip=-5pt}
\centering 
\subfloat{\includegraphics[width=\linewidth]{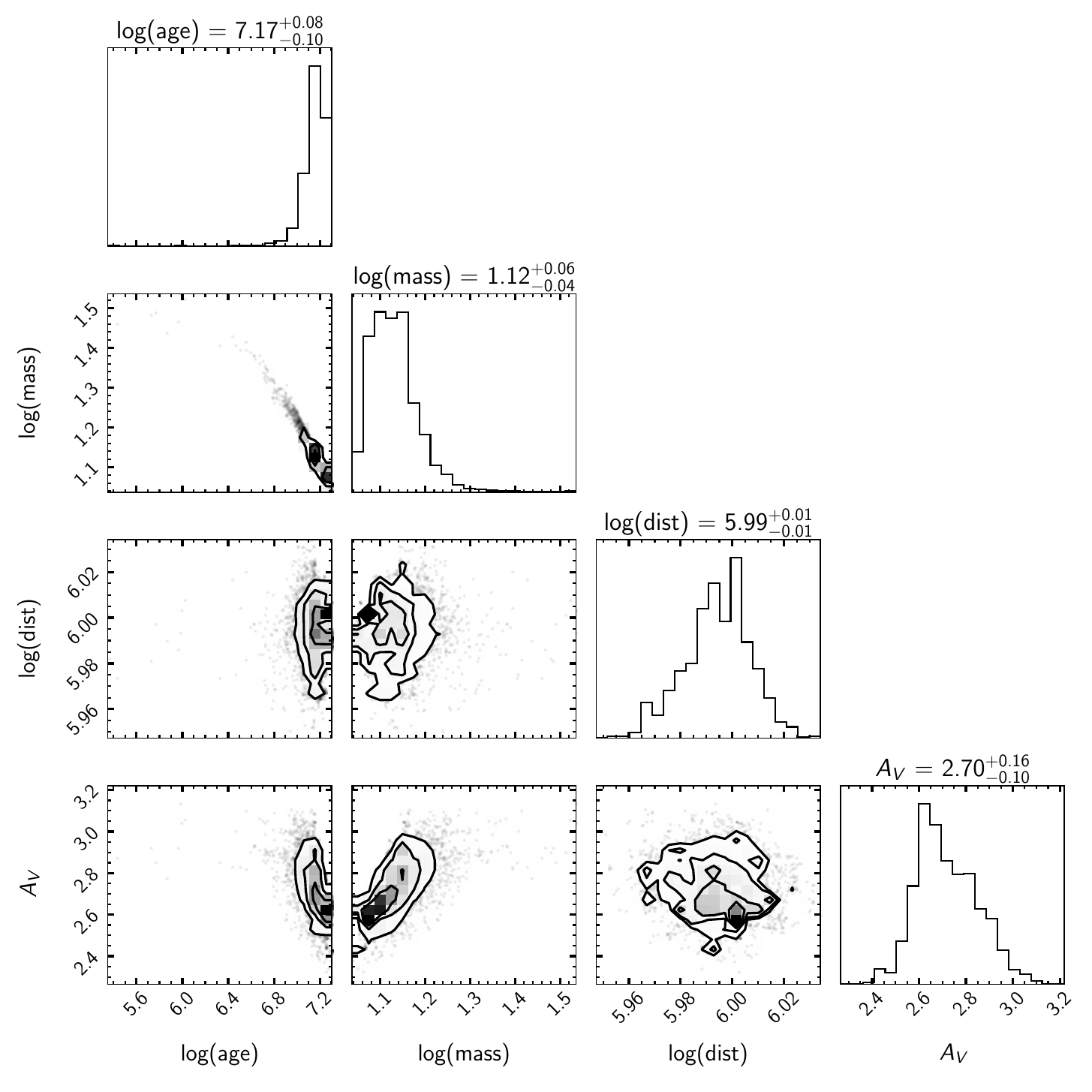}}
\caption{Corner plot of samples drawn from the posterior distribution for Star 6 in Table \ref{tab:awayPDRawayCO_no275_sed} of stars away from the PDR and away from the CO cores not detected in the F275W filter demonstrating the degeneracies between mass, age, distance, and dust.
\label{fig:corner_awayPDRawayCO_no275}}
\end{figure*}

Corner plots of the SED fits for a representative star from each of the seven categories based on their proximity to the PDR and CO cores, along with whether or not they were detected in the F275W filter, are show in Figures \ref{fig:corner_inPDRinCO}, \ref{fig:corner_inPDRinCO_no275}, \ref{fig:corner_inPDRawayCO}, \ref{fig:corner_inPDRawayCO_no275},
\ref{fig:corner_awayPDRinCO},
\ref{fig:corner_awayPDRawayCO}, and \ref{fig:corner_awayPDRawayCO_no275}. No additional stars were detected away from the PDR and projected inside the CO cores when excluding the F275W filter. The values shown in the plots are the posterior median (50th percentile) along with the 84th and 16th percentiles as the upper and lower errors respectively for each parameter. Figure \ref{fig:corner_inPDRawayCO_no275} in particular shows how degeneracies between mass and age can result in multiple solutions.

\begin{deluxetable*}{R|DDDDDDDDDD}
\tablecaption{The sharpness and chi parameters for the five \textit{HST} filters for sources inside the PDR and projected inside the CO cores \label{tab:inPDRinCO_sharpChi}}
\tablehead{\colhead{} & \multicolumn2c{F275W} & \multicolumn2c{F275W} & \multicolumn2c{F336W} & \multicolumn2c{F336W} & \multicolumn2c{F438W} & \multicolumn2c{F438W} & \multicolumn2c{F555W} & \multicolumn2c{F555W} & \multicolumn2c{F625W} & \multicolumn2c{F625W} \\ 
\colhead{} & \multicolumn2c{Sharpness} & \multicolumn2c{$\chi$} & \multicolumn2c{Sharpness} & \multicolumn2c{$\chi$} & \multicolumn2c{Sharpness} & \multicolumn2c{$\chi$} & \multicolumn2c{Sharpness} & \multicolumn2c{$\chi$} & \multicolumn2c{Sharpness} & \multicolumn2c{$\chi$} } 
\decimals
\startdata
1 & 0.12 & 0.25 & -0.07 & 0.23 & -0.06 & 0.5 & -0.06 & 0.57 & -0.43 & 0.89 \\
2 & 0.01 & 0.21 & -0.03 & 0.34 & 0.01 & 0.37 & 0.02 & 0.38 & -0.08 & 0.54 \\
\enddata
\end{deluxetable*}

\begin{deluxetable*}{R|DDDDDDDDDD}
\tablecaption{The sharpness and chi parameters for the five \textit{HST} filters for sources inside the PDR and outside the CO cores \label{tab:inPDRoutCO_sharpChi}}
\tablehead{\colhead{} & \multicolumn2c{F275W} & \multicolumn2c{F275W} & \multicolumn2c{F336W} & \multicolumn2c{F336W} & \multicolumn2c{F438W} & \multicolumn2c{F438W} & \multicolumn2c{F555W} & \multicolumn2c{F555W} & \multicolumn2c{F625W} & \multicolumn2c{F625W} \\ 
\colhead{} & \multicolumn2c{Sharpness} & \multicolumn2c{$\chi$} & \multicolumn2c{Sharpness} & \multicolumn2c{$\chi$} & \multicolumn2c{Sharpness} & \multicolumn2c{$\chi$} & \multicolumn2c{Sharpness} & \multicolumn2c{$\chi$} & \multicolumn2c{Sharpness} & \multicolumn2c{$\chi$} } 
\decimals
\startdata
1 & -0.07 & 0.33 & -0.21 & 0.35 & -0.18 & 0.70 & -0.11 & 0.82 & -0.32 & 0.85 \\
2 & -0.01 & 0.11 & -0.16 & 0.26 & -0.18 & 0.96 & -0.10 & 1.08 & -0.35 & 2.75 \\
3 & -0.09 & 1.19 & 0.09 & 0.86 & -0.09 & 1.14 & -0.21 & 0.89 & -0.07 & 1.92 \\
4 & -0.07 & 0.64 & -0.22 & 0.86 & -0.01 & 0.75 & -0.09 & 1.40 & -0.56 & 1.69 \\
5 & -0.16 & 0.53 & -0.34 & 0.76 & -0.22 & 1.20 & -0.12 & 1.22 & -0.37 & 2.17 \\
6 & -0.16 & 0.44 & 0.11 & 0.41 & -0.04 & 0.48 & -0.09 & 0.63 & -0.44 & 1.08 \\
7 & -0.16 & 0.28 & -0.11 & 0.33 & -0.33 & 0.62 & -0.30 & 0.63 & -0.21 & 0.62 \\
8 & 0.01 & 0.20 & -0.17 & 0.54 & -0.14 & 1.93 & -0.04 & 2.09 & -0.31 & 3.81 \\
9 & 0.10 & 0.77 & 0.23 & 1.09 & 0.24 & 3.10 & 0.25 & 2.52 & -0.44 & 3.44 \\
10 & -0.08 & 0.43 & -0.28 & 0.49 & -0.18 & 0.59 & -0.42 & 0.68 & -0.28 & 1.03 \\
\enddata
\tablecomments{Only a portion of this table is shown here to demonstrate its form and content. A machine-readable version is available for all 443 sources detected inside the PDR and outside the CO cores.}
\end{deluxetable*}

\begin{deluxetable*}{R|DDDDDDDDDD}
\tablecaption{The sharpness and chi parameters for the five \textit{HST} filters for sources outside the PDR and projected inside the CO cores \label{tab:outPDRinCO_sharpChi}}
\tablehead{\colhead{} & \multicolumn2c{F275W} & \multicolumn2c{F275W} & \multicolumn2c{F336W} & \multicolumn2c{F336W} & \multicolumn2c{F438W} & \multicolumn2c{F438W} & \multicolumn2c{F555W} & \multicolumn2c{F555W} & \multicolumn2c{F625W} & \multicolumn2c{F625W} \\ 
\colhead{} & \multicolumn2c{Sharpness} & \multicolumn2c{$\chi$} & \multicolumn2c{Sharpness} & \multicolumn2c{$\chi$} & \multicolumn2c{Sharpness} & \multicolumn2c{$\chi$} & \multicolumn2c{Sharpness} & \multicolumn2c{$\chi$} & \multicolumn2c{Sharpness} & \multicolumn2c{$\chi$} } 
\decimals
\startdata
1 & 0.11 & 0.70 & 0.05 & 0.83 & 0.14 & 1.41 & 0.77 & 4.14 & -0.02 & 1.15 \\
2 & 0.19 & 0.41 & 0.24 & 0.34 & 0.41 & 0.62 & 0.24 & 1.16 & -0.13 & 0.63 \\
\enddata
\end{deluxetable*}

\begin{deluxetable*}{R|DDDDDDDDDD}
\tablecaption{The sharpness and chi parameters for the five \textit{HST} filters for sources outside the PDR and outside the CO cores \label{tab:outPDRoutCO_sharpChi}}
\tablehead{\colhead{} & \multicolumn2c{F275W} & \multicolumn2c{F275W} & \multicolumn2c{F336W} & \multicolumn2c{F336W} & \multicolumn2c{F438W} & \multicolumn2c{F438W} & \multicolumn2c{F555W} & \multicolumn2c{F555W} & \multicolumn2c{F625W} & \multicolumn2c{F625W} \\ 
\colhead{} & \multicolumn2c{Sharpness} & \multicolumn2c{$\chi$} & \multicolumn2c{Sharpness} & \multicolumn2c{$\chi$} & \multicolumn2c{Sharpness} & \multicolumn2c{$\chi$} & \multicolumn2c{Sharpness} & \multicolumn2c{$\chi$} & \multicolumn2c{Sharpness} & \multicolumn2c{$\chi$} }
\decimals
\startdata
1 & -0.07 & 0.25 & -0.22 & 0.31 & 0.02 & 0.32 & 0.01 & 0.49 & -0.17 & 0.63 \\
2 & -0.11 & 0.32 & -0.01 & 0.33 & -0.11 & 0.50 & -0.02 & 0.52 & -0.43 & 1.62 \\
3 & 0.08 & 0.11 & -0.05 & 0.22 & -0.01 & 0.40 & 0.05 & 0.51 & -0.49 & 0.96 \\
4 & 0.09 & 0.13 & -0.11 & 0.34 & 0.57 & 2.00 & 0.65 & 2.02 & -0.56 & 2.19 \\
5 & -0.05 & 0.26 & -0.21 & 0.42 & -0.01 & 0.46 & -0.03 & 0.57 & -0.14 & 0.65 \\
6 & 0.01 & 0.26 & -0.01 & 0.58 & 0.28 & 1.03 & -0.11 & 0.63 & -0.38 & 1.44 \\
7 & -0.62 & 0.17 & -0.21 & 0.27 & 0.05 & 0.51 & -0.04 & 0.85 & -0.38 & 1.15 \\
8 & -0.25 & 0.20 & -0.18 & 0.44 & -0.10 & 1.51 & -0.06 & 1.72 & -0.38 & 2.86 \\
9 & -0.12 & 0.93 & -0.06 & 0.92 & 0.10 & 1.34 & 0.16 & 1.08 & -0.06 & 1.15 \\
10 & -0.05 & 0.17 & -0.02 & 0.29 & 0.14 & 0.86 & 0.14 & 0.71 & -0.22 & 1.24 \\
\enddata
\tablecomments{Only a portion of this table is shown here to demonstrate its form and content. A machine-readable version is available for all 566 sources detected outside the PDR and outside the CO cores.}
\end{deluxetable*}

\begin{deluxetable*}{R|DDDDDDDD}
\tablecaption{The sharpness and chi parameters for the four \textit{HST} filters for the source inside the PDR and projected inside the CO cores not detected in the F275W filter \label{tab:inPDRinCO_sharpChi_no275}}
\tablehead{\colhead{} & \multicolumn2c{F336W} & \multicolumn2c{F336W} & \multicolumn2c{F438W} & \multicolumn2c{F438W} & \multicolumn2c{F555W} & \multicolumn2c{F555W} & \multicolumn2c{F625W} & \multicolumn2c{F625W} \\ 
\colhead{} & \multicolumn2c{Sharpness} & \multicolumn2c{$\chi$} & \multicolumn2c{Sharpness} & \multicolumn2c{$\chi$} & \multicolumn2c{Sharpness} & \multicolumn2c{$\chi$} & \multicolumn2c{Sharpness} & \multicolumn2c{$\chi$} }
\decimals
\startdata
1 & -0.23 & 0.27 & 0.04 & 0.24 & 0.46 & 0.60 & -0.24 & 0.61 \\
\enddata
\end{deluxetable*}

\begin{deluxetable*}{R|DDDDDDDD}
\tablecaption{The sharpness and chi parameters for the four \textit{HST} filters for sources inside the PDR and outside the CO cores not detected in the F275W filter \label{tab:inPDRoutCO_sharpChi_no275}}
\tablehead{\colhead{} & \multicolumn2c{F336W} & \multicolumn2c{F336W} & \multicolumn2c{F438W} & \multicolumn2c{F438W} & \multicolumn2c{F555W} & \multicolumn2c{F555W} & \multicolumn2c{F625W} & \multicolumn2c{F625W} \\ 
\colhead{} & \multicolumn2c{Sharpness} & \multicolumn2c{$\chi$} & \multicolumn2c{Sharpness} & \multicolumn2c{$\chi$} & \multicolumn2c{Sharpness} & \multicolumn2c{$\chi$} & \multicolumn2c{Sharpness} & \multicolumn2c{$\chi$} }
\decimals
\startdata
1 & -0.05 & 0.22 & -0.00 & 0.40 & 0.05 & 0.51 & -0.49 & 0.95 \\
2 & -0.11 & 0.34 & 0.57 & 2.00 & 0.65 & 2.02 & -0.56 & 2.19 \\
3 & -0.16 & 0.49 & -0.14 & 0.61 & -0.27 & 0.85 & -0.28 & 0.98 \\
4 & -0.06 & 0.92 & 0.10 & 1.34 & 0.16 & 1.07 & -0.06 & 1.15 \\
5 & -0.19 & 2.60 & 0.28 & 2.31 & 0.14 & 2.12 & -0.51 & 3.68 \\
6 & 0.03 & 0.35 & 0.04 & 0.47 & 0.10 & 0.80 & -0.16 & 0.83 \\
7 & -0.29 & 0.21 & 0.10 & 0.53 & 0.13 & 0.53 & -0.56 & 1.57 \\
8 & -0.18 & 0.34 & 0.07 & 0.94 & 0.05 & 0.96 & -0.42 & 1.37 \\
9 & -0.38 & 0.29 & -0.03 & 0.49 & -0.02 & 0.54 & -0.31 & 1.28 \\
10 & -0.13 & 0.20 & -0.14 & 0.55 & -0.05 & 1.23 & -0.62 & 3.40 \\
\enddata
\tablecomments{Only a portion of this table is shown here to demonstrate its form and content. A machine-readable version is available for all 144 sources detected inside the PDR and outside the CO cores.}
\end{deluxetable*}

\begin{deluxetable*}{R|DDDDDDDD}
\tablecaption{The sharpness and chi parameters for the four \textit{HST} filters for sources outside the PDR and outside the CO cores not detected in the F275W filter \label{tab:outPDRoutCO_sharpChi_no275}}
\tablehead{\colhead{} & \multicolumn2c{F336W} & \multicolumn2c{F336W} & \multicolumn2c{F438W} & \multicolumn2c{F438W} & \multicolumn2c{F555W} & \multicolumn2c{F555W} & \multicolumn2c{F625W} & \multicolumn2c{F625W} \\ 
\colhead{} & \multicolumn2c{Sharpness} & \multicolumn2c{$\chi$} & \multicolumn2c{Sharpness} & \multicolumn2c{$\chi$} & \multicolumn2c{Sharpness} & \multicolumn2c{$\chi$} & \multicolumn2c{Sharpness} & \multicolumn2c{$\chi$} }
\decimals
\startdata
1 & 0.16 & 0.20 & -0.01 & 0.63 & 0.05 & 1.00 & -0.44 & 2.84 \\
2 & -0.30 & 0.30 & -0.10 & 0.51 & 0.09 & 0.77 & -0.28 & 1.88 \\
3 & -0.10 & 0.30 & -0.05 & 1.27 & -0.13 & 1.76 & -0.21 & 4.00 \\
4 & -0.03 & 0.38 & 0.09 & 0.72 & -0.04 & 1.25 & -0.35 & 3.83 \\
5 & -0.13 & 0.29 & 0.28 & 0.90 & 0.17 & 1.08 & -0.37 & 1.32 \\
6 & -0.37 & 0.28 & -0.20 & 0.40 & -0.10 & 0.50 & -0.13 & 0.93 \\
7 & -0.07 & 0.21 & -0.05 & 0.63 & 0.07 & 0.68 & -0.13 & 1.55 \\
8 & -0.05 & 0.49 & 0.29 & 3.19 & 0.05 & 4.45 & -0.23 & 5.18 \\
9 & -0.09 & 0.23 & -0.02 & 0.55 & 0.06 & 0.68 & -0.13 & 2.05 \\
10 & 0.17 & 0.18 & -0.13 & 0.78 & 0.05 & 1.35 & -0.09 & 3.05 \\
\enddata
\tablecomments{Only a portion of this table is shown here to demonstrate its form and content. A machine-readable version is available for all 78 sources detected outside the PDR and outside the CO cores.}
\end{deluxetable*}

\bibliography{hst_co}{}

\begin{thebibliography}{}
\expandafter\ifx\csname natexlab\endcsname\relax\def\natexlab#1{#1}\fi
\providecommand{\url}[1]{\href{#1}{#1}}
\providecommand{\dodoi}[1]{doi:~\href{http://doi.org/#1}{\nolinkurl{#1}}}
\providecommand{\doeprint}[1]{\href{http://ascl.net/#1}{\nolinkurl{http://ascl.net/#1}}}
\providecommand{\doarXiv}[1]{\href{https://arxiv.org/abs/#1}{\nolinkurl{https://arxiv.org/abs/#1}}}

\bibitem[{{Albers} {et~al.}(2019){Albers}, {Weisz}, {Cole}, {Dolphin}, {Skillman}, {Williams}, {Boylan-Kolchin}, {Bullock}, {Dalcanton}, {Hopkins}, {Leaman}, {McConnachie}, {Vogelsberger}, \& {Wetzel}}]{Albers_2019}
{Albers}, S.~M., {Weisz}, D.~R., {Cole}, A.~A., {et~al.} 2019, \mnras, 490, 5538, \dodoi{10.1093/mnras/stz2903}

\bibitem[{{Anderson} {et~al.}(2021){Anderson}, {Baggett}, \& {Kuhn}}]{Anderson_2021}
{Anderson}, J., {Baggett}, S., \& {Kuhn}, B. 2021, {Updating the WFC3/UVIS CTE model and Mitigation Strategies}, Instrument Science Report 2021-9, 44 pages

\bibitem[{{Annunziatella} {et~al.}(2013){Annunziatella}, {Mercurio}, {Brescia}, {Cavuoti}, \& {Longo}}]{annunziatella_2013}
{Annunziatella}, M., {Mercurio}, A., {Brescia}, M., {Cavuoti}, S., \& {Longo}, G. 2013, \pasp, 125, 68, \dodoi{10.1086/669333}

\bibitem[{{Archer} {et~al.}(2022{\natexlab{a}}){Archer}, {Cigan}, {Elmegreen}, {Hunt}, {Hunter}, {Jansen}, {Rubio}, \& {Windhorst}}]{archer_hst_prop}
{Archer}, H.~N., {Cigan}, P., {Elmegreen}, B., {et~al.} 2022{\natexlab{a}}, {Young Stars and Gas Structure within the ALMA Coverage of Dwarf Irregular Galaxy WLM}, HST Proposal. Cycle 30, ID. \#17068

\bibitem[{{Archer} {et~al.}(2022{\natexlab{b}}){Archer}, {Hunter}, {Elmegreen}, {Cigan}, {Jansen}, {Windhorst}, {Hunt}, \& {Rubio}}]{archer_2022}
{Archer}, H.~N., {Hunter}, D.~A., {Elmegreen}, B.~G., {et~al.} 2022{\natexlab{b}}, \aj, 163, 141, \dodoi{10.3847/1538-3881/ac4e88}

\bibitem[{Bezanson {et~al.}(2017)Bezanson, Edelman, Karpinski, \& Shah}]{bezanson_2017}
Bezanson, J., Edelman, A., Karpinski, S., \& Shah, V.~B. 2017, SIAM {R}eview, 59, 65, \dodoi{10.1137/141000671}

\bibitem[{{Bolatto} {et~al.}(2013){Bolatto}, {Wolfire}, \& {Leroy}}]{Bolatto_2013}
{Bolatto}, A.~D., {Wolfire}, M., \& {Leroy}, A.~K. 2013, \araa, 51, 207, \dodoi{10.1146/annurev-astro-082812-140944}

\bibitem[{{Boyer} {et~al.}(2024){Boyer}, {Pastorelli}, {Girardi}, {Marigo}, {Dolphin}, {McQuinn}, {Newman}, {Savino}, {Weisz}, {Williams}, {Anderson}, {Cohen}, {Correnti}, {Cole}, {Geha}, {Gennaro}, {Kallivayalil}, {Kirby}, {Sandstrom}, {Skillman}, {Garling}, {Richstein}, \& {Warfield}}]{Boyer_2024}
{Boyer}, M.~L., {Pastorelli}, G., {Girardi}, L., {et~al.} 2024, \apj, 973, 120, \dodoi{10.3847/1538-4357/ad6449}

\bibitem[{{Bressan} {et~al.}(2012){Bressan}, {Marigo}, {Girardi}, {Salasnich}, {Dal Cero}, {Rubele}, \& {Nanni}}]{Bressan_2012}
{Bressan}, A., {Marigo}, P., {Girardi}, L., {et~al.} 2012, \mnras, 427, 127, \dodoi{10.1111/j.1365-2966.2012.21948.x}

\bibitem[{{Brosch} {et~al.}(1998){Brosch}, {Heller}, \& {Almoznino}}]{brosch_1998}
{Brosch}, N., {Heller}, A., \& {Almoznino}, E. 1998, ApJ, 504, 720, \dodoi{10.1086/306127}

\bibitem[{{Chen} {et~al.}(2015){Chen}, {Bressan}, {Girardi}, {Marigo}, {Kong}, \& {Lanza}}]{Chen_2015}
{Chen}, Y., {Bressan}, A., {Girardi}, L., {et~al.} 2015, \mnras, 452, 1068, \dodoi{10.1093/mnras/stv1281}

\bibitem[{{Chen} {et~al.}(2014){Chen}, {Girardi}, {Bressan}, {Marigo}, {Barbieri}, \& {Kong}}]{chen_2014}
{Chen}, Y., {Girardi}, L., {Bressan}, A., {et~al.} 2014, \mnras, 444, 2525, \dodoi{10.1093/mnras/stu1605}

\bibitem[{{Chevance} {et~al.}(2020{\natexlab{a}}){Chevance}, {Kruijssen}, {Hygate}, {Schruba}, {Longmore}, {Groves}, {Henshaw}, {Herrera}, {Hughes}, {Jeffreson}, {Lang}, {Leroy}, {Meidt}, {Pety}, {Razza}, {Rosolowsky}, {Schinnerer}, {Bigiel}, {Blanc}, {Emsellem}, {Faesi}, {Glover}, {Haydon}, {Ho}, {Kreckel}, {Lee}, {Liu}, {Querejeta}, {Saito}, {Sun}, {Usero}, \& {Utomo}}]{Chevance_2020_1}
{Chevance}, M., {Kruijssen}, J. M.~D., {Hygate}, A., {et~al.} 2020{\natexlab{a}}, MNRAS, 493, 2872, \dodoi{10.1093/mnras/stz3525}

\bibitem[{{Chevance} {et~al.}(2020{\natexlab{b}}){Chevance}, {Madden}, {Fischer}, {Vacca}, {Lebouteiller}, {Fadda}, {Galliano}, {Indebetouw}, {Kruijssen}, {Lee}, {Poglitsch}, {Polles}, {Cormier}, {Hony}, {Iserlohe}, {Krabbe}, {Meixner}, {Sabbi}, \& {Zinnecker}}]{Chevance_2020_2}
{Chevance}, M., {Madden}, S.~C., {Fischer}, C., {et~al.} 2020{\natexlab{b}}, \mnras, 494, 5279, \dodoi{10.1093/mnras/staa1106}

\bibitem[{{Choi} {et~al.}(2016){Choi}, {Dotter}, {Conroy}, {Cantiello}, {Paxton}, \& {Johnson}}]{Choi_2016}
{Choi}, J., {Dotter}, A., {Conroy}, C., {et~al.} 2016, \apj, 823, 102, \dodoi{10.3847/0004-637X/823/2/102}

\bibitem[{{Cigan} {et~al.}(2016){Cigan}, {Young}, {Cormier}, {Lebouteiller}, {Madden}, {Hunter}, {Brinks}, {Elmegreen}, {Schruba}, {Heesen}, \& {the <SPAN CLASS=''sml''>Little Things</SPAN> Team}}]{Cigan_2016}
{Cigan}, P., {Young}, L., {Cormier}, D., {et~al.} 2016, \aj, 151, 14, \dodoi{10.3847/0004-6256/151/1/14}

\bibitem[{{Cormier} {et~al.}(2017){Cormier}, {Bendo}, {Hony}, {Lebouteiller}, {Madden}, {Galliano}, {Glover}, {Klessen}, {Abel}, {Bigiel}, \& {Clark}}]{Cormier_2017}
{Cormier}, D., {Bendo}, G.~J., {Hony}, S., {et~al.} 2017, \mnras, 468, L87, \dodoi{10.1093/mnrasl/slx034}

\bibitem[{{Dale} {et~al.}(2009){Dale}, {Cohen}, {Johnson}, {Schuster}, {Calzetti}, {Engelbracht}, {Gil de Paz}, {Kennicutt}, {Lee}, {Begum}, {Block}, {Dalcanton}, {Funes}, {Gordon}, {Johnson}, {Marble}, {Sakai}, {Skillman}, {van Zee}, {Walter}, {Weisz}, {Williams}, {Wu}, \& {Wu}}]{dale_2009}
{Dale}, D.~A., {Cohen}, S.~A., {Johnson}, L.~C., {et~al.} 2009, \apj, 703, 517, \dodoi{10.1088/0004-637X/703/1/517}

\bibitem[{{Draine} \& {Li}(2007)}]{drain_2007}
{Draine}, B.~T., \& {Li}, A. 2007, ApJ, 657, 810, \dodoi{10.1086/511055}

\bibitem[{{Elmegreen}(1989)}]{Elmegreen_1989}
{Elmegreen}, B.~G. 1989, \apj, 338, 178, \dodoi{10.1086/167192}

\bibitem[{{Elmegreen}(2000)}]{Elmegreen_2000}
---. 2000, \apj, 530, 277, \dodoi{10.1086/308361}

\bibitem[{{Elmegreen} {et~al.}(1980){Elmegreen}, {Morris}, \& {Elmegreen}}]{Elmegreen_1980}
{Elmegreen}, B.~G., {Morris}, M., \& {Elmegreen}, D.~M. 1980, \apj, 240, 455, \dodoi{10.1086/158251}

\bibitem[{{Elmegreen} {et~al.}(2013){Elmegreen}, {Rubio}, {Hunter}, {Verdugo}, {Brinks}, \& {Schruba}}]{Elmegreen_2013}
{Elmegreen}, B.~G., {Rubio}, M., {Hunter}, D.~A., {et~al.} 2013, \nat, 495, 487, \dodoi{10.1038/nature11933}

\bibitem[{{Fukui} \& {Kawamura}(2010)}]{fukui_2010}
{Fukui}, Y., \& {Kawamura}, A. 2010, AARA, 48, 547, \dodoi{10.1146/annurev-astro-081309-130854}

\bibitem[{{Gordon} {et~al.}(2003){Gordon}, {Clayton}, {Misselt}, {Landolt}, \& {Wolff}}]{gordon_2003}
{Gordon}, K.~D., {Clayton}, G.~C., {Misselt}, K.~A., {Landolt}, A.~U., \& {Wolff}, M.~J. 2003, \apj, 594, 279, \dodoi{10.1086/376774}

\bibitem[{{Gordon} {et~al.}(2016){Gordon}, {Fouesneau}, {Arab}, {Tchernyshyov}, {Weisz}, {Dalcanton}, {Williams}, {Bell}, {Bianchi}, {Boyer}, {Choi}, {Dolphin}, {Girardi}, {Hogg}, {Kalirai}, {Kapala}, {Lewis}, {Rix}, {Sandstrom}, \& {Skillman}}]{Gordon_2016}
{Gordon}, K.~D., {Fouesneau}, M., {Arab}, H., {et~al.} 2016, \apj, 826, 104, \dodoi{10.3847/0004-637X/826/2/104}

\bibitem[{Harris {et~al.}(2020)Harris, Millman, van~der Walt, Gommers, Virtanen, Cournapeau, Wieser, Taylor, Berg, Smith, Kern, Picus, Hoyer, van Kerkwijk, Brett, Haldane, del R{\'{i}}o, Wiebe, Peterson, G{\'{e}}rard-Marchant, Sheppard, Reddy, Weckesser, Abbasi, Gohlke, \& Oliphant}]{harris_2020}
Harris, C.~R., Millman, K.~J., van~der Walt, S.~J., {et~al.} 2020, Nature, 585, 357, \dodoi{10.1038/s41586-020-2649-2}

\bibitem[{Hoffman \& Gelman(2011)}]{hoffman_2011}
Hoffman, M.~D., \& Gelman, A. 2011, The No-U-Turn Sampler: Adaptively Setting Path Lengths in Hamiltonian Monte Carlo.
\newblock \doarXiv{1111.4246}

\bibitem[{{Hoffmann} {et~al.}(2021){Hoffmann}, {Mack}, {Avila}, {Martlin}, {Cohen}, \& {Bajaj}}]{hoffmann_2021}
{Hoffmann}, S.~L., {Mack}, J., {Avila}, R., {et~al.} 2021, in American Astronomical Society Meeting Abstracts, Vol.~53, American Astronomical Society Meeting Abstracts, 216.02

\bibitem[{{Hu} {et~al.}(2022){Hu}, {Schruba}, {Sternberg}, \& {van Dishoeck}}]{Hu_2022}
{Hu}, C.-Y., {Schruba}, A., {Sternberg}, A., \& {van Dishoeck}, E.~F. 2022, \apj, 931, 28, \dodoi{10.3847/1538-4357/ac65fd}

\bibitem[{{Hu} {et~al.}(2021){Hu}, {Sternberg}, \& {van Dishoeck}}]{Hu_2021}
{Hu}, C.-Y., {Sternberg}, A., \& {van Dishoeck}, E.~F. 2021, \apj, 920, 44, \dodoi{10.3847/1538-4357/ac0dbd}

\bibitem[{{Hu} {et~al.}(2023){Hu}, {Sternberg}, \& {van Dishoeck}}]{Hu_2023}
---. 2023, \apj, 952, 140, \dodoi{10.3847/1538-4357/acdcfa}

\bibitem[{{Hunt} {et~al.}(2023){Hunt}, {Belfiore}, {Lelli}, {Draine}, {Marasco}, {Garc{\'\i}a-Burillo}, {Venturi}, {Combes}, {Wei{\ss}}, {Henkel}, {Menten}, {Annibali}, {Casasola}, {Cignoni}, {McLeod}, {Tosi}, {Beltr{\'a}n}, {Concas}, {Cresci}, {Ginolfi}, {Kumari}, \& {Mannucci}}]{Hunt_2023}
{Hunt}, L.~K., {Belfiore}, F., {Lelli}, F., {et~al.} 2023, \aap, 675, A64, \dodoi{10.1051/0004-6361/202245805}

\bibitem[{{Hunter} {et~al.}(2024){Hunter}, {Elmegreen}, \& {Madden}}]{Hunter_2024}
{Hunter}, D.~A., {Elmegreen}, B.~G., \& {Madden}, S.~C. 2024, AARA, 62, 113, \dodoi{10.1146/annurev-astro-052722-104109}

\bibitem[{{Hunter} {et~al.}(1998){Hunter}, {Wilcots}, {van Woerden}, {Gallagher}, \& {Kohle}}]{Hunter_1998}
{Hunter}, D.~A., {Wilcots}, E.~M., {van Woerden}, H., {Gallagher}, J.~S., \& {Kohle}, S. 1998, ApJL, 495, L47, \dodoi{10.1086/311213}

\bibitem[{{Hunter} {et~al.}(2012){Hunter}, {Ficut-Vicas}, {Ashley}, {Brinks}, {Cigan}, {Elmegreen}, {Heesen}, {Herrmann}, {Johnson}, {Oh}, {Rupen}, {Schruba}, {Simpson}, {Walter}, {Westpfahl}, {Young}, \& {Zhang}}]{Hunter_2012}
{Hunter}, D.~A., {Ficut-Vicas}, D., {Ashley}, T., {et~al.} 2012, \aj, 144, 134, \dodoi{10.1088/0004-6256/144/5/134}

\bibitem[{Hunter(2007)}]{hunter_2007}
Hunter, J.~D. 2007, Computing in Science \& Engineering, 9, 90, \dodoi{10.1109/MCSE.2007.55}

\bibitem[{{Iorio} {et~al.}(2017){Iorio}, {Fraternali}, {Nipoti}, {Di Teodoro}, {Read}, \& {Battaglia}}]{iorio_2017}
{Iorio}, G., {Fraternali}, F., {Nipoti}, C., {et~al.} 2017, \mnras, 466, 4159, \dodoi{10.1093/mnras/stw3285}

\bibitem[{Kingma \& Ba(2017)}]{kingma_2017}
Kingma, D.~P., \& Ba, J. 2017, Adam: A Method for Stochastic Optimization.
\newblock \doarXiv{1412.6980}

\bibitem[{{Kroupa}(2002)}]{kroupa_2002}
{Kroupa}, P. 2002, Science, 295, 82, \dodoi{10.1126/science.1067524}

\bibitem[{{Krumholz} {et~al.}(2012){Krumholz}, {Dekel}, \& {McKee}}]{krumholz_2012}
{Krumholz}, M.~R., {Dekel}, A., \& {McKee}, C.~F. 2012, \apj, 745, 69, \dodoi{10.1088/0004-637X/745/1/69}

\bibitem[{{Leaman} {et~al.}(2012){Leaman}, {Venn}, {Brooks}, {Battaglia}, {Cole}, {Ibata}, {Irwin}, {McConnachie}, {Mendel}, \& {Tolstoy}}]{Leaman_2012}
{Leaman}, R., {Venn}, K.~A., {Brooks}, A.~M., {et~al.} 2012, \apj, 750, 33, \dodoi{10.1088/0004-637X/750/1/33}

\bibitem[{{Lee} {et~al.}(2021){Lee}, {Freedman}, {Madore}, {Owens}, {Monson}, \& {Hoyt}}]{Lee_2021}
{Lee}, A.~J., {Freedman}, W.~L., {Madore}, B.~F., {et~al.} 2021, \apj, 907, 112, \dodoi{10.3847/1538-4357/abd253}

\bibitem[{{Lee} {et~al.}(2005){Lee}, {Skillman}, \& {Venn}}]{Lee_2005}
{Lee}, H., {Skillman}, E.~D., \& {Venn}, K.~A. 2005, \apj, 620, 223, \dodoi{10.1086/427019}

\bibitem[{{Leja} {et~al.}(2019){Leja}, {Johnson}, {Conroy}, {van Dokkum}, {Speagle}, {Brammer}, {Momcheva}, {Skelton}, {Whitaker}, {Franx}, \& {Nelson}}]{leja_19}
{Leja}, J., {Johnson}, B.~D., {Conroy}, C., {et~al.} 2019, \apj, 877, 140, \dodoi{10.3847/1538-4357/ab1d5a}

\bibitem[{{Leroy} {et~al.}(2008){Leroy}, {Walter}, {Brinks}, {Bigiel}, {de Blok}, {Madore}, \& {Thornley}}]{Leroy_2008}
{Leroy}, A.~K., {Walter}, F., {Brinks}, E., {et~al.} 2008, AJ, 136, 2782, \dodoi{10.1088/0004-6256/136/6/2782}

\bibitem[{{Madden}(2022)}]{madden_2022}
{Madden}, S.~C. 2022, in European Physical Journal Web of Conferences, Vol. 265, European Physical Journal Web of Conferences, 00011, \dodoi{10.1051/epjconf/202226500011}

\bibitem[{{Madden} {et~al.}(2020){Madden}, {Cormier}, {Hony}, {Lebouteiller}, {Abel}, {Galametz}, {De Looze}, {Chevance}, {Polles}, {Lee}, {Galliano}, {Lambert-Huyghe}, {Hu}, \& {Ramambason}}]{madden_2020}
{Madden}, S.~C., {Cormier}, D., {Hony}, S., {et~al.} 2020, \aap, 643, A141, \dodoi{10.1051/0004-6361/202038860}

\bibitem[{{Marigo} {et~al.}(2017){Marigo}, {Girardi}, {Bressan}, {Rosenfield}, {Aringer}, {Chen}, {Dussin}, {Nanni}, {Pastorelli}, {Rodrigues}, {Trabucchi}, {Bladh}, {Dalcanton}, {Groenewegen}, {Montalb{\'a}n}, \& {Wood}}]{marigo_2017}
{Marigo}, P., {Girardi}, L., {Bressan}, A., {et~al.} 2017, \apj, 835, 77, \dodoi{10.3847/1538-4357/835/1/77}

\bibitem[{{Martin} {et~al.}(2005){Martin}, {Fanson}, {Schiminovich}, {Morrissey}, {Friedman}, {Barlow}, {Conrow}, {Grange}, {Jelinsky}, {Milliard}, {Siegmund}, {Bianchi}, {Byun}, {Donas}, {Forster}, {Heckman}, {Lee}, {Madore}, {Malina}, {Neff}, {Rich}, {Small}, {Surber}, {Szalay}, {Welsh}, \& {Wyder}}]{galex}
{Martin}, D.~C., {Fanson}, J., {Schiminovich}, D., {et~al.} 2005, \apjl, 619, L1, \dodoi{10.1086/426387}

\bibitem[{{McQuinn} {et~al.}(2024){McQuinn}, {B. Newman}, {Savino}, {Dolphin}, {Weisz}, {Williams}, {Boyer}, {Cohen}, {Correnti}, {Cole}, {Geha}, {Gennaro}, {Kallivayalil}, {Sandstrom}, {Skillman}, {Anderson}, {Bolatto}, {Boylan-Kolchin}, {Garling}, {Gilbert}, {Girardi}, {Kalirai}, {Mazzi}, {Pastorelli}, {Richstein}, \& {Warfield}}]{McQuinn_2024}
{McQuinn}, K. B.~W., {B. Newman}, M.~J., {Savino}, A., {et~al.} 2024, \apj, 961, 16, \dodoi{10.3847/1538-4357/ad1105}

\bibitem[{Mogensen \& Riseth(2018)}]{mogensen_2018}
Mogensen, P.~K., \& Riseth, A.~N. 2018, Journal of Open Source Software, 3, 615, \dodoi{10.21105/joss.00615}

\bibitem[{{Newman} {et~al.}(2024){Newman}, {McQuinn}, {Skillman}, {Boyer}, {Cohen}, {Dolphin}, \& {Telford}}]{Newman_2024}
{Newman}, M. J.~B., {McQuinn}, K. B.~W., {Skillman}, E.~D., {et~al.} 2024, \apj, 975, 195, \dodoi{10.3847/1538-4357/ad79f8}

\bibitem[{{Nicolet}(1980)}]{nicolet_1980}
{Nicolet}, B. 1980, \aaps, 42, 283

\bibitem[{{Osman} {et~al.}(2020){Osman}, {Bekki}, \& {Cortese}}]{osman_2020}
{Osman}, O., {Bekki}, K., \& {Cortese}, L. 2020, MNRAS, 497, 2002, \dodoi{10.1093/mnras/staa1554}

\bibitem[{{Pastorelli} {et~al.}(2019){Pastorelli}, {Marigo}, {Girardi}, {Chen}, {Rubele}, {Trabucchi}, {Aringer}, {Bladh}, {Bressan}, {Montalb{\'a}n}, {Boyer}, {Dalcanton}, {Eriksson}, {Groenewegen}, {H{\"o}fner}, {Lebzelter}, {Nanni}, {Rosenfield}, {Wood}, \& {Cioni}}]{pastorelli_2019}
{Pastorelli}, G., {Marigo}, P., {Girardi}, L., {et~al.} 2019, \mnras, 485, 5666, \dodoi{10.1093/mnras/stz725}

\bibitem[{{Pastorelli} {et~al.}(2020){Pastorelli}, {Marigo}, {Girardi}, {Aringer}, {Chen}, {Rubele}, {Trabucchi}, {Bladh}, {Boyer}, {Bressan}, {Dalcanton}, {Groenewegen}, {Lebzelter}, {Mowlavi}, {Chubb}, {Cioni}, {de Grijs}, {Ivanov}, {Nanni}, {van Loon}, \& {Zaggia}}]{pastorelli_2020}
---. 2020, \mnras, 498, 3283, \dodoi{10.1093/mnras/staa2565}

\bibitem[{{Pedregosa} {et~al.}(2011){Pedregosa}, {Varoquaux}, {Gramfort}, {Michel}, {Thirion}, {Grisel}, {Blondel}, {M{\"u}ller}, {Nothman}, {Louppe}, {Prettenhofer}, {Weiss}, {Dubourg}, {Vanderplas}, {Passos}, {Cournapeau}, {Brucher}, {Perrot}, \& {Duchesnay}}]{sklearn}
{Pedregosa}, F., {Varoquaux}, G., {Gramfort}, A., {et~al.} 2011, Journal of Machine Learning Research, 12, 2825, \dodoi{10.48550/arXiv.1201.0490}

\bibitem[{{Peltonen} {et~al.}(2024){Peltonen}, {Rosolowsky}, {Williams}, {Koch}, {Dolphin}, {Chastenet}, {Dalcanton}, {Ginsburg}, {Johnson}, {Leroy}, {Richardson}, {Sandstrom}, {Sarbadhicary}, {Smercina}, {Wainer}, \& {Williams}}]{Peltonen_2024}
{Peltonen}, J., {Rosolowsky}, E., {Williams}, T.~G., {et~al.} 2024, \mnras, 527, 10668, \dodoi{10.1093/mnras/stad3879}

\bibitem[{{Pineda} {et~al.}(2014){Pineda}, {Langer}, \& {Goldsmith}}]{Pineda_2014}
{Pineda}, J.~L., {Langer}, W.~D., \& {Goldsmith}, P.~F. 2014, \aap, 570, A121, \dodoi{10.1051/0004-6361/201424054}

\bibitem[{{Planck Collaboration} {et~al.}(2011){Planck Collaboration}, {Ade}, {Aghanim}, {Arnaud}, {Ashdown}, {Aumont}, {Baccigalupi}, {Balbi}, {Banday}, {Barreiro}, {Bartlett}, {Battaner}, {Benabed}, {Beno{\^\i}t}, {Bernard}, {Bersanelli}, {Bhatia}, {Bock}, {Bonaldi}, {Bond}, {Borrill}, {Bouchet}, {Boulanger}, {Bucher}, {Burigana}, {Cabella}, {Cardoso}, {Catalano}, {Cay{\'o}n}, {Challinor}, {Chamballu}, {Chiang}, {Chiang}, {Christensen}, {Clements}, {Colombi}, {Couchot}, {Coulais}, {Crill}, {Cuttaia}, {Dame}, {Danese}, {Davies}, {Davis}, {de Bernardis}, {de Gasperis}, {de Rosa}, {de Zotti}, {Delabrouille}, {Delouis}, {D{\'e}sert}, {Dickinson}, {Dobashi}, {Donzelli}, {Dor{\'e}}, {D{\"o}rl}, {Douspis}, {Dupac}, {Efstathiou}, {En{\ss}lin}, {Eriksen}, {Falgarone}, {Finelli}, {Forni}, {Fosalba}, {Frailis}, {Franceschi}, {Fukui}, {Galeotta}, {Ganga}, {Giard}, {Giardino}, {Giraud-H{\'e}raud}, {Gonz{\'a}lez-Nuevo}, {G{\'o}rski}, {Gratton}, {Gregorio}, {Grenier}, {Gruppuso}, {Hansen}, {Harrison}, {Helou},
  {Henrot-Versill{\'e}}, {Herranz}, {Hildebrandt}, {Hivon}, {Hobson}, {Holmes}, {Hovest}, {Hoyland}, {Huffenberger}, {Jaffe}, {Jones}, {Juvela}, {Kawamura}, {Keih{\"a}nen}, {Keskitalo}, {Kisner}, {Kneissl}, {Knox}, {Kurki-Suonio}, {Lagache}, {Lamarre}, {Lasenby}, {Laureijs}, {Lawrence}, {Leach}, {Leonardi}, {Leroy}, {Lilje}, {Linden-V{\o}rnle}, {L{\'o}pez-Caniego}, {Lubin}, {Mac{\'\i}as-P{\'e}rez}, {MacTavish}, {Maffei}, {Maino}, {Mandolesi}, {Mann}, {Maris}, {Martin}, {Mart{\'\i}nez-Gonz{\'a}lez}, {Masi}, {Matarrese}, {Matthai}, {Mazzotta}, {McGehee}, {Meinhold}, {Melchiorri}, {Mendes}, {Mennella}, {Miville-Desch{\^e}nes}, {Moneti}, {Montier}, {Morgante}, {Mortlock}, {Munshi}, {Murphy}, {Naselsky}, {Natoli}, {Netterfield}, {N{\o}rgaard-Nielsen}, {Noviello}, {Novikov}, {Novikov}, {O'Dwyer}, {Onishi}, {Osborne}, {Pajot}, {Paladini}, {Paradis}, {Pasian}, {Patanchon}, {Perdereau}, {Perotto}, {Perrotta}, {Piacentini}, {Piat}, {Plaszczynski}, {Pointecouteau}, {Polenta}, {Ponthieu}, {Poutanen}, {Pr{\'e}zeau},
  {Prunet}, {Puget}, {Reach}, {Reinecke}, {Renault}, {Ricciardi}, {Riller}, {Ristorcelli}, {Rocha}, {Rosset}, {Rowan-Robinson}, {Rubi{\~n}o-Mart{\'\i}n}, {Rusholme}, {Sandri}, {Santos}, {Savini}, {Scott}, {Seiffert}, {Shellard}, {Smoot}, {Starck}, {Stivoli}, {Stolyarov}, {Stompor}, {Sudiwala}, {Sygnet}, {Tauber}, {Terenzi}, {Toffolatti}, {Tomasi}, {Torre}, {Tristram}, {Tuovinen}, {Umana}, {Valenziano}, {Vielva}, {Villa}, {Vittorio}, {Wade}, {Wandelt}, {Wilkinson}, {Yvon}, {Zacchei}, \& {Zonca}}]{planck_2011}
{Planck Collaboration}, {Ade}, P.~A.~R., {Aghanim}, N., {et~al.} 2011, \aap, 536, A19, \dodoi{10.1051/0004-6361/201116479}

\bibitem[{{Ramambason} {et~al.}(2024){Ramambason}, {Lebouteiller}, {Madden}, {Galliano}, {Richardson}, {Saintonge}, {De Looze}, {Chevance}, {Abel}, {Hernandez}, \& {Braine}}]{Ramambason_2024}
{Ramambason}, L., {Lebouteiller}, V., {Madden}, S.~C., {et~al.} 2024, \aap, 681, A14, \dodoi{10.1051/0004-6361/202347280}

\bibitem[{{Requena-Torres} {et~al.}(2016){Requena-Torres}, {Israel}, {Okada}, {G{\"u}sten}, {Stutzki}, {Risacher}, {Simon}, \& {Zinnecker}}]{raquena_2016}
{Requena-Torres}, M.~A., {Israel}, F.~P., {Okada}, Y., {et~al.} 2016, \aap, 589, A28, \dodoi{10.1051/0004-6361/201526244}

\bibitem[{{Rubio} {et~al.}(2015){Rubio}, {Elmegreen}, {Hunter}, {Brinks}, {Cort{\'e}s}, \& {Cigan}}]{Rubio_2015}
{Rubio}, M., {Elmegreen}, B.~G., {Hunter}, D.~A., {et~al.} 2015, \nat, 525, 218, \dodoi{10.1038/nature14901}

\bibitem[{{Salda{\~n}o} {et~al.}(2023){Salda{\~n}o}, {Rubio}, {Bolatto}, {Verdugo}, {Jameson}, \& {Leroy}}]{Saldano_2023}
{Salda{\~n}o}, H.~P., {Rubio}, M., {Bolatto}, A.~D., {et~al.} 2023, \aap, 672, A153, \dodoi{10.1051/0004-6361/202142217}

\bibitem[{{Stan Development Team}(2024)}]{stan_2024}
{Stan Development Team}. 2024, {RStan}: the {R} interface to {Stan}.
\newblock \url{https://mc-stan.org/}

\bibitem[{{Stetson}(1987)}]{stetson_1987}
{Stetson}, P.~B. 1987, \pasp, 99, 191, \dodoi{10.1086/131977}

\bibitem[{{Tang} {et~al.}(2014){Tang}, {Bressan}, {Rosenfield}, {Slemer}, {Marigo}, {Girardi}, \& {Bianchi}}]{tang_2014}
{Tang}, J., {Bressan}, A., {Rosenfield}, P., {et~al.} 2014, \mnras, 445, 4287, \dodoi{10.1093/mnras/stu2029}

\bibitem[{{Taylor} {et~al.}(1998){Taylor}, {Kobulnicky}, \& {Skillman}}]{Taylor_1998}
{Taylor}, C.~L., {Kobulnicky}, H.~A., \& {Skillman}, E.~D. 1998, \aj, 116, 2746, \dodoi{10.1086/300655}

\bibitem[{{Teyssier} {et~al.}(2012){Teyssier}, {Johnston}, \& {Kuhlen}}]{Teyssier_2012}
{Teyssier}, M., {Johnston}, K.~V., \& {Kuhlen}, M. 2012, \mnras, 426, 1808, \dodoi{10.1111/j.1365-2966.2012.21793.x}

\bibitem[{{Tody}(1986)}]{Tody_1986}
{Tody}, D. 1986, in Society of Photo-Optical Instrumentation Engineers (SPIE) Conference Series, Vol. 627, Instrumentation in astronomy VI, ed. D.~L. {Crawford}, 733, \dodoi{10.1117/12.968154}

\bibitem[{{Wakelam} {et~al.}(2017){Wakelam}, {Bron}, {Cazaux}, {Dulieu}, {Gry}, {Guillard}, {Habart}, {Hornek{\ae}r}, {Morisset}, {Nyman}, {Pirronello}, {Price}, {Valdivia}, {Vidali}, \& {Watanabe}}]{wakelan_2017}
{Wakelam}, V., {Bron}, E., {Cazaux}, S., {et~al.} 2017, Molecular Astrophysics, 9, 1, \dodoi{10.1016/j.molap.2017.11.001}

\bibitem[{{Wang} {et~al.}(2022){Wang}, {Gao}, {Ren}, \& {Chen}}]{Wang_2022}
{Wang}, Y., {Gao}, J., {Ren}, Y., \& {Chen}, B. 2022, \apjs, 260, 41, \dodoi{10.3847/1538-4365/ac63c1}

\bibitem[{{Weisz} {et~al.}(2023){Weisz}, {McQuinn}, {Savino}, {Kallivayalil}, {Anderson}, {Boyer}, {Correnti}, {Geha}, {Dolphin}, {Sandstrom}, {Cole}, {Williams}, {Skillman}, {Cohen}, {Newman}, {Beaton}, {Bressan}, {Bolatto}, {Boylan-Kolchin}, {Brooks}, {Bullock}, {Conroy}, {Cooper}, {Dalcanton}, {Dotter}, {Fritz}, {Garling}, {Gennaro}, {Gilbert}, {Girardi}, {Johnson}, {Johnson}, {Kalirai}, {Kirby}, {Lang}, {Marigo}, {Richstein}, {Schlafly}, {Schmidt}, {Tollerud}, {Warfield}, \& {Wetzel}}]{Weisz_2023}
{Weisz}, D.~R., {McQuinn}, K. B.~W., {Savino}, A., {et~al.} 2023, \apjs, 268, 15, \dodoi{10.3847/1538-4365/acdcfd}

\bibitem[{{Windhorst} {et~al.}(2011){Windhorst}, {Cohen}, {Hathi}, {McCarthy}, {Ryan}, {Yan}, {Baldry}, {Driver}, {Frogel}, {Hill}, {Kelvin}, {Koekemoer}, {Mechtley}, {O'Connell}, {Robotham}, {Rutkowski}, {Seibert}, {Straughn}, {Tuffs}, {Balick}, {Bond}, {Bushouse}, {Calzetti}, {Crockett}, {Disney}, {Dopita}, {Hall}, {Holtzman}, {Kaviraj}, {Kimble}, {MacKenty}, {Mutchler}, {Paresce}, {Saha}, {Silk}, {Trauger}, {Walker}, {Whitmore}, \& {Young}}]{Windhorst_2011}
{Windhorst}, R.~A., {Cohen}, S.~H., {Hathi}, N.~P., {et~al.} 2011, \apjs, 193, 27, \dodoi{10.1088/0067-0049/193/2/27}

\bibitem[{{Windhorst} {et~al.}(2022){Windhorst}, {Carleton}, {O'Brien}, {Cohen}, {Carter}, {Jansen}, {Tompkins}, {Arendt}, {Caddy}, {Grogin}, {Koekemoer}, {MacKenty}, {Casertano}, {Davies}, {Driver}, {Dwek}, {Kashlinsky}, {Kenyon}, {Miles}, {Pirzkal}, {Robotham}, {Ryan}, {Abate}, {Andras-Letanovszky}, {Berkheimer}, {Chambers}, {Gelb}, {Goisman}, {Henningsen}, {Huckabee}, {Kramer}, {Patel}, {Pawnikar}, {Pringle}, {Rogers}, {Sherman}, {Swirbul}, \& {Webber}}]{Windhorst_2022}
{Windhorst}, R.~A., {Carleton}, T., {O'Brien}, R., {et~al.} 2022, \aj, 164, 141, \dodoi{10.3847/1538-3881/ac82af}

\bibitem[{{Wolfire} {et~al.}(2010){Wolfire}, {Hollenbach}, \& {McKee}}]{Wolfire_2010}
{Wolfire}, M.~G., {Hollenbach}, D., \& {McKee}, C.~F. 2010, \apj, 716, 1191, \dodoi{10.1088/0004-637X/716/2/1191}

\bibitem[{Xu {et~al.}(2020)Xu, Ge, Tebbutt, Tarek, Trapp, \& Ghahramani}]{xu_2020}
Xu, K., Ge, H., Tebbutt, W., {et~al.} 2020, in Symposium on Advances in Approximate Bayesian Inference, PMLR, 1--10

\bibitem[{{Zhang} {et~al.}(2012){Zhang}, {Hunter}, {Elmegreen}, {Gao}, \& {Schruba}}]{Zhang_2012}
{Zhang}, H.-X., {Hunter}, D.~A., {Elmegreen}, B.~G., {Gao}, Y., \& {Schruba}, A. 2012, \aj, 143, 47, \dodoi{10.1088/0004-6256/143/2/47}

\end{thebibliography}
\bibliographystyle{aasjournal}



\end{document}